%% file: Monster_BL_LacV0.tex
\documentclass[longauth]{aa} 

\usepackage{graphicx}
\usepackage{txfonts}
\usepackage{hyperref}

\hypersetup{colorlinks=true,linkcolor=blue,filecolor=blue,urlcolor=blue,citecolor=blue}
\bibpunct{(}{)}{;}{a}{}{,} 

\usepackage{arydshln}
\usepackage{xcolor}
\usepackage{multirow}
\usepackage[switch]{lineno}

\makeatletter
\renewcommand*\aa@pageof{, page \thepage{} of \pageref*{LastPage}}
\makeatother

\begin{document}

\title{Testing two-component models on very-high-energy gamma-ray emitting BL~Lac objects}
\titlerunning{Two-component models for VHE BL Lacs}


\input{authors.tex}

\date{Received XXX XX, 2019; accepted XXX XX, 20XX}
 
\abstract
{It has become evident that one-zone synchrotron self-Compton (SSC) models are not always adequate for very-high-energy (VHE) gamma-ray emitting blazars. While two-component models are performing better, they are difficult to constrain due to the large number of free parameters.}
{In this work, we make a first attempt to take into account the observational constraints from Very Long Baseline Interferometry (VLBI) data, long-term light curves (radio, optical, and X-rays) and optical polarisation to limit the parameter space for a two-component model and test if it can still reproduce the observed spectral energy distribution (SED) of the blazars.}
{We selected five TeV BL Lac objects based on the availability of VHE gamma-ray and optical polarisation data. We collected constraints for the jet parameters from VLBI observations. We evaluated the contributions of the two components to the optical flux by means of decomposition of long-term radio and optical light curves as well as modelling of the optical polarisation variability of the objects. We selected eight epochs for these five objects, based on the variability observed at VHE gamma rays, for which we constructed the SEDs that we then modelled with a two-component model.}
{We found parameter sets which can reproduce the broadband SED of the sources in the framework of two-component models considering all available observational constraints from VLBI observations. Moreover, the constraints obtained from the long-term behaviour of the sources in the lower energy bands could be used to determine {the region where the emission in each band originates.} 
Finally, {we attempted to use} optical polarisation data to shed new light on the behaviour of the {two} components in the optical band. 
{Our observationally constrained two zone model allows explanation of the entire SED from radio to VHE with two co-located emission regions.}}
{}

\maketitle
%
\section{Introduction}
{Blazars are a subclass of active galactic nuclei, with their jet axes oriented close to the observer’s line of sight. They are divided into two sub classes,} flat-spectrum radio quasars (FSRQs) and BL Lac objects (BL Lacs), which are thought to be intrinsically different. FSRQs show broad emission lines in their optical spectra while BL Lacs have featureless spectra with weak or no emission lines \citep{1991ApJS...76..813S,1991ApJ...374..431S}. The spectral energy distribution (SED) of blazars exhibit a generic two-bump structure: one peak with a maximum in the spectral range from radio to X-rays and a second one in the interval from X-rays to gamma rays. The radiation is produced in a highly-beamed plasma jet and the double-peaked SED is often explained by a single population of relativistic electrons. The low-energy SED bump is thought to arise from synchrotron emission of particles within the magnetic field of the jet. The origin of the high-energy SED bump is less certain. It is commonly attributed to inverse Compton (IC) scattering of low-energy photons \citep{1967MNRAS.135..345R}. The low-energy photons can originate externally to the jet \citep[external Compton scattering,][]{1993ApJ...416..458D} or be produced within the jet via synchrotron radiation \citep[synchrotron self-Compton scattering, SSC,][]{1981ApJ...243..700K, 1992ApJ...397L...5M}. As there are no observational evidence for strong external photon fields present in BL Lacs, the main population of seed photons for Compton scattering should originate from the synchrotron emission. As a confirmation of this hypothesis, most of the SEDs of BL Lacs are well described with a simple one-zone SSC model {\citep{1996ApJ...461..657B, 1998ApJ...509..608T, 2019ApJ...874...47V}}. An alternative framework to explain the high-energy emission is the acceleration of hadrons, along with leptons \citep{1989A&A...221..211M}. In the following we will focus on leptonic models.

{Blazars} are classified according to the frequency of the first peak of their SED into low- (LSP, $\nu_{syn}<10^{14}$ Hz), intermediate- (ISP, $10^{14}\leq\nu_{syn}<10^{15}$ Hz), and high- (HSP, $\nu_{syn}\geq10^{15}$ Hz) synchrotron-peaked {objects} \citep{0004-637X-716-1-30}. Within the very-high-energy (VHE; $>100$\,GeV) gamma-ray emitting extragalactic objects, the most numerous sources are the HSP BL Lacs. With a large number of multi-wavelength (MWL) campaigns performed {since the launch of the \textit{Fermi Gamma-Ray Space Telescope} (\textit{Fermi})}, there is growing evidence that a one-zone SSC model, typically with a single spherical blob dominating the emission from optical to VHE gamma rays, is too simple to describe the SEDs of these objects \citep[e.g.][]{2017A&A...603A..31A}. Two component models, such as the spine-layer model by \citet{2005A&A...432..401G}, have gained popularity. Two-component models, however, require a {larger} number of free parameters (twice as {many} as in single-zone ones) and therefore they often end up with a large degeneracy for the parameters involved \citep[see e.g.][]{2014MNRAS.441.2885B}. Also the nature and the location within the jet of these two components is still unclear. 

One way to constrain the two-component model is to derive the contribution of the different components from long-term variability. \citet{2014A&A...567A.135A} found a common {increasing} trend in radio and optical light curves of PKS~1424+240 and used this to constrain the contribution of the two components to the optical part of the SED. \citet{2016A&A...593A..98L} found a similar increasing/decreasing trend in {radio and optical light curves of} additional 12  sources, when analysing radio and optical light curves of 32 northern-sky BL Lacs. The authors argued that as the radio variability very closely traces the variability of the core flux in Very Large Baseline Interferometry (VLBI) images, also the slowly varying optical flux originates from the core. The fast varying component of the optical flux could instead originate from a distinct, smaller emission region. In this work we have selected a sub-sample of five of the BL Lacs from \citet{2016A&A...593A..98L}, based on the availability of MWL data (VER~J0521+211, PKS~1424+240, 1ES~1727+502, 1ES~1959+650, 1ES~2344+514) with the aim of placing observational constraints on two-component SSC models. Independent from these common trends on long-term light curves, we also use optical polarisation data to disentangle the contribution of the two components and we take into account constraints on jet parameters from the VLBI observations. 

The paper is organised as follows: the observations, analysis methods and wavelength-specific results of our sub-sample are described in Section \ref{sec:obs}. The observational constraints for SED modelling from VLBI data, MWL light curves and optical polarisation observations are derived in Section \ref{sec:constraints}. In Section \ref{sec:sed} the SED modelling of all five sources are described. Section~\ref{sec:discuss} includes the discussions of the results of the SED modelling. Finally, in Section~\ref{sec:sum} we present the summary and conclusions of the main results of the paper. 

\section{\label{sec:obs}Observations, data analysis and results}

The general properties of our sample are listed in Table \ref{tab_general}. Power law (PL) and log-parabola (LP) are the two mathematical functions which are employed for our spectral analysis in different bands. They are defined as follows: \\
A simple power law
\begin{equation}
    \frac{dF}{dE}(E) = F_0 \bigg(\frac{E}{E_0}\bigg)^{-\Gamma},
\end{equation}
and a log-parabola
\begin{equation}
    \frac{dF}{dE} (E) = F_0 \bigg(\frac{E}{E_0}\bigg)^{-\Gamma-\beta(\log_{10}(E/E_0))},
\end{equation} 
where $dF/dE$ is the differential flux as a function of the energy $E$. $F_0$, $\Gamma$, and $\beta$ are the flux at the normalisation energy  $E_0$, the spectral index, and the curvature parameter of the spectrum at $E_0$, respectively.

\input{tables/tab_general.tex}

\subsection{\label{magic}Very-high-energy gamma rays (MAGIC)}
The Major Atmospheric Gamma-ray Imaging Cherenkov experiment \citep[MAGIC, ][]{2016APh....72...76A} is a system of two, 17-m diameter telescopes located at the Observatorio del Roque de los Muchachos (ORM), La Palma, Canary islands, Spain. The objects of our sample were observed by MAGIC between 2013  and 2016 as part of different observation campaigns (see Table~\ref{tab_magic_flux} for a detailed list of included epochs for each source). The data have been analysed using the MAGIC Standard Analysis Software \citep[MARS,][]{2009arXiv0907.0943M, zanin13} taking into account the instrument performance under different observation conditions \citep{2016APh....72...76A, 2017MNRAS.468.1534A}.

We calculated the VHE gamma-ray integral flux of each object and searched for variability at different timescales (from 10 minutes to a week). The constant-flux hypothesis on 1-day timescale is rejected at the 3-$\sigma$ confidence level for 1ES~1727+502, 1ES~1959+650 and 1ES~2344+514 (see below). For PKS~1424+240, no variability was found during 2014 (MJD 56740-56826) and 2015 (MJD 57045-57187) campaigns. However, the VHE gamma-ray flux of the 2015 campaign was $\sim60\%$ of the one observed during the 2014 campaign. Therefore, the data are divided into the 2014 and 2015 campaigns. In the case of VER~J0521+211, we do not find any significant variability {during 4 nights of MAGIC observation in 2013 \citep{2015ICRC...34..864P}}. 

For 1ES~1727+502, there is one night (MJD 57309, 2015 October 14) when the VHE gamma-ray flux was 52\% of the average flux. The VHE gamma-ray flux during MJD 57309 was $3.3\sigma$ away from the average flux computed using all of the five nights of observation. However, the VHE gamma-ray spectrum could not be computed using the observations of this single night. Therefore, we reproduced the VHE gamma-ray spectrum of this source using all available observations. Exclusion/inclusion of the observation on MJD 57309 did not affect the parameters describing the VHE gamma-ray spectrum.

1ES~1959+650 was in a flaring state during 2016. We selected three different nights based on the level of VHE gamma-ray flux of the source during 2016 and availability of the simultaneous MWL observations at lower energy bands. The highest, intermediate and lowest VHE gamma-ray flux was observed on 2016 June 14 (MJD 57553), June 8 (MJD 57547), and November 20 (MJD 57711), respectively. No intra-night variability was detected in the data of these selected observations. This is in agreement with the results reported by \citet{1959flare}, where a detailed variability analysis was performed on three nights of the highest detected fluxes (including MJD 57553) during the 2016 campaign and intra-night variability (with a timescale of 35 minutes) was found on the nights of 2016 June 13  and 2016 July 1 (MJD 57552 and 57570; see Table~3 in \citealp{1959flare}). 

1ES~2344+514 showed variability on daily timescale but no shorter variability timescale was detected in the VHE gamma-ray band. \citet{2344fact} performed a detailed analysis on different emission states of this source and found the spectral shape to be similar during different observational epochs despite different levels of the VHE gamma-ray flux. The results of the VHE gamma-ray flux study of our sample are summarised in Table~\ref{tab_magic_flux}. The derived variability timescales are further used in Section~\ref{sec:sed}. 

The VHE gamma-ray spectra are computed for each source and epoch separately in case the source showed variability. The effect of the extragalactic background light (EBL) to VHE gamma-ray spectra was taken into account by using the model of \citet{2011MNRAS.410.2556D}. Then, two different models (PL and LP) were tested. The LP model was preferred over the PL model at $3\sigma$ confidence level if the F-test probability value was less than $0.27\%$.
The results of the spectral analysis in the VHE gamma-ray band are summarised in Table~\ref{tab_magic_spec}. 
 
\input{tables/tab_magic_flux.tex}

\input{tables/tab_magic_spec.tex}

\subsection{High-energy gamma rays (\textit{Fermi}-LAT)}

The Large Area Telescope (LAT), based on the pair conversion technique, is the high-energy instrument on-board the \textit{Fermi}. It {has continuously monitored} the high-energy (HE, 100\,MeV $<E<300$\,GeV) gamma-ray sky \citep{2009ApJ...697.1071A} since its launch in 2008. The 6-year, MJD 56200 (2012 September 4) to 58340 (2018 August 9), light curve for each source was obtained by applying a weekly binning to the events collected by the LAT with an energy higher than 100\,MeV over a region of interest of $10\degr$ centred on the selected sources. {Time intervals coinciding with bright solar flares and gamma}-ray burst were excised from the data set as it is done in the fourth \textit{Fermi}-LAT source catalogue \citep[4FGL,][]{2019arXiv190210045T}. The data reduction and analysis of the events belonging to the Pass8 source class was performed with the FermiTools software package version 11-07-00 and fermipy \citep{2017ICRC...35..824W} version 0.17.4. To reduce the Earth limb contamination, a zenith angle cut of $90\degr$ was applied to the data. To calculate the weekly flux of the selected sources, a likelihood fit to the data was performed including 
{each source of interest, modelled with a power-law spectrum}, the Galactic diffuse-emission model\footnote{\url{https://fermi.gsfc.nasa.gov/ssc/data/analysis/software/aux/4fgl/Galactic\_Diffuse\_Emission\_Model\_for\_the\_4FGL\_Catalog\_Analysis.pdf}} (gll\_iem\_v07.fits), and isotropic component (iso\_P8R3\_SOURCE\_V2\_v1.txt) recommended for the Pass8 Source event class as well as the sources of the \textit{Fermi}-LAT 4FGL within $15\degr$ from the position of the source of interest. The normalisation of both diffuse components in the source model were allowed to vary during the spectral fitting procedure. The normalisation were allowed to vary for the sources located within a distance smaller than $2\degr$ from the source of interest and with a detection test statistics (TS\footnote{The square root of the TS is approximately equal to the detection significance for a given source.}) higher than 50 integrated over the full data set. The sources located at the distance between $2\degr$ and $7\degr$ had their normalisation set as a free parameter if their variability index was higher than 18.48\footnote{The level of 18.48 was chosen according to the 4FGL catalogue.}. {The spectral indices of all the sources with free normalisation were left as free parameter if the source showed a TS value higher than 25 over an integration time of one week, in all the other cases the indexes where frozen to the value obtained in the overall fit \footnote{We performed this check using the "shape\_ts\_threshold" option in the fermipy light curve tool.}.}  We apply the correction for the energy dispersion to all sources except for the isotropic background. The {HE light curves} are shown in Figures \ref{0521mwlfig} to \ref{2344mwlfig}. The spectrum was obtained only analysing data collected over the selected epochs, which were \mbox{(quasi-)simultaneous} to MAGIC data and had sufficient statistics to compute at least 2 spectral data points per decade in energy range between 100\,MeV and 300\,GeV (Appendix~\ref{app:sedepoch}, Tab.~\ref{sedepochs}). In all of the cases, the LP model can describe the spectra of the sources better than PL model at $4\sigma$ confidence level, except for 1ES~1727+502 where {the LP model was not statistically preferred over a PL model}. These findings are inline with the results reported in {the} 4FGL catalogue. Moreover, {except for PKS~1424+240 that showed a harder spectrum during the 2015 campaign, the spectral parameters were in agreement to those reported in the 4FGL at $3\sigma$ confidence level.}

\subsection{\label{sec:xray}X-ray and UV (\textit{Swift})}
\begin{figure}
\includegraphics[width=9.cm, height=18cm]{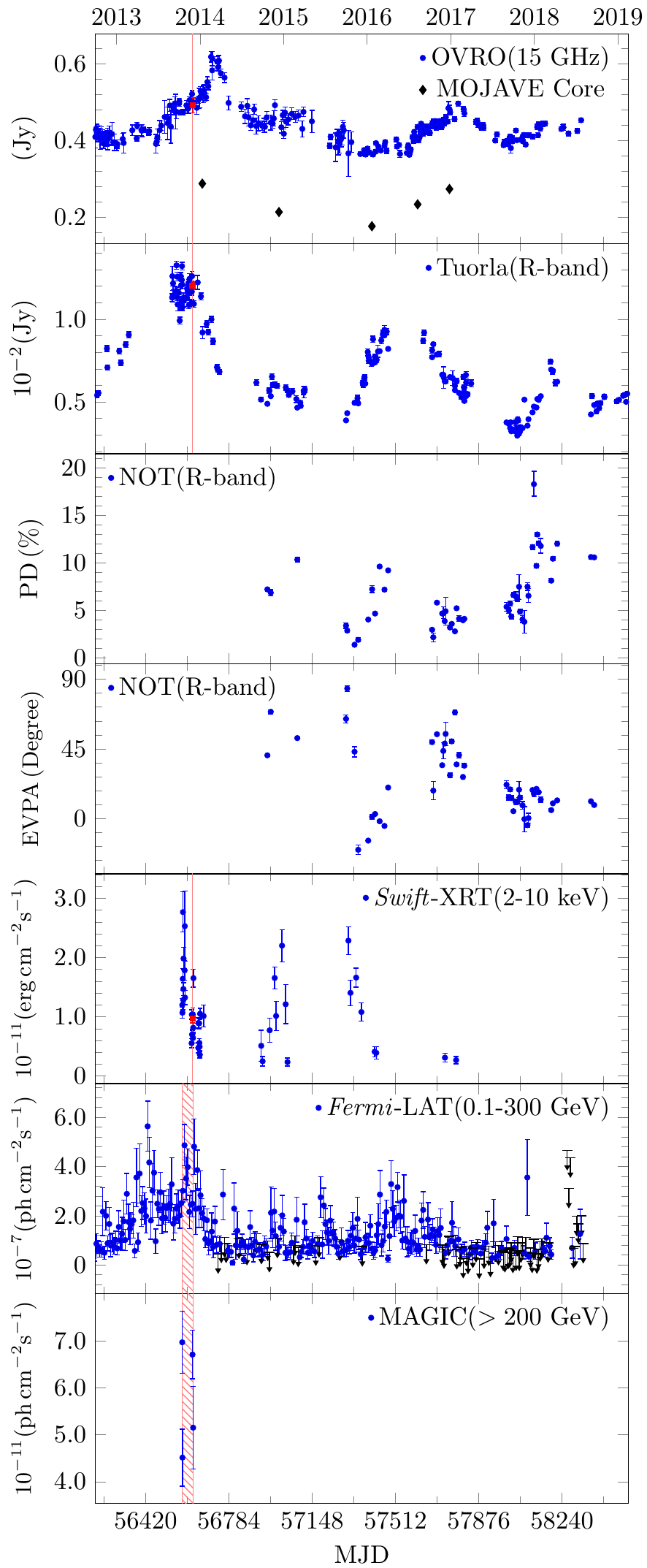}
\caption{\label{0521mwlfig}MWL light curves of VER~J0521+211 in the range from MJD 56200 (2012 September 30) to 58400 (2018 October 9). \textit{From top to bottom panels}: Radio and VLBI core flux (15\,GHz), optical \mbox{(R-band)}, optical polarisation degree, electric vector polarisation angle, X-ray {flux} (2-10\,keV), HE gamma-ray {photon flux} (0.1-300\,GeV), and VHE gamma-ray {photon flux above the threshold energy given in the panel}. Black arrows show the 95\% confidence level upper limits. The data, {which are marked with vertical lines/area and squares in different bands, are used in the SED modelling.}}
\end{figure}

\begin{figure}
\includegraphics[width=9.cm, height=18cm]{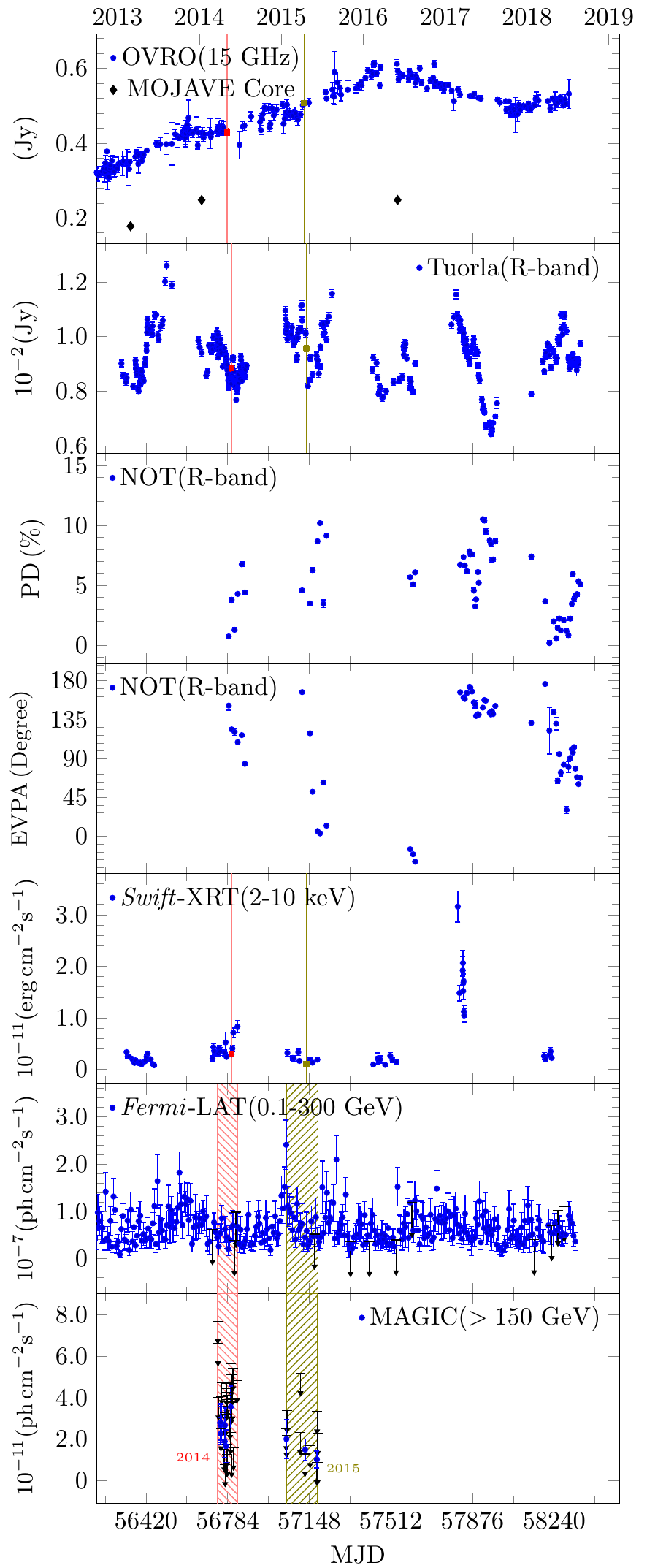}
\caption{\label{1424mwlfig}{Same description as in Figure~\ref{0521mwlfig} for PKS~1424+240.}
}
\end{figure}

\begin{figure}
\includegraphics[width=9.cm, height=18cm]{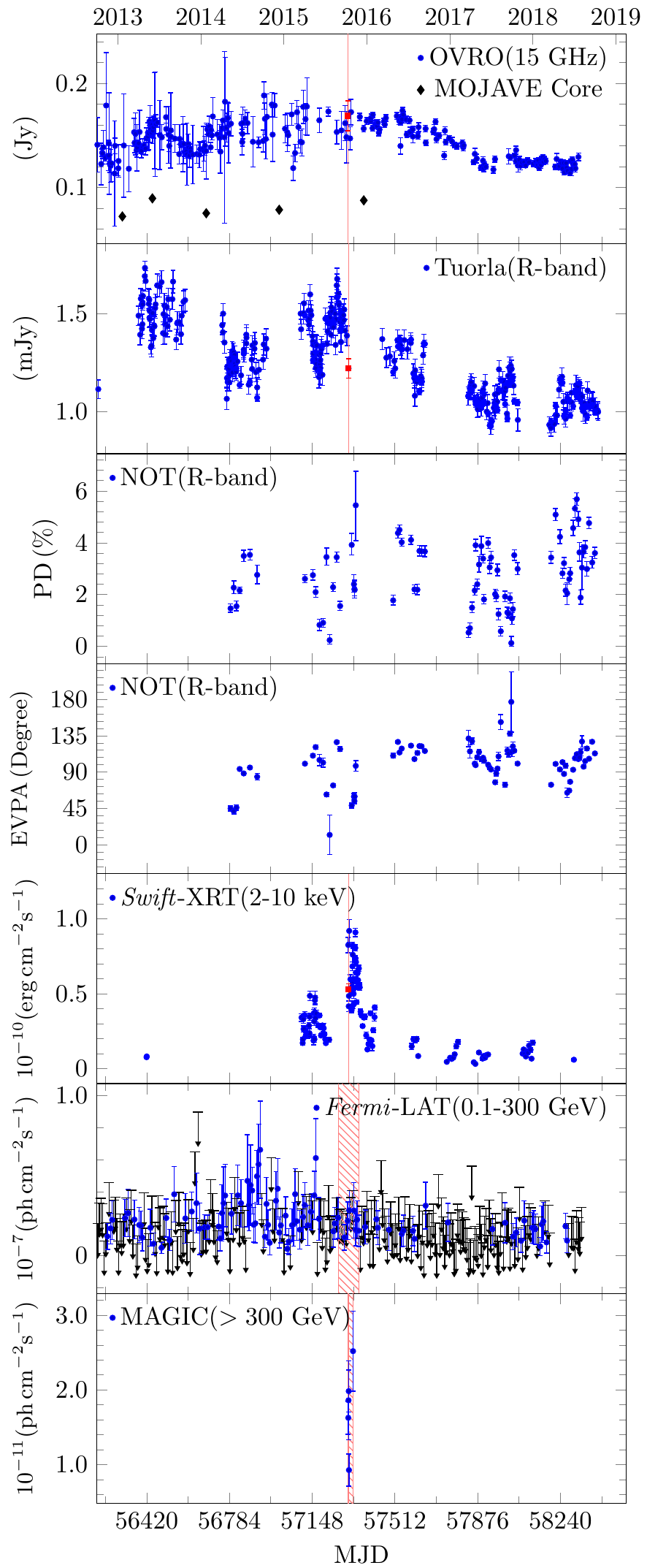}
\caption{\label{1727mwlfig}{Same description as in Figure~\ref{0521mwlfig} for 1ES~1727+502.}
}
\end{figure}

\begin{figure}
\includegraphics[width=9.cm, height=18cm]{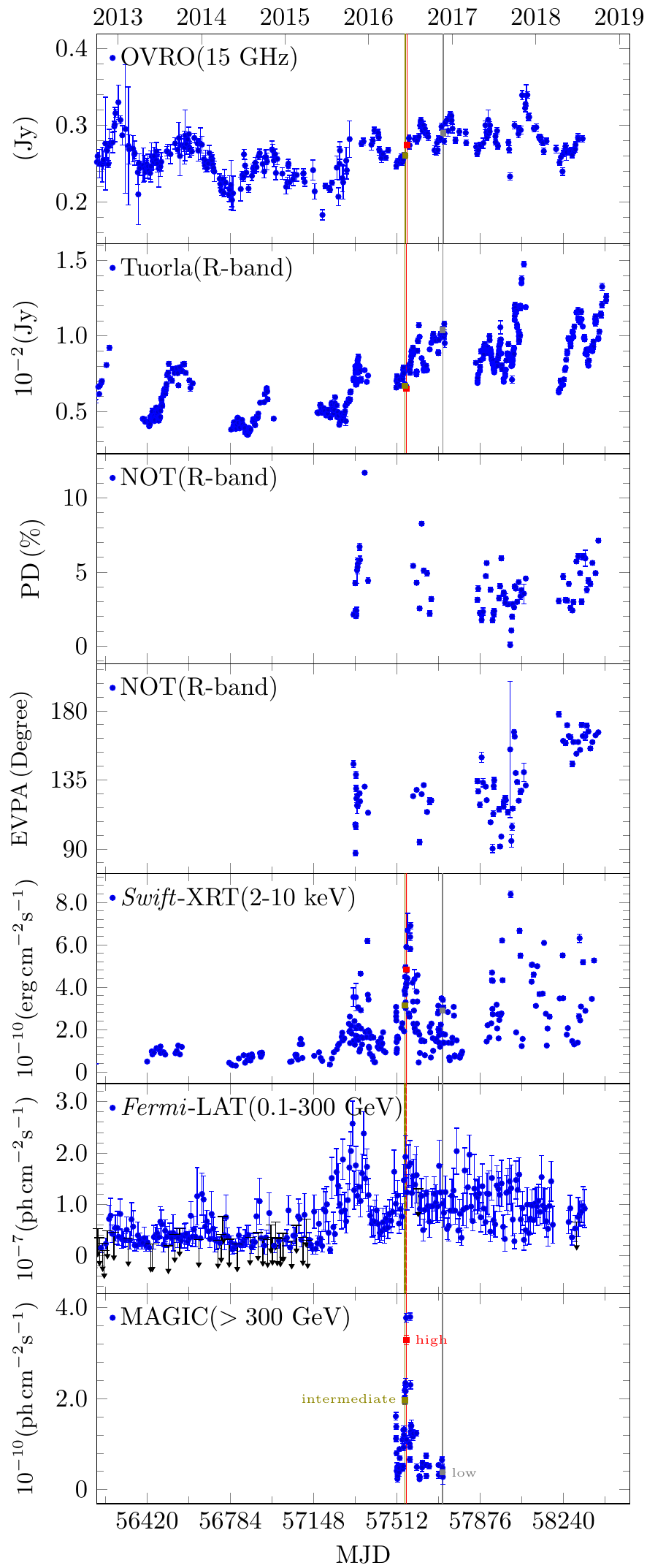}
\caption{\label{1959mwlfig}{Same description as in Figure~\ref{0521mwlfig} for 1ES~1959+650.}
}
\end{figure}

\begin{figure}[ht!]
\includegraphics[width=9.cm, height=18cm]{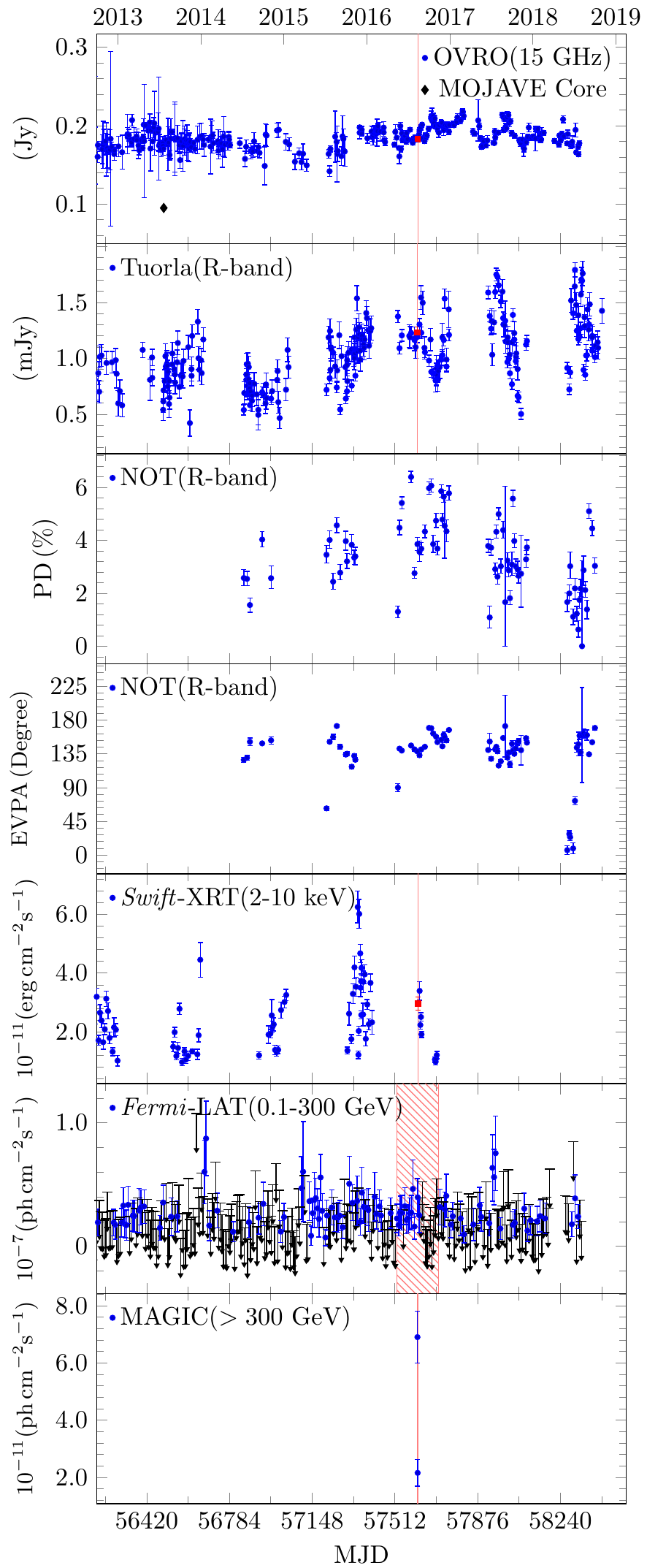}
\caption{\label{2344mwlfig}{Same description as in Figure~\ref{0521mwlfig} for 1ES~2344+514.}
}
\end{figure}

The X-ray Telescope \citep[XRT,][]{2004SPIE.5165..201B} on-board the \textit{Neil Gehrels Swift observatory (Swift)} has been observing the sources in the sample since 2004 in both photon-counting (PC) and window-timing (WT) modes. The multi-epoch event lists for the period from 2012 September 30 to 2018 October 9 were downloaded from the publicly available SWIFTXRLOG (\textit{Swift}-XRT Instrument Log)\footnote{\url{https://heasarc.gsfc.nasa.gov/W3Browse/swift/swiftxrlog.html}}. {Following the standard \textit{Swift}-XRT analysis procedure described by \citet{2009MNRAS.397.1177E},} the PC observation data were processed using the configuration described by \citet{2017A&A...608A..68F} for blazars. For the WT observations data, we defined the source region as a box with a length of 40 pixels centred on the source position and aligned to the telescope roll angle. The background region is defined by a box with a length of 40 pixel aligned to the telescope roll angle and 100 pixel away from the centre of the source.  
For both modes of observation, due to the open issues for analysing the \textit{Swift}-XRT data\footnote{These open issues mostly affect the data obtained with the WT mode. However, some of them (charge Traps) still can affect the spectra observed during PC mode. More details are available at:\\ \url{http://www.swift.ac.uk/analysis/xrt/digest_cal.php} and\\ \url{http://www.swift.ac.uk/analysis/xrt/rmfs.php}}, we fitted the spectra of each observation {in the 0.3-10 keV energy range} assuming all possible combinations of pixel-clipping and point-spread-function together with two mathematical models (i.e. PL and LP), a normalisation energy $E_0=0.3$\,keV, and the fixed equivalent Galactic hydrogen column density reported by \citet{2005A&A...440..775K} and listed in Table~\ref{tab_general}. In total, for each XRT observation 6 and 16 spectra (for PC and WT modes, respectively) were extracted and the best-fitted model was selected using least $\chi^2$ and F-test methods. The results of this analysis are partially (only X-ray flux in range of 2-10\,keV) presented in Figures~\ref{0521mwlfig} to \ref{2344mwlfig} for each source. An example of full version of the results is presented in Table~\ref{tab:xray}, while the complete version of the results is available online\footnote{The complete version is available online at: a link to CDS}. All sources are variable in the X-ray band in the studied time period. 

During the \textit{Swift} pointings, the UVOT instrument observed the sources in our sample in its optical (V, B, and U) and UV (W1, M2, and W2) photometric bands \citep{2008MNRAS.383..627P,2010MNRAS.406.1687B}. We selected the UVOT data \mbox{(quasi-)simultaneous} to the MAGIC observations and analysed the data using the \texttt{uvotsource} task included in the \texttt{HEAsoft} package (v6.25)\footnote{\url{https://heasarc.gsfc.nasa.gov/docs/software/heasoft/}}. Source counts were extracted from a circular region of $5\arcsec$ radius centred on the source, while background counts were derived from a circular region of $20\arcsec$ radius in a nearby source-free region. The contribution of the host-galaxy flux in the UVOT bands are derived from the R-band values reported by \citet{2007A&A...475..199N} and the conversion factors reported by \citet{1995PASP..107..945F}. The host-galaxy subtracted (when applicable) UVOT flux densities, corrected for extinction using the E(B--V) values from \citet{2011ApJ...737..103S} and the extinction laws from \citet{1989ApJ...345..245C}, are used in the SED modelling (Sec.~\ref{sec:sed}). 

\subsection{\label{sec:opt_radio}Optical and Radio ({Tuorla, OVRO, and MOJAVE})}

The optical (R-band) data from MJD 56200 (2012 September 30) to 58320 (2018 July 21) obtained in the framework of the Tuorla blazar monitoring programme\footnote{\url{http://users.utu.fi/kani/1m/}} using the 50-cm Searchlight Observatory Network telescope (San Pedro de Atacama, Chile), the 40-cm Searchlight Observatory Network telescope (New Mexico, USA), and the 60-cm telescope at Belogradchik (Bulgaria) in addition to the Kungliga Vetenskapsakademien (KVA) telescope (ORM{, La Palma, Canary islands, Spain}). Most of the observations have been performed with the KVA telescope. The data are analysed and calibrated using the method described by \citet{2018A&A...620A.185N}. The data are corrected for Galactic extinction and host galaxy emission for aperture photometry. The correction coefficients and the aperture used for each individual source are summarised in Table~\ref{tab_general}. The results of these observations are presented in Figures~\ref{0521mwlfig} to \ref{2344mwlfig}. An example of the results is presented in Table~\ref{tab:tuorla}, while the complete version of the results is available online\footnote{The complete version is available online at: a link to CDS}.

We have used the long-term light curves from the Owens Valley Radio Telescope (OVRO, 15\,GHz). This programme, the observations, and the analysis methods are described in \citet{2011ApJS..194...29R}. In this work we have included the data from the time interval MJD 56200-58320 (2012 September 30 -- 2018 July 21). We note that the light curves cover data from the period between 2015 August 1 and November 24 when the instrument was not working optimally. Therefore, the noise in the data is higher during this period. {Moreover, we collected the core flux at 15\,GHz using the data from the Monitoring Of Jets in Active galactic nuclei with VLBA Experiments (MOJAVE) programme \citep{2019ApJ...874...43L}.}

\subsection{\label{poldata}Optical polarisation (NOT)}

The sources have been monitored with the Nordic Optical Telescope (NOT). The ALFOSC\footnote{\url{http://www.not.iac.es/instruments/alfosc}} instrument was used in the standard setup for linear-polarisation observations ($\lambda/2$ retarder followed by a calcite plate). The observations were performed in the R-band between 2014 and 2018 two to four times per month. The data were analysed as described by \citet{2016A&A...596A..78H} and \citet{2018MNRAS.480..879M} with a semi-automatic pipeline using standard aperture photometry procedures. The data were corrected for Galactic extinction and host-galaxy emission using the values listed in Table \ref{tab_general}. The results of these observations are presented in Figures~\ref{0521mwlfig} to \ref{2344mwlfig}.

\section{\label{sec:constraints}Observational constraints for two-component emission modelling}

Two-component models are models where two emission regions are responsible for the observed radiation. There is some evidence showing that there is a correlation between the X-ray and VHE gamma-ray bands in HSP BL Lacs \citep{2011ApJ...738...25A, 2015A&A...576A.126A}. This suggests that the observable emission in these two wavebands originates from the same region. The second component is then the one we see dominating in the radio band. In the optical band, we presumably see a mixture of these two components. In this section we use the radio-to-X-ray data to obtain constraints for these two emission regions to be used in the SED modelling.

As discussed in the introduction, \citet{2016A&A...593A..98L} found common trends in the long-term optical and radio variability for all five sources of our sample. They also showed that the brightness of the core in the 15\,GHz VLBI images (hereafter VLBI core) closely follows the 15\,GHz  light  curve,  as  had been  previously  found  at higher frequencies \citep[37 and 43\,GHz][]{2002A&A...394..851S}, and therefore suggested that the common slowly varying radio-optical emission region is located at the VLBI core. We therefore collect the observational constraints on the jet parameters from VLBI observations to be used directly in the SED modelling (Sec.~\ref{sec:sed_2comp}). 

On top of the slow variability, the optical band also shows a fast variability, which could originate from a second emission component. For simplicity, we assumed that this component is the one dominating the X-ray and VHE gamma-ray emission. In order to constrain the contribution of these two emission regions to the optical flux, we use the long-term light curves and optical polarisation data described in Section \ref{sec:obs} by implementing two independent procedures described in Sections~\ref{sec:longterm} and \ref{sec:pol}. We also searched for the correlations between different long-term light curves to determine if the same region produces the emission observed at different wavelengths.


\subsection{\label{sec:VLBA}Constraints on jet parameters from VLBI}

The arguably strongest observational evidence for two-component models comes from VLBI observations. \citet{attridge99} discovered in polarimetric VLBA observations a clear difference in the polarisation direction of the inner jet and outer layer of FSRQ 1055+018 and similar polarisation structures have been observed in several sources after that \citep{pushkarev05, gabuzda14}. Another indication of a spine-sheath structure of the jets is the so-called limb brightening, where the edges of the jet appear brighter than the central spine which has been observed in several radio galaxies and blazars \citep{giroletti04, nagai14, gabuzda14}. In particular, such limb brightening has been observed in Mrk~501 \citep{piner09} which is a source rather similar to the sources in our sample (in terms of VLBI jet properties and synchrotron peak frequency).

The sources in our sample are rather weak in the radio band and therefore potential spine-sheath structures would be impossible to resolve. Their VLBI images all show compact jets in which the core is the brightest component. The core fluxes follow the total intensity variations observed at 15\,GHz (see Figs.~\ref{0521mwlfig}-\ref{1727mwlfig}, the two other sources had no or only one simultaneous core flux measurement), which agrees with the results found in larger samples, that the radio emitting component is located at the 15\,GHz core. We used VLBI data to constrain some of the jet parameters: the apparent speed of the jet, the size of the VLBI core, the jet position angle and the core polarisation properties. The jet velocities and the size of the core are used directly in the SED modelling. The jet position angle, and polarisation properties are only used for comparison with the optical polarisation data in Section~\ref{sec:pol}. These were collected from \citet{2018ApJ...862..151H}. They report an uncertainty in the fractional polarisation to be approximately 7\% of the given values and the electric vector polarisation angle (EVPA) is accurate within 5$^\circ$, while no error is given for the jet position angle.
 
The major fraction of the jet parameter constraints are from the MOJAVE programme \citep{2005AJ....130.1389L, 2009AJ....137.3718L, 2016AJ....152...12L, 2018ApJ...862..151H, 2019ApJ...874...43L}, during which observations have been performed at 15\,GHz. All of the sources in our sample have been observed as part of this programme. However, not all of the collected data were obtained between 2013 and 2018, and most of the sources have been observed only few times. Additionally, we collected the results reported by \citet{2004ApJ...600..115P, 2018ApJ...853...68P, 2008ApJ...678...64P, 2010ApJ...723.1150P, 2012arXiv1205.2399T}. 

VER~J0521+211 has been observed seven times in the framework of the MOJAVE programme between 2014 and 2018. \citet{2019ApJ...874...43L} reported several moving components in the jet. The fastest component has a maximum jet speed of $\mu= 774 \pm 45\,\mu$as\,yr$^{-1}$ which corresponds to an apparent projected speed of $\beta_{\text{app}}= 7.72 \pm 0.42$ considering $z=0.18$. Assuming a viewing angle of $3\degr$ and $5\degr$ these give Doppler factors $\delta\sim15$ and $\delta\sim11$. The median fractional polarisation and the EVPA of the core correspond to 0.5\% and $\sim200\degr$ respectively, and any significant variability is ruled out within the observations. The jet position angle is $250\degr$ \citep{2018ApJ...862..151H}.

PKS~1424+240 has been observed three times in 2013-2016 in the framework of the MOJAVE programme. Significant motion was detected for two components. The  fastest  component has  a  maximum  apparent speed of $\beta_{\text{app}}=2.83 \pm 0.89$. The latter corresponds to a Doppler factor of 10 or 7 depending on the assumptions of the viewing angle being $3\degr$ or $5\degr$, respectively. The core is polarised with a median fractional polarisation of 2.3\% and the EVPA lies between $140-150\degr$. The jet position angle is $140\degr$ \citep{2018ApJ...862..151H}.

1ES~1727+502 has been observed five times between 2013 and 2015. \citet{2018ApJ...853...68P} fitted four components to the MOJAVE data of 1ES~1727+502 and they were all consistent with no motion (\citealt{2019ApJ...874...43L} reports  $0.041\pm0.043$ as maximum jet speed). The polarisation of the core is not significant, and therefore the EVPA cannot be derived. The position angle of the jet is $270\degr$ \citep{2018ApJ...862..151H}.

1ES~1959+650 was dropped from the MOJAVE programme in 2009 as it is too compact and weak at 15\,GHz and there is no data from the source between 2013 and 2018. In the earlier data the source showed a polarisation degree of 2.6-4.5\% and its polarisation angle was pretty stable at $144-160\degr$. The position angle of the jet is $140-160\degr$. The apparent speeds are in agreement with no motion \citep{2004ApJ...600..115P, 2010ApJ...723.1150P} which is also confirmed by \citet{2018ApJ...862..151H}  and \citet{2019ApJ...874...43L}.

1ES~2344+514 has been observed 3 times between 2013 and 2018. \citet{2016AJ....152...12L} reported one component with $\beta_{\text{app}}=0.055\pm0.036$ and the most recent measurements are in line with this result \citep[$\beta_{\text{app}}=0.037\pm0.012$,][]{2019ApJ...874...43L}. The polarisation of the core is not significant \citep{2018ApJ...862..151H}, while the jet position angle is $130-145\degr$ \citep{2004ApJ...600..115P}.

Finally, we collected the measured full-width-half-maximum values of the major axis core region to estimate the size of the VLBI core. For each source we selected a MOJAVE epoch at which the core was resolved and if there were several, we selected the one closest to the epoch used for the SED modelling. As discussed, for 1ES~1959+650 the latest MOJAVE observation epoch was in 2009, so 7 years before our SED data, but the values reported for 2000-2009 were all very similar \citep{2019ApJ...874...43L}. We used the measured full-width-half-maximum values of the major axis values 0.08, 0.14, 0.09, 0.09 and 0.13 mas as the diameter of the core emission region for VER~J0521+211, PKS~1424+240, 1ES~1727+502, 1ES~1959+650, and 1ES~2344+514, respectively.

\subsection{\label{sec:longterm}Long-term light curves}

\citet{2016A&A...593A..98L} analysed the long-term radio (15 GHz) and optical (R-band) light curves {of the sources studied in this paper using the data from 2008-2013}. We repeated the same analysis procedure using {data collected} between 2013 and 2018 to investigate whether the results obtained by \citet{2016A&A...593A..98L} are temporary or not and to use these results for the two{-}zone SED modelling, in particular to constrain the contribution of the two components in the optical band.

\subsubsection{\label{sec:component}Common emission component at radio and optical frequencies}
Following \citet{2016A&A...593A..98L}, we analysed the long-term radio and optical light curves to separate the slowly varying component from the optical light curves and estimate its minimum contribution to the optical flux.

\begin{figure*}
\includegraphics[width=0.95\textwidth, height=18cm, angle=-90]{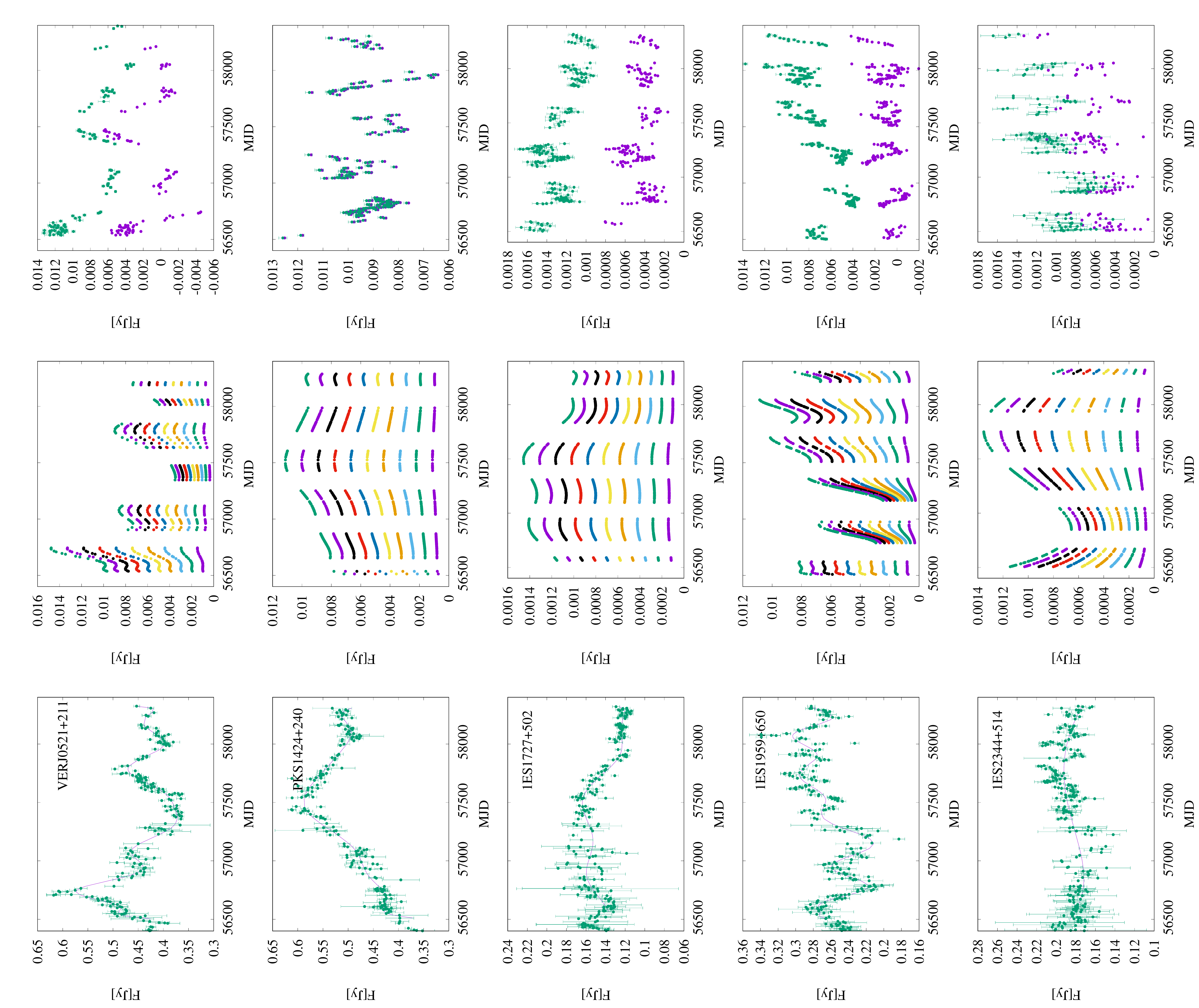}
\caption{\label{polysubtract} The analysis steps of determining the contribution of the common emission component at radio and optical frequencies. \textit{Left:} Fitting a polynomial to the radio light curve. \textit{Middle:} The polynomial fit scaled to the average flux of the optical light curve and multiplied with a scaling factor (different colors correspond to different scaling factors, see text). \textit{Right:} The scaled polynomial is subtracted from the optical data (green filled circles with errors). The RMS of the resulting light curve (purple filled circles) is compared with the RMS of the original data. These analysis steps are shown for all sources from bottom to top: VER~J0521+211, PKS~1424+240, 1ES~1727+502, 1ES~1959+650 and 1ES~2344+514. In the case of PKS~1424+240 subtracting the polynomial did not decrease the RMS of the optical light curve and therefore the purple dots are under the green symbols and not well visible in the right most panel.}
\end{figure*}

In short the analysis steps are:
\begin{enumerate}
    \item Testing if there are linear correlations between time and flux density in radio and optical light curves. The Spearman r-values for optical and radio light curves are reported in Table~\ref{radio-opt}.
    \item Fitting a polynomial to the radio light curve (see Fig.~\ref{polysubtract}, left panel). The polynomial is determined by  adding  new  orders  until  the {$\chi^2$/d.o.f} of the fit does not improve anymore. The resulting polynomial does not describe all of the radio variability, in particular short flares are not described by this polynomial.
    \item The polynomial fit is scaled to the average flux of the optical light curve (see Fig.~\ref{polysubtract}, middle panel). Then it is multiplied with a scaling factor assuming values from 0.1, 0.2, 0.3, ... to 1 and the resulting curve is subtracted from the optical data. The root mean square (RMS){\footnote{calculated around the average flux $\sqrt((\sum(x_i-\mathrm{mean_{flux}})^2/N)$}} of the resulting light curves are calculated and the one that minimises the RMS is selected as best-fit.
    \item To estimate the fractional contribution of this slowly varying component to the optical flux density, we divide the RMS of the best-fit-subtracted data (RMS1) with the RMS of the original data (RMS2) (see Fig.~\ref{polysubtract}, right panel). The contribution of the slowly varying component to the optical {variability} is then 1-RMS1/RMS2 and is given in Table~\ref{radio-opt}. As discussed in \citet{2016A&A...593A..98L}, this fraction represents the minimum contribution as in addition to this slow variation, there can be flares in the radio band that are not reproduced by this polynomial.
    The minimum fraction is then used {to guide} the two-component modelling in the next section. 
\end{enumerate}

We find that, in all of the five sources, the increasing or decreasing trends in the radio light curves have persisted and are in the same direction as found in \citet{2016A&A...593A..98L}. This is interesting, because the time span of the light curves used in these works are different. This means that the increasing or decreasing trends in the radio light curves have persisted for a timescale of $\sim10$ years.

In the optical, significant trends have persisted in four sources out of five, the exception being PKS~1424+240, where there is no significant rising or decaying trend. Accordingly, the minimum fraction of a slowly varying component (common with radio) contributing to the optical flux is zero for this source. For other sources the fraction varies from 6\% to 52\% (see Tab.~\ref{radio-opt}).

\input{tables/tab_radio_opt.tex}

\subsubsection{\label{sec:lccf}Correlation studies}

We calculated the cross-correlation function between three pairs of light curves (radio -- optical, radio -- X-rays, and optical -- X-rays) for each source following the method described by \citet{2014MNRAS.445..428M} and \citet{2016A&A...593A..98L}. We did not include the HE gamma-ray light curves as the uncertainties of the data points are rather large. Moreover, \citet{2018MNRAS.480.5517L} performed a cross-correlation analysis between the radio/optical and HE gamma-ray bands on a sample of 145 blazars which includes four of the sources in our sample (except 1ES~1727+502). They found only one significant correlation ($>3\sigma$ confidence level) between optical and HE gamma-ray bands for VER~J0521+211. The VHE gamma-ray light curves were too sparse to be included in the correlation study. 

We used the Discrete Correlation Function \citep[DCF;][]{1988ApJ...333..646E} with local normalisation \citep[LCCF;][]{1999PASP..111.1347W}. We used temporal binning of 10 days and require that each LCCF bin has at least 10 elements. Following \citet{2014MNRAS.445..437M}, the significance of the correlation was estimated using simulated light curves. In the simulations, we used power spectral density (PSD) indices of -2.35, -1.55, and -1.7 for the radio light curves of PKS~1424+240, 1ES~1959+650, and the three other sources in our sample, respectively \citep[determined from the radio data: Max-Moerbeck, private communications;][]{2016A&A...593A..98L}. For the optical, we used PSD indices reported by \citet{2018A&A...620A.185N} for each source except for VER~J0521+211 which was not included in their sample. We used the same method as described by \citet{2018A&A...620A.185N} and calculated the PSD index of the optical light curve of VER~J0521+211 to be -1.6. For the X-ray light curves we used the PSD index value of -1.4 \citep{2015A&A...576A.126A}.

The results of the cross-correlation analysis are illustrated in {Appendix~\ref{app:lccf}}. While there are several peaks (or rather wide features) in the LCCFs which reach $2\sigma$ significance level, only the radio -- optical data sets of two sources (1ES~1727+502 and 1ES~1959+650) show significant correlations ($3\sigma$ level of confidence). The two significant radio -- optical correlations are rather wide (90 and 60 days for 1ES~1727+502 and 1ES~1959+650, respectively) and compatible with zero-days time lag. The LCCF peaks are located at -10 and +40 days for 1ES~1727+502, while they are located at -30 and -20 days for 1ES~1959+650\footnote{{Positive significant lags show that the flare at radio is preceding the one in optical.}}. Again, PKS~1424+240 is an exception, as there are no $2\sigma$ peaks in the radio-optical LCCF, which is in agreement with the results in Section~\ref{sec:component} and different from results in \citet{2016A&A...593A..98L}. In general we find our radio -- optical results in a good agreement with those reported by  \citet{2016A&A...593A..98L} and \citet{2018MNRAS.480.5517L}. Only one radio -- X-ray correlation is found for the case of 1ES~1727+502 with the time lag of $680\pm20$ days where the radio flare is leading the X-ray outburst. However, the length of the X-ray light curve is rather short (1200 days) and this correlation could be the artefact of associating the X-ray outburst with one of the previous flares in radio when the X-ray data was not available. It is notable that the optical -- X-ray correlation for 1ES~1727+502 shows many features. However, these features are the result of a single dominating outburst in X-rays which results in a time-delay peak with every optical flare which is one of the limitations of the LCCF method \citep{2013MNRAS.433..907E}.

Correlations are generally used to probe if the emission regions in different energy bands are causally connected. Our results support that at least in the case of 1ES~1727+502 and 1ES~1959+650 the radio and optical emission would partially originate from the same emission region (as the time lag is consistent with zero), which is in line with the result in Section~\ref{sec:component}.

\subsection{\label{sec:pol}Polarisation analysis}

The observed optical polarisation of blazars usually contains signatures of two components: an optical polarisation core and a {stochastic} component \citep[see e.g.][]{valtaoja91,2010MNRAS.402.2087V, 2010MNRAS.408.1778B}. \citet{2014MNRAS.441.2885B} made a first attempt to separate the two components and evaluate their relative strengths from the optical polarisation data. We follow this idea, but instead of the iterative fitting applied there, we used a physical model and Bayesian fitting methods.  

To do this, we assumed that the R-band flux originates from two components, referred to as the ``constant'' and ``variable'' components in the following. We thus have for the Stokes parameters
\begin{eqnarray}
   \nonumber
   I &=& I_C + I_V\\
   Q &=& Q_C + Q_V\\ \nonumber
   U &=& U_C + U_V,
\end{eqnarray}
where the subscripts $C$ and $V$ refer to the constant and variable components, respectively. {The observed degree of polarisation (PD) and EVPA are then
\begin{equation}
{\rm PD} = \sqrt{ \left( \frac{Q}{I} \right)^2 + \left( \frac{U}{I} \right)^2 }
\end{equation}
and
\begin{equation}
{\rm EVPA} = \tan^{-1} \left( \frac{U}{Q} \right),
\end{equation}
where $-\pi \leq {\rm EVPA} \leq \pi$.} The constant component was modelled directly by letting $I_{\text{C}}$, $Q_{\text{C}}$ and $U_{\text{C}}$ be free parameters, whereas the variable component had nine free parameters, (see below and  Appendix~\ref{app:pol}). We modelled the variable component as a homogeneous cylindrical emission region in a jet with a helical magnetic field and computed the Stokes parameters using the formulae described by \cite{2005MNRAS.360..869L}. We assumed that the orientation of the variable component remains constant with respect to the observer, which means that any change in the polarisation of the source must arise from the change of the relative flux ratio between the constant and variable component. This is because in the formulation by \cite{2005MNRAS.360..869L} the EVPA of the radiation is always either parallel or perpendicular to the direction of the relativistic outflow. 

We describe the parameters of the model, the assumptions we made, and the details of the fitting procedure in Appendix~\ref{app:pol}. In short, the model has 12 free parameters. Most of the parameters in the model cannot be constrained with monochromatic observational data due to a high degree of degeneracy. We fixed 5 of the parameters {(of the variable component)}: index of the electron spectrum, $p$, to 2.1, radius of the emitting region, $r$, to $2.5 \times 10^{15}$ cm, length of the emission region, $l$, to $5 \times 10^{15}$\,cm, magnetic field strength, $B_0$, to  0.1\,Gauss and apparent speed $\beta$ to 0.99. These values are similar to those applied for the SED modelling in the next section (see Section~\ref{sec:sed_2comp}). {This model was fitted to the observed R-band polarisation data (Sec.~\ref{poldata}) in the $Q-U$ plane. One important ingredient of the model is $\sigma$, the standard deviation on random variations of $Q$ and $U$. This parameter adjusts itself according to the predictive power of the other parameters. This parameter is discussed in more detail in Appendix~\ref{app:pol}}.

The results of the fitting procedure are reported in Table~\ref{poltable}. The errors represent the 68\% confidence intervals derived from marginalised distributions. One of our main goals of this fitting procedure was to obtain some constraints on the flux ratio of the two emission components in the optical band {to be compared to the ratio derived in Section~\ref{sec:component}}. This unfortunately was not achieved in all cases. For instance, in the case of VER~J0521+211 the priori range for $I_C$ was from 0 to 3.0\,mJy (see column 3 in Table \ref{poltable}) and the posteriori averages in the middle of this range with errors that fill the priori completely. The flux ratio $I/I_{\text{C}}$ is best constrained in the case of PKS~1424+240. Therefore, the polarisation study performed here provides limited additional constraints for the SED modelling in this work, but we intend to perform a more detailed study of this method in future work. The results on the flux ratio of the two components in the optical band are compared to those obtained with the decomposition of the long-term light curves in Section~\ref{sec:twocomp}. For that purpose, we calculated $I/I_{\text{C}}$ for the SED modelling epochs, i.e. $I$ is the total optical flux in the periods reported in Appendix~\ref{app:sedepoch}. These values are reported in Section~\ref{sec:twocomp}.

\input{tables/tab_pol.tex}

Finally, we compare the observed optical EVPAs and jet position angles that we derive with our fitting to those from VLBI observations (see Section~\ref{sec:VLBA}). {If the radio and optical emission originate from same region, one would expect agreement between the optical and VLBI results.} BL Lacs objects have a preferred orientation of position angle, i.e. the EVPA is often stable. This feature can be interpreted as the stability of the emission region geometry in the optical band \citep{1978bllo.conf..117A, 1994ApJ...428..130J, 2007AJ....134..799J} and is also seen in our optical polarisation data (see middle panel in Figs.~\ref{0521mwlfig}-\ref{2344mwlfig}). In two cases (1ES~1959+650 and 1ES~2344+514) our jet position angle agrees well with the VLBI angle ($\sim50$\% probability of being the same), indicating that the EVPA is parallel to the jet for these sources. The EVPA of the radio core in 1ES~1959+650 is similarly aligned. In the case of  VER~J0521+211, if we pick the solution with $\varphi_0 = 13\degr$ and take into account the $180\degr$ ambiguity, a good agreement is again achieved with the radio core EVPA = $20\degr$. For PKS~1424+240 a similarly good agreement is found if we assume the EVPA to be perpendicular to the jet. For 1ES~1727+502 the agreement is not so clear. Out of two solutions for $\varphi_0$, one is too noisy to draw conclusions and the other one can not be made compatible with the jet radio position angle. As a general conclusion, there appears to be a good correlation between the radio and optical results, which is in line with the results from previous comparisons on TeV BL Lacs \citet{2016A&A...596A..78H}. They found that the difference between the EVPA and the jet position angle is less than $20\degr$ (i.e. the magnetic field is perpendicular to the jet direction) for two-third of the sources within a sample of 9 TeV BL Lacs. 
{Given} that our errors are approximately $10\degr$ and that we can choose from 4 different angles in the range from 0 to $360\degr$, it is not clear if this agreement is statistically significant in our case, {but certainly in line with common origin of radio and optical emission in these sources}.

\section{\label{sec:sed}SED modelling}

\subsection{\label{sec:sed_2comp}Two-component model}

\begin{figure}
\begin{center}
\includegraphics[width=.49\textwidth]{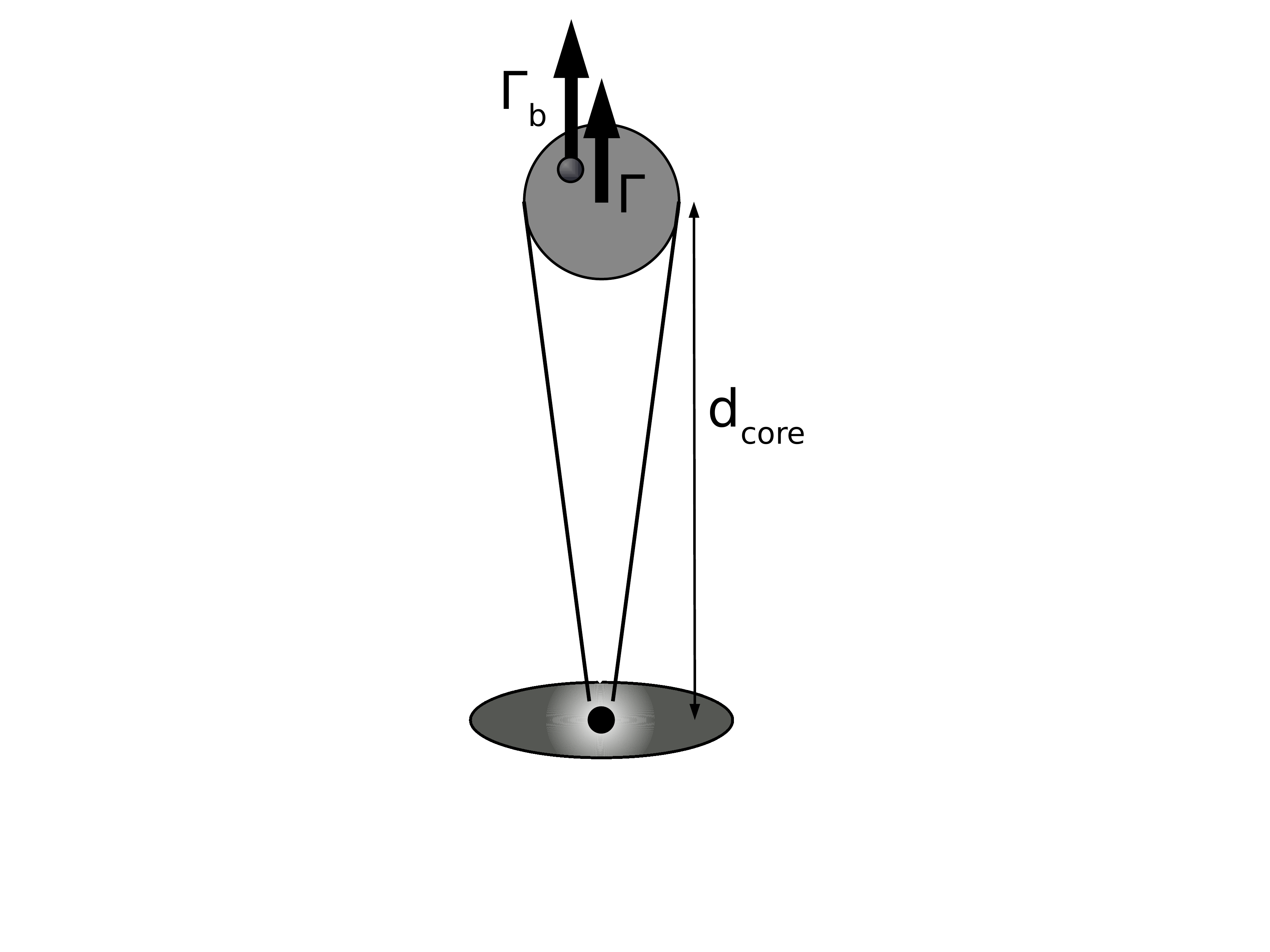}
\vspace{-1.6cm}\caption{\label{fig:cartoon}Sketch of the geometrical modelling setup. The two emission regions are located several parsecs from the central black hole (at $d_{\text{core}}$). The smaller emission region (blob) is embedded in the larger emission region (core) and the interaction of the two emission regions provides additional seed photons for the Compton scattering, see discussion in \citet[][Appendix B]{2011A&A...534A..86T}.}
\end{center}
\end{figure}

{The SEDs are modelled with a two-component model based on \citet{2011A&A...534A..86T} which calculates synchrotron and SSC emission for spherical emission regions and takes also into account synchrotron-self absorption. It is similar to the one used in \citet{2014A&A...567A.135A}, but the two emission regions are considered to be co-spatial and interacting as in \citet{2018A&A...619A..45M,2019A&A...623A.175M} to mimic a simple spine-sheath model (see Section~\ref{sec:VLBA}).
We call the two emission regions "core" and "blob", with sizes $R_{\text{core}}>R_{\text{blob}}$ (see Fig.~\ref{fig:cartoon}). These two regions correspond to the constant and variable components defined in Section \ref{sec:pol}, respectively. 

The regions are filled with electrons distributed in Lorentz factor according to a smoothed broken power law (in the following, physical quantities are expressed in the co-moving frame of each individual region):
\begin{equation}
N(\gamma)=K\gamma^{-n_1}\left(1+\frac{\gamma}{\gamma_{b}}\right)^{n_1-n_2},   \gamma_{min}<\gamma<\gamma_{max}.
\end{equation}

The distribution has a normalisation $K$ between $\gamma_{min}$ and $\gamma_{max}$ and slopes $n_1$ and $n_2$ below and above the break in the electron distribution, $\gamma_{b}$ \citep{2003ApJ...593..667M}. Each of the emission regions has size $R$, Doppler factor $\delta$ and magnetic field strength $B$, for which we searched for constraints from observations}:

\begin{itemize}

\item The sizes of the core emission region were derived from VLBI observations (see Sec.~\ref{sec:VLBA}). The sizes are of the order of  $10^{17}$\,cm. We note that the derived sizes would suggest variability timescales shorter than what we obtain for the slowly varying component from the data. This means that the origin of the slow variability cannot be the delay caused by a core-size (unlike for the blob, see below) or acceleration/cooling processes that are generally assumed as origin of the faster variability, but rather traces e.g. injection/decay phases of the central engine.

\item The existence of strong correlation between X-rays and VHE gamma-ray bands indicates that the observable emission in these two wavebands originates from a single emission region. Therefore, the maximum size of the blob emission region was calculated from the VHE gamma-ray or X-ray variability timescale using the causality relation, $R< ct_{var}\delta/(1+z)$. The VHE gamma-ray variability timescale for 1ES~1727+502 and 1ES~2344+514 is 24\,h, while for 1ES~1959+650 this timescale is 35 minutes (Sec.~\ref{magic}). The X-ray variability timescale for the case of VER~J0521+211 and PKS~1424+240 is 24\,h (Appendix~\ref{app:sedepoch}).

\item The apparent speeds of the jets can be used to derive the Doppler factor of the core, assuming the viewing angle to be known. We did this for VER~J0521+211 and PKS~1424+240 assuming viewing angles equal to $3\degr$ and $5\degr$. As discussed in Section~\ref{sec:VLBA}, three  of our sources show sub-luminal speeds or even no motion, which is common for TeV BL Lacs \citep[][and references therein]{2018ApJ...853...68P}. Therefore, we use the result from \citet{2018ApJ...853...68P}, who suggest bulk Lorentz factors with values up to 4. We convert this to Doppler factor assuming a jet viewing angle $\sim$1/$\Gamma$ and thus $\delta\sim\Gamma$.

\item The magnetic field strength of the core can be estimated from the VLBI "core shift"-measurements, assuming {a conical jet \citep{1979ApJ...232...34B} and} equal energy to be carried by particles and the magnetic field as done in \citet{2012A&A...545A.113P}. The median of the magnetic field strength of the core in their sample of 18 BL Lacs is $B_{\text{core}}=0.10\pm0.01$\,G. This sample includes 6 TeV BL Lacs. They are S5~0716+714, OJ~287, BL~Lac, OT~081, Mrk~421 and Mrk501. The first four objects have the magnetic field strength of $B_{\text{core}}\sim0.1$\,G and the last two sources have $B_{\text{core}}\sim0.4$\,G. Another way to estimate the magnetic field strength is to consider the cooling timescale of the electrons, which provides a lower limit to magnetic field strength. 
In {high-synchrotron-peaked sources}, the observed emission in the hard X-ray band is due to the high energy tail of the synchrotron emission. Therefore, the variability timescales are directly linked with particle cooling timescales. \citet{2018A&A...619A..93B} studied the variability timescale of 13 blazars in hard X-ray. They reported the hard X-ray variability timescale between $\sim5$\,min and $\sim5$\,h for six TeV BL Lacs. These timescales were calculated using 18 observation epochs. The average of estimated variability timescales in their work is $\sim1$\,h. We use equation 11 described by \citet{2018A&A...619A..93B} to calculate the magnetic field strength. We find that for the variability timescale $\sim1$\,h the magnetic field strength varies between 0.1 and 0.3\,G depending on the assumed Doppler factor and redshift. 
Therefore, we assumed a magnetic field strength to have value between 0.1 and 0.4\,G.

\end{itemize}

As the emission regions are co-spatial, we use the same magnetic field strength for the blob and core component. The magnetic field strength is generally assumed to scale with distance from central engine as $d^{-1}$, so if the blob was closer to the central engine than the core, it would nominally need to have a stronger magnetic field than the core component, of the order of $\sim1$\,G. \citet{2016MNRAS.456.2374T} showed that for single-zone models the magnetic field strengths tend to be significantly \textit{lower} than the values required for equipartition values and even in two-component models it is very difficult to reproduce the observed SED with the magnetic field strength values of the order of 1\,G. There are ways to invoke reduced local magnetic field strengths in jets such as re-connection layers and radial structures of magnetic fields across the jets \citep[see discussion in][]{2014ApJ...796L...5N}, but here we are interested to test if the observations can be modelled with magnetic field strengths obtained from the core shift measurements (see also Section~\ref{sec:twoone}) and without such reduced local magnetic field strengths. We note, however, that different effects can change the blob magnetic field (e.g. internal shocks responsible for the emission from the blob, relativistic movement of the blob w.r.t the core) and therefore it is not likely that the magnetic field strengths of the two components would in reality be exactly the same.

Finally, we also try to take into account the derived estimations of the relative strengths of the core and blob components in the optical as derived in Sections~\ref{sec:longterm} and \ref{sec:pol}. As the first method only gives a lower limit to the contribution and the second method did not converge in all cases, these constraints are not very strong, but give us a clue which of these two components dominates the emission in this waveband. We discuss the comparison of the two methods in Section~\ref{sec:twocomp}.

Based on these assumptions, the parameters let free to vary in the SED fit are:
\begin{itemize}
    \item $\gamma_{\text{min}}$, $\gamma_{\text{b}}$ and $\gamma_{\text{max}}$ of the two components: We limited the range for the values to a physically reasonable regime parameter space. i,e, $\gamma_{\text{min}}<10{^4}$, $10^{3}<\gamma_{\text{b}}<10^{5}$, $\gamma_{\text{max}}<3\times 10{^6}$, and the values for the core to be always lower than those for the blob.
    \item $n_{\text{1, blob}}$ and $n_{\text{2, blob}}$: we considered $n_{\text{1, blob}}$ to be always $\sim$2. Lower spectral index values are traditionally disfavoured as for lower values the strong radiative losses of the dominant high-energy electrons would lead to substantial pressure decrease along the jet and prevent the shock to propagate far out \citep{1985ApJ...298..114M}. We also assumed $n_{\text{2, blob}}-n_{\text{1, blob}}>0.5$.
    \item $n_{\text{1, core}}$ and $n_{\text{2, core}}$: we first considered $n_{\text{1, core}}$ to be always $\sim$2, but this did not reproduce the shape in the radio part of the SED. Radio observations \citep[e.g.][]{1988A&A...203....1V,1989ApJ...341...68H} suggest that hard spectral indices are common in AGN and there are also theoretical models \citep[e.g.][]{2003MNRAS.345..590S,2005ApJ...621..313V} that can produce indices significantly harder than 2, so we decided to consider values $n_{\text{1, core}}$ $>1.6$, which seemed to reproduce the shape of the archival data better. We assumed $n_{\text{2, core}}-n_{\text{1, core}}>1.0$.
    \item The electron energy density normalisation factor $K$: we limited the range for the values to $10^2-10^4$\,cm$^{-3}$ and considered only models where K$_{\text{blob}}$>K$_{\text{core}}$. 
    \item Doppler factor of the blob: we limit ourselves to $\delta_{\text{blob}}<30$.
\end{itemize}

\begin{figure*}
    \centering
    \includegraphics{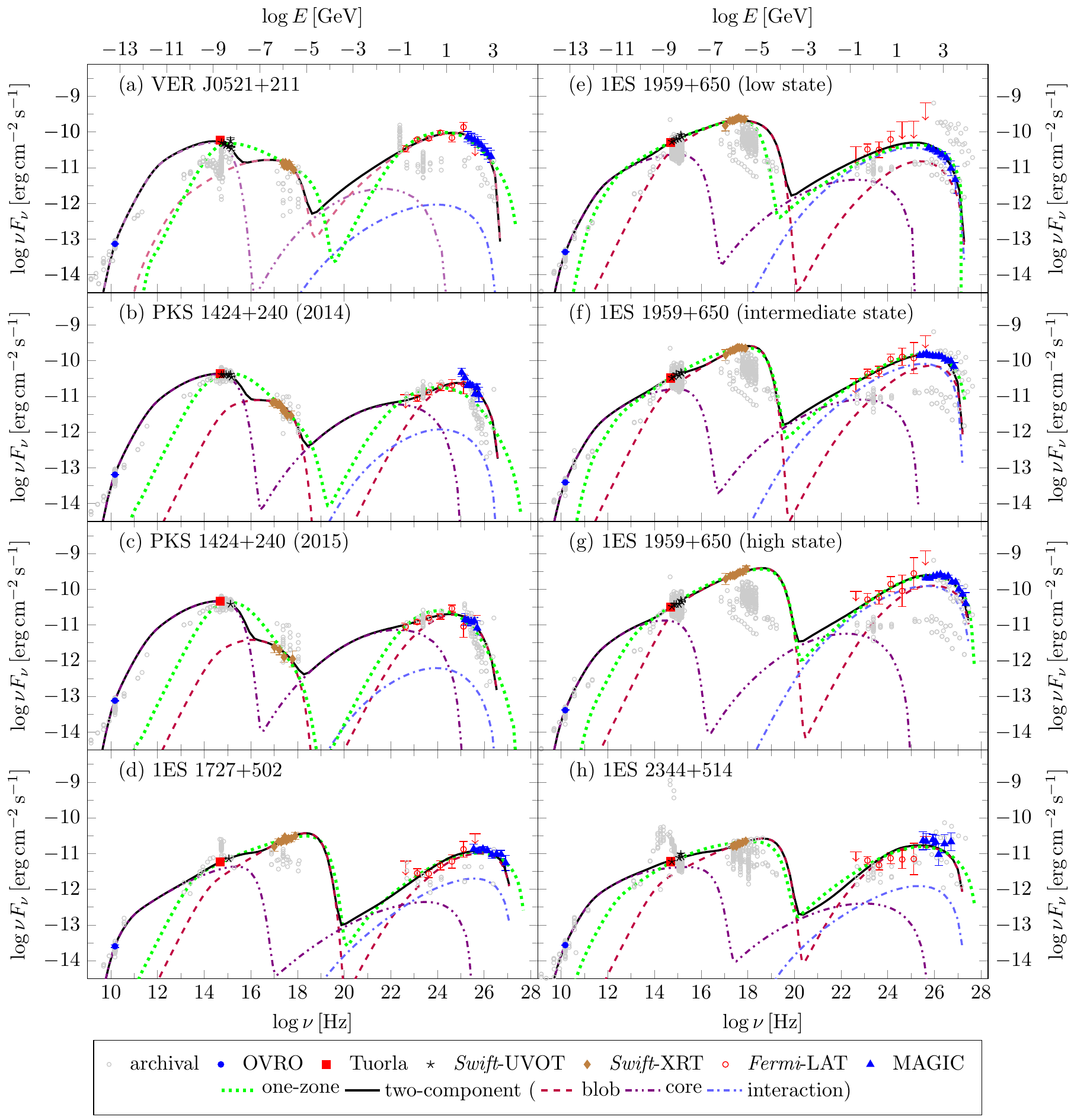}
    \caption{\label{fig:seds}The broadband SED of the source sample during the selected observation epochs/time. The details of the data selection for each SED is presented in Appendix~\ref{app:sedepoch}. The spectral data points in the panels are: archival non-simultaneous data from the ASI Space Science Data Centre, grey open circles; radio data (15\,GHz) from OVRO, blue circle; optical (R-band) from Tuorla, red square; optical and UV data from \textit{Swift}-UVOT, black stars; X-ray data from  \textit{Swift}-XRT, brown diamonds; HE gamma-ray data from \textit{Fermi}-LAT, red open circles; and {de-absorbed} VHE gamma-ray data from MAGIC, blue triangles. The SEDs are modelled within the {one-zone SSC (green dotted lines) and} two-component scenario (black lines). {Within the two-component scenario,} the violet {dash double dotted} and purple dashed lines show the emission from the core and the blob, respectively. {Moreover,} the result of interaction between emissions from the blob and the core are plotted with blue dash-dotted lines.}
\end{figure*}

Based on the availability of \mbox{(quasi-)simultaneous} data and the observed flux variability at VHE gamma rays, we selected eight SED data sets. The details of the MWL data selection for each data set is presented in Appendix~\ref{app:sedepoch}. With the observational and theoretical constraints listed above, we check if we can find a set of parameters that reproduces these observed SEDs. Figure~\ref{fig:seds} shows that in all of the eight cases we find a set of parameters (listed in Tab.~\ref{sedparams}) which produces a two-component model in a good agreement with the (quasi)-simultaneous observational data. There are some common "trends" in these parameters. In all cases $\gamma_{\text{min, blob}}$ is high ($>10^3$) and $n_{\text{1, blob}}$ is hard. The $\gamma_{\text{b, blob}}$ varies from $3\times10^{4}$ to $9.5\times10^{4}$, while the $\gamma_{\text{max, blob}}$ values fall into a range of one order of magnitude higher. Also the $n_{\text{2, blob}}$ values spread over a large range from 2.45 to 3.85. Both for the cores and the blobs $\gamma_{\text{min}}$ and $n_1$ values used in all sources are similar. Interestingly, for the cores, in all but one case a power-law electron distribution without a break was used. 

The applied model is not time dependent, so all epochs were modelled independently. We only aim at testing the model on "snapshot SEDs" and acknowledge that the fast blob would exit the core region at some point. Therefore, in this model setup the observed changes in the SED can be produced for example by exiting of one blob component and entering of a new one. There are two sources for which we had multiple epoch SEDs. For PKS~1424+240 the two different SEDs are characterised by a lower synchrotron flux in 2015 compared to 2014, while the gamma-ray flux was not changing. We modelled this by decreasing the $\gamma_{\text{min, blob}}$  and $K_{\text{blob}}$ and by softening the electron energy density spectral index $n_{\text{1, blob}}$. In the case of 1ES~1959+650, the X-ray and the VHE gamma-ray data of the SED changed significantly. In our models, these parts are largely dominated by the blob emission, for which we altered almost all parameters between the different states, but we also had to alter the core parameters to achieve good representation of the observed SEDs. Finally, we note that the co-spatiality of the emission regions means that we have to consider possible gamma-gamma absorption between the core seed photon field and the highest energy photons emitted by the blob. Our calculations showed that the absorption is negligible.

\input{tables/tab_sedparams.tex}

\subsection{\label{sec:sedonezone}One-zone model}

{In previous works, SEDs of the sources of our sample have all been modelled with one-zone SSC models. The data sets used in those works are not always the same as the ones presented here. As the sources are variable, also parameters to reproduce the SEDs would vary from one epoch to another. Therefore, for comparison purposes, we also modelled the same SEDs using the one-zone SSC model. The model applied here is the one of \citet{2003ApJ...593..667M}, i.e. the same model used as the basis of the two-component model. We kept the same physically motivated range (but not the same parameters) for electron distribution of the emission region as we used for "blob" in the two component model. We also applied the same constraint from the variability timescale for the size of the emission region as we applied for the "blob" component in the two-component model. We then searched for a set of parameters that described the optical to VHE gamma-ray part of the SED well. The resulting parameters are shown in Table~\ref{sedparams} and are similar to those derived in previous one-zone modelling for these sources (see Appendix~\ref{app:sedepoch}). The broadband SEDs, including the one-zone SSC models, are illustrated in Figure~\ref{fig:seds}. We discuss the differences in parameters and appearance of the SEDs in the next section.}

\section{\label{sec:discuss}Discussion}


In the first part of the discussion we compare our observationally-constrained two-component models to {other} modellings of the SEDs (Sec.~\ref{sec:twoone}, \ref{sec:equipartition}, and \ref{sec:others}) then we discuss the SED classification of our sample sources (Sec.~\ref{sec:peaks}) and finally compare the flux ratios of the two components with different methods (Sec.~\ref{sec:twocomp}).  

\subsection{\label{sec:twoone}Comparison of the two-component and one-zone models}

{In Section~\ref{sec:sed}, we have modelled the observed SEDs with the two-component model and one-zone SSC model for comparison purposes. One has to be careful when comparing the two models as we did not perform extensive scans of parameter space that could reproduce the SED.  However, some general comparison can be done. For the two-component model we have selected the parameters as observationally motivated as possible, while for the one-zone SSC model, we looked into previous one-zone SSC models and modified the parameters to fit the data in our epoch
(\citealp[VER~J0521+211:][]{2013ApJ...776...69A}; \citealp[PKS~1424+240:][]{2014A&A...567A.135A}, \citealp{2016MNRAS.461.1862K}, \citealp{2017A&A...606A..68C};  \citealp[1ES~1727+502:][]{2015ApJ...808..110A}; \citealp[1ES~1959+650:][]{2003A&A...412..711T}, \citealp{2006ApJ...644..742G}, \citealp{2008ApJ...679.1029T}, \citealp{1959flare}; \citealp[1ES~2344+514:][]{2013A&A...556A..67A}, \citealp{2344fact}).

The one-zone SSC models describe the observational data from optical to VHE gamma-rays well, while the radio part of the SED is ignored (the radio part originates from another component as small emission regions are optically thick to radio emission). 
In all cases the magnetic field strength is smaller than the one in two-component models, which is in line with very low magnetic field strengths ($\sim$ 0.01-0.001\,G) typically used in one-zone SSC models in the literature. \citet{2016MNRAS.456.2374T} showed that this is a general result for one-zone SSC models (see also Section~\ref{sec:equipartition}). Also the Doppler factors used in one-zone SSC models are in all cases higher than two-component models, again in line with values ($20 \leq \delta \leq 150$) reported in the literature for these sources.}

In our two-component model, the blob dominates the emission for all sources from X-ray to VHE  gamma-rays, while the core dominates the emission in the radio band. There are two important consequences of taking the core component into account for the modelling of the total SED. Firstly, the core component extends always to the optical band, which then constraints the flux of the low energy part of the blob component to lower values. This in general forces us to use rather a narrow electron energy range for the blob component with high $\gamma_{\text{min}}$ values. Secondly, if the two zones are co-spatial, as it is the case here, the core component provides seed photons for the Compton scattering, which then relaxes the requirement of very low $B$ and high $\delta$ values for the blob component. 

As discussed in Section~\ref{sec:sed_2comp}, the magnetic field strength \mbox{$B\sim0.1$\,G} is used for both core and blob component in our model as it is difficult to physically motivate that the magnetic field of the small blob would be orders of magnitude weaker than that of the core in the case the blob is either co-spatial to the core or closer to the central engine than the core. We emphasise that we have assumed the same $B$ for the two components, but we cannot exclude $B$ values significantly lower than those used for the blob. {On the other hand, $B$ values significantly higher are easy to imagine, they could be caused by some dynamo effect or simply compression of the field in shocks as the faster blob moves through the slower core, but with high $B$-values we could not reproduce the SEDs}. However, we demonstrate that with $B$ values close to those derived by \mbox{\citet{2012A&A...545A.113P}} from the VLBI observations, we can actually reproduce the observed SEDs well from radio to VHE gamma rays.

As one of the main advances of {the observationally constrained two-component model} with respect to the one-zone SSC model is the inclusion of the radio part of the SED into the modelling, it is rather unfortunate that each SED in this part is constrained only with one quasi-simultaneous data point at 15\,GHz. There are some archival data at higher radio frequencies from the Planck satellite at 100-217\,GHz. As shown in Figure~\ref{fig:seds}, our {two-component} modelling reproduces also this part acceptably for two sources (1ES~1727+502 and 1ES~2344+514) while in the case of the three remaining sources (VER~J0521+211, PKS~1424+240 and 1ES~1959+650) the model overproduces the flux in this range. As discussed in Section~\ref{sec:sed_2comp}, it seems that to produce the shape of the SED correctly in this range, $n_{\text{1, core}}\lesssim 1.6$ {(or inclusion of further components)} is required. However, for further investigation of this issue simultaneous data in this band would be required, and therefore it is beyond the scope of this paper.

\subsection{\label{sec:equipartition}Equipartition in the two-component model}

\citet{2016MNRAS.456.2374T} recently showed that typical SED parameters of one-zone models for BL Lacs are far from equipartition, with the kinetic energy of the particles dominating the energy carried by the magnetic field by several orders of magnitude. They also showed that for two-component models solutions close to equipartition can be found. Therefore we check, using equations 5, 9, 16 and A1 from \citet{2016MNRAS.456.2374T}, the ratio between the magnetic-field and accelerated relativistic-electron energy densities $U'_{\text{B}}/U'_{\text{e}}$ for our models. We note that {for the two-component model} this is calculated for the whole system \citep[see][]{2016MNRAS.456.2374T}. The values are reported in Table~\ref{sedresults}. In all cases, the model parameters {of the two-component model} suggest solutions close to equipartition, with $U'_{\text{B}}$ slightly dominating. The only exception to this is found for the VER~J0521+211, where $U'_{\text{e}}$ is slightly dominating. {For one-zone SSC, the solutions are far from equipartition in all cases, with the kinetic energy of the particles dominating the energy carried by the magnetic field by several orders of magnitude. In conclusion,} the results are in agreement with the conclusion of \citet{2016MNRAS.456.2374T}, i.e., in two-component models solutions close to equipartition can be found. 

We calculated the kinetic energy associated with the electrons, cold protons and magnetic field of the core (Tab.~\ref{sedresults}) using the equations 1 to 3 reported by \citet{2008MNRAS.385..283C}. 
{We performed the calculation for one-zone and two component models.}
We note that the two ISP sources (VER~J0521+211 and PKS~1424+240) have significantly higher kinetic energies than the three HSP sources (1ES~1727+502, 1ES~1959+650 and 1ES~2344+504). Comparing the results of our sample to the sample reported by \citet{2008MNRAS.385..283C}, we found that this is in line with results on larger samples and that the values we estimated for our sample are similar to those obtained for larger BL~Lac samples.

\input{tables/tab_sedresults.tex}

\subsection{\label{sec:others}Comparison to other models}

Among the sources of the sample, there has been multiple attempts of modelling the broadband SED of PKS~1424+240 and 1ES~1959+650 in several frameworks different from that of one-zone SSC modelling which include the VHE gamma-ray data. In this section, these results are briefly discussed. 

\subsubsection*{PKS~1424+240}
Since the discovery of PKS~1424+240 at VHE gamma rays in 2009 \citep{2009ATel.2084....1O,2009ATel.2098....1T} and the first firm lower limit for the redshift of the source \citep{2013ApJ...768L..31F}, it has become an interesting source for many authors, being the most distant TeV BL Lac object so far detected {\citep[z=0.6;][]{2016A&A...589A..92R, 2017ApJ...837..144P}}. The source has been monitored since 2009 and there have been several attempts to model the broadband SED using different sets of VHE gamma-ray observations. \citet{2016MNRAS.461.1862K} and \citet{2017A&A...606A..68C} used the VHE gamma-ray data obtained with VERITAS during 2009 and 2013  \citep{2014ApJ...785L..16A}. MAGIC observations during 2009-2011 were used by \citet{2014A&A...567A.135A} to {build} the broadband SED of the source. 

In the leptonic scenario, external IC models were tested assuming external photons from a broad-line region \citep{2016MNRAS.461.1862K} and/or a dusty torus \citep{2016MNRAS.461.1862K,2017A&A...606A..68C}. While the model reproduces the observed SED acceptably, there is no observational evidence for a broad-line region and/or a dusty torus in the optical spectra of PKS~1424+240. Hadronic models, with the gamma-ray component dominated by pure proton-synchrotron emissions, did not lead to any reasonable solution. However, in the framework of a lepto-hadronic scenario, good solutions were found assuming that synchrotron emission from secondary particles was responsible for producing the VHE gamma-ray emission and proton-synchrotron emission produces the radiation at lower energies \citep{2017A&A...606A..68C}. 

\citet{2014A&A...567A.135A} used a two-component model for PKS~1424+240. The model is the same as used in our work, but the two emission zones are not assumed to be co-spatial. The differences in the parameters are the following: i) a lower Doppler factor for the blob is used in this work (18-20 vs. 30). ii) the magnetic field strength is by a factor of three higher than the one used in \citet{2014A&A...567A.135A}. This comparison demonstrates that while in this work we have sought for solutions close to equipartition, the two-component model can reproduce the SED also with parameters that are closer to parameters typically needed for one-zone models (low magnetic field strengths and high Doppler factors).  

\subsubsection*{1ES~1959+650}
The source was first detected at VHE gamma rays by the Utah Seven Telescope Array in 1998 \citep{1999ICRC....3..370N}. Being one of the nearby, bright objects among the extragalactic VHE gamma-ray emitters, the source has been observed frequently since its discovery. After the detection of an orphan  VHE gamma-ray flare during 2002 \citep{2004ApJ...601..151K}, hadronic models were motivated for describing the broadband SED of the source. \citet{1959flare} applied the proton-synchrotron scenario to describe the broadband SED of the source during the 2016 flaring activity (on MJD 57552 and 57553). They found that this model requires a high value of the magnetic field strength ($B=150$\,G) and an acceleration efficiency close to the theoretical limit ($\eta_{\text{acc}}=1$). Unlike the pure proton-synchrotron scenario, the lepto-hadronic models can use lower magnetic field strengths to describe the broadband SED of the source during the flaring activity \citep{2005ApJ...630..186R, 2010ApJ...719L.162B, 2013PhRvD..87j3015S, 1959flare}.

In the leptonic scenario, the external IC model was tested by \citet{2013ApJ...775....3A}. The external Compton component was motivated by the existence of dust in the central environment of 1ES~1959+650 \citep{2012MNRAS.424.2276F}. The external-Compton scenario could describe the observational data during the low VHE gamma-ray state. Moreover, this setup was able to generate the anti-correlated X-ray variability seen during the 2007-2011 observation campaign. However, this model needed a relatively low magnetic field ($B=0.02$\,G) and needed a high Doppler factor ($\delta=30$) as typical for one-zone models (see Sect.~\ref{sec:twoone}). 

It has been suggested that the simple one-zone SSC model cannot reproduce the multiple flaring activity of 1ES~1959+650 and multiple-zone SSC models are favoured \citep{2004ApJ...601..151K, 2018A&A...611A..44P}. Those models are different from the one discussed here. \citet{2004ApJ...601..151K} did not take into account the interaction between the two emission zones while \citet{2018A&A...611A..44P} did not assume the co-spatiality of the emission regions. Finally, in both of these studies, the strength of the magnetic field was one order of magnitude lower than the value obtained from VLBI observations assuming equipartition. 

\subsection{\label{sec:peaks}SED peak frequencies}
All of the sources in our sample are TeV BL Lacs, but actually have quite different SED peak frequencies. The {observed} synchrotron and IC peak frequencies are estimated from the broad-band SED models (local maxima in the model, first peak corresponding to $\nu_{\text{sync}}$ and second to $\nu_{\text{IC}}$, in case of $\nu_{\text{sync}}$ we consider the peak to be the higher of the two) presented in Section~\ref{sec:sed}. The results are summarised in Table~\ref{sedresults}. 

\textit{VER~J0521+211:} The synchrotron and IC peaks are located at $\nu_{\text{sync}}=2.40\times10^{14}$ and $\nu_{\text{IC}}=3.65 \times 10^{24}$\,Hz, respectively. The ratio of the luminosity at the IC peak to that at the synchrotron peak (CD, Compton dominance parameter) is equal to CD $=1.71$. These values are in line with the current classification of the source to be an ISP BL Lac object \citep{2011ApJ...743..171A}, while \citet{4LAC} classifies the source as HSP (with $\nu_{\text{sync}}=1.40\times10^{15}$, so very close to ISP-HSP borderline value).

\textit{PKS~1424+240:} The source is reported to be a HSP \citep{2014ApJ...785L..16A,4LAC}. However, most of the observations which led to this conclusion were obtained during the relatively high state of the source. We found that the source is rather an ISP BL Lac object with $\nu_{\text{sync}}=(4.68-4.99) \times10^{14}$\,Hz during its quiescent state in 2014 and 2015. This is in line with the location of the synchrotron peak reported by \citet[][See Fig. 3]{2014A&A...567A.135A} using the data set of the 2010 campaign. Such a transition synchrotron peak occurs also in other BL Lacs \citep{2018MNRAS.480..879M}.

\textit{1ES~1727+502:} In our SED the synchrotron and IC peaks are located at $\nu_{\text{sync}}=1.92\times10^{18}$ and $\nu_{\text{IC}}=7.48 \times 10^{25}$\,Hz, respectively. It is notable that the source was in high state at VHE gamma rays during the 2015 campaign. This is significantly higher than what was derived by \citet{2018A&A...620A.185N,4LAC} for this source ($\nu_{\text{sync}}=2.2\times10^{16}$\,Hz {and $\nu_{\text{sync}}=7.1\times10^{15}$), respectively} using archival data. It is also clearly visible in Figure~\ref{fig:seds} that the $\nu_{\text{sync}}$ has clearly shifted to higher frequency during this high state. 

\textit{1ES~1959+650:} As discussed in Section~\ref{sec:obs}, all of the SEDs were observed during high states with the $\nu_{\text{sync}}$ evolving from $4.02\times10^{17}$ to $3.58\times10^{18}$\,Hz from the lowest to the highest state. As clearly visible in Figure~\ref{fig:seds}, in the archival data the $\nu_{\text{sync}}$ was clearly lower, \citet{2018A&A...620A.185N} and \mbox{\citet{4LAC}} estimated $5.0\times10^{16}${\,Hz} {and $\nu_{\text{sync}}=9.0\times10^{15}$, respectively}.

\textit{1ES~2344+514:} Also for this source, it is clearly visible in Figure~\ref{fig:seds} that in the archival data $\nu_{\text{sync}}$ was lower than during the campaign modelled in our work. For the archival data \citet{2018A&A...620A.185N} estimated $5.0\times10^{16}${\,Hz} and \mbox{\citet{4LAC}} $\nu_{\text{sync}}=1.6\times10^{16}$, while from our modelling we get $\nu_{\text{sync}}=4.61\times10^{18}${\,Hz}.

In summary, our sample includes two ISPs and three HSPs, with peak frequencies shifted to higher frequencies during the flaring states studied in this paper. This is a typical behaviour for blazars \citep{2000ApJ...536..742P, 2013A&A...559A.136H, 2016MNRAS.459.2286A, 2018MNRAS.480..879M}. Even if the sample is small and therefore not conclusive, we note that the SED parameters $\gamma_{\text max}$ and $n_2$ we used are more similar between the two ISPs and the three HSPs. Also the two ISPs are the most luminous sources in our sample (see Table~\ref{sedresults}). The CD parameters of the SEDs are typical for BL Lacs \mbox{\citep[CD$\sim 1.0$,][]{2017A&A...606A..44N}} and there is no significant differences, in terms of CD, between the ISPs and HSPs in the sample. 

Even if the sample is small, only five sources, it still represents different SED peak frequencies, and both flaring and quiescent states. This implies that the model applied here should be applicable to a wide range of BL Lacs, with the exception of very fast, very bright VHE gamma-ray flares \citep{2019A&A...623A.175M}, where acceptable representation of the observed SED was only found with low magnetic field strength values even within the two-component model.

\subsection{\label{sec:twocomp}Contribution of the two components in the optical band}

One of the main goals of this paper was to constrain the contribution of the two components from the long-term data and use that as an input in the SED modelling to limit the parameter space for two-component models even further. This was not entirely successful. Instead we decided to compare the resulting ratio of the two components given by the long-term light curve decomposition method (Sec.~\ref{sec:longterm}), polarisation study (Sec.~\ref{sec:pol}) and the SED modelling (Sec.~\ref{sec:sed_2comp}). For the SED modelling, the ratio is calculated as the ratio between the core optical flux and the observed optical flux. In the SED modelling, the constant component usually dominates in the optical band, 1ES~1959+650 being an exception having a rather even ratio. The comparison between the three ratios is presented in Table~\ref{corblob}. 

For four sources (PKS~1424+240, 1ES~1727+502, 1ES~1959+650 and 1ES~2344+514) the lower limit derived by the first method matches the range given by the polarisation study, while for VER~J0521+211 it is larger. As can be seen from Table~\ref{poltable}, the flux of the constant component is not well constrained for this source, the posteriori (column 3) just fills the priori (column 2). The low ratio in Table~\ref{corblob} is then the result of the priori (i.e. the minimum optical flux in the whole period), being much lower than the optical flux during the flare in 2013 for which the SED is constructed (see  Appendix~\ref{app:sedepoch}). In all cases the ratios derived from the SEDs are larger than or comparable to the lower limits derived with the first method (see below). Only for two sources (1ES~1727+502 and 1ES~1959+650) the ratio derived from the SED modelling is within the range derived from the optical polarisation study. For all other sources, the ratio derived from the SED modelling is always larger than the one from the optical polarisation. 

It is important to understand that the three derived fractions are not directly comparable. The first method gives the overall minimum fraction \textit{throughout the 5-year period} that of the slowly variable component contribution to the {variability}. As discussed in \citet{2016A&A...593A..98L} it is not possible to constrain the contribution at a given time to the snapshot SED, while that is exactly what is calculated from the polarisation study and the SED modelling. Moreover, the SED modelling obviously does not give an independent result, as we indeed tried to use the results from the other two methods to guide us on the relative strengths of the two components in the optical band. Also, as discussed in Section~\ref{sec:pol}, for the polarisation study the posteriori was very wide in several cases, which of course results in a very large range for the ratio. Unfortunately for the source PKS~1424+240 for which the polarisation study was performing best, we failed to derive constraints from the radio-optical long-term study. We intend to further investigate these methods with larger sample of sources in future work.

\input{tables/tab_core_blob.tex}

\section{\label{sec:sum}Summary and conclusions}

In this work, we have studied the broadband SED of five VHE gamma-ray emitting BL Lacs. We studied the broadband SEDs of these sources during eight observation epochs based on the variability of the VHE gamma-ray flux of the objects. Traditionally, in view of their relative simplicity and agreement with the data, single-zone SSC models have been used to describe TeV BL Lac SEDs \citep{2003A&A...412..711T, 2006ApJ...644..742G, 2008ApJ...679.1029T, 2013ApJ...776...69A, 2014A&A...567A.135A, 2015ApJ...808..110A, 2016MNRAS.461.1862K, 2017A&A...606A..68C}. However, in almost all of the cases the radio parts of the SEDs are excluded in the one-zone SSC model arguing that the radio emission originates from an outer region. Moreover, the application of a one-zone SSC model requires a low magnetic field strength and a high Doppler factor in order to reproduce the IC part of the SEDs. This results in jets which are far from equipartion \citep{2016MNRAS.456.2374T}. Therefore, the application of  multiple-component SSC models are suggested by many authors \citep[e.g.][]{2004ApJ...601..151K, 2014A&A...567A.135A, 2016MNRAS.456.2374T, 2018A&A...611A..44P}. We used a two-component model with two interacting, co-spatial emission regions (core and blob) in order to mimic the so-called \textit{spine-layer} SSC model described by \citet{2016MNRAS.456.2374T}.

We collected MWL data of the sample in the time span between 2012 September 30 and 2018~October~9. We tried to constrain the contribution of each emission zone to the total optical flux employing two different approaches. In the first approach we used the method described in \citet{2016A&A...593A..98L} to see if the observed trend in radio and optical light curves were temporary and calculated the minimum fraction of the optical (R-band) flux that originates from the same region as the radio emission at 15\,GHz. In the second approach, we used optical polarisation data to distinguish between constant and variable components. Moreover, we searched for the existence and time lag of the flaring activity between the radio, optical, and X-ray flux of the sample following the method described by \citet{2014MNRAS.445..428M} and \citet{2016A&A...593A..98L}. 

In order to reduce the parameter space of our two-component model, we used observational constrains from VLBI data. In particular, the magnetic field strength, the size, and the Doppler factor of the core emission region are derived from the VLBI data using the values obtained from the MOJAVE programme \citep{2004ApJ...600..115P,2008ApJ...678...64P,2010ApJ...723.1150P,2012arXiv1205.2399T,2012A&A...545A.113P,2018ApJ...853...68P,2019ApJ...874...43L}.

We find parameter sets to describe the broadband SEDs of the sample during the selected epochs. The results of our study can be summarised as follows:
\begin{itemize}
\item All of the sources of the sample show variability in at least one of the studied bands. Intra-night variability on the nights of 2016 June 13 and 2016 July 1 was detected at VHE gamma rays during the 2016 observation campaign of 1ES~1959+650 (See Tab. 3 in \citealt{1959flare} for details). Two of the sources (1ES~1727+502 and 1ES~2344+514) show significant flux variability in the timescale of one day at VHE gamma rays. The other two sources, VER~J0521+211 and PKS~1424+240, show daily variability in the X-ray and optical bands, respectively. These variability timescale constraints were applied to the SED modelling.

\item The uncertainty of the data points in the HE gamma-ray light curves of the sources were rather large and prevented us from cross-correlation studies. The HE gamma-ray spectral parameters (for the selected epochs) are generally similar to those reported by \citet{2015ApJS..218...23A} and \citet{2019arXiv190210045T}.

\item We found that the significant increasing/decreasing trends in the optical band have persisted in four sources out of five (except for PKS~1424+240). Moreover, the {increasing/decreasing} trends in the radio light curves have persisted for all of the sources in the sample. Taking into the account similar findings reported by \citet{2016A&A...593A..98L} for the time span of 5 years before the starting date of our light curves, the radio light curves trends have persisted for $\sim10$ years.

\item Among the 15 cross-correlation pairs, we only find two significant correlations between radio and optical for 1ES~1727+502 and 1ES~1959+650 with the time lag compatible with zero days, suggesting that in these two sources the emission in these two bands originates from the same region.

\item {Optical and VLBI core polarisation angles seem to align also in our sources, which is in agreement with the earlier studies \citep[e.g.][]{2016A&A...596A..78H} and with common origin of optical and radio emission as assumed in our two-component SED modelling.} 

\item The polarisation analysis provided limited additional constraints to the contribution of the two emission components in the optical band. We attempted to model the optical polarisation with simple "variable" and "constant" component model, but the limitation of the available data lead to a large range on the calculated flux of the constant component. Except for VER~J0521+211, the lower limit derived by the long-term radio -- optical analysis method matches the range given by the polarisation study. 

\item In all cases the ratios between the two components in optical derived from the SEDs are larger than or comparable to the lower limits derived with the long-term radio -- optical analysis method. 

\end{itemize}

We presented the first systematic attempt to model the broadband SEDs of BL Lacs considering all the observational constraints provided by radio and optical observations: VLBI, long-term variability and polarisation. The modelled SEDs are associated with ISP and HSP sources, in low and flaring states, and in all of the cases we find a model that is in good agreement with the observed SED. This implies that the model should be applicable to a large fraction of the BL Lacs. Moreover, within the selected two-component scenario, where the emission regions are co-spatial and located at the VLBI core, i.e. several parsecs away from the central engine, it is possible to reproduce the SEDs with magnetic field strengths and Doppler factors that are well in agreement with the values derived directly from the VLBI observations. This demonstrates that the usually neglected radio component does not have to originate from a region far away from the region dominating the emission in the X-rays and VHE gamma-ray bands and when this component is properly taken into account, we can reproduce the observed SEDs with parameters where the whole system are close to equipartition. 

{The two-component model presented in this work can be further validated by future observations. Even if the data presented in this work are comprehensive, they are still sparse. The MWL data set and therefore the model can be improved by new MWL simultaneous observation from radio to VHE gamma-ray band. In particular, simultaneous radio, mm, IR and optical data are needed to better constrain the contribution of the emission regions in lower energy bands. Moreover, the energy band where the IC component starts dominating the synchrotron component (hard X-ray to MeV) has to be studied in more details. Furthermore, better estimations of the IC emission in the energy range between several GeV to 100 GeV are needed to constrain accurately the IC peak frequency. Finally, the connection between radio, optical and X-ray polarisation should be studied to understand the topography of the magnetic field. In order to achieve these purposes, simultaneous data including radio-mm, mid-IR (\textit{JWST, Euclid}), X-ray (\textit{NuSTAR, IXPE, Spektr-RG}), MeV (\textit{e-Astrogam or AMEGO}), and HE and VHE gamma-ray (CTA) are of interest.}


\begin{acknowledgements}
{We would like to thank the anonymous reviewer whose comments have helped us improving this manuscript.}\\
We would like to thank the Instituto de Astrof\'{\i}sica de Canarias for the excellent working conditions at the Observatorio del Roque de los Muchachos in La Palma. The financial support of the German BMBF and MPG; the Italian INFN and INAF; the Swiss National Fund SNF; the ERDF under the Spanish MINECO (FPA2017-87859-P, FPA2017-85668-P, FPA2017-82729-C6-2-R, FPA2017-82729-C6-6-R, FPA2017-82729-C6-5-R, AYA2015-71042-P, AYA2016-76012-C3-1-P, ESP2017-87055-C2-2-P, FPA2017‐90566‐REDC); the Indian Department of Atomic Energy; the Japanese ICRR, the University of Tokyo, JSPS, and MEXT;  the Bulgarian Ministry of Education and Science, National RI Roadmap Project DO1-268/16.12.2019 and the Academy of Finland grant nr. 320045 is gratefully acknowledged. This work was also supported by the Spanish Centro de Excelencia ``Severo Ochoa'' SEV-2016-0588 and SEV-2015-0548, the Unidad de Excelencia ``Mar\'{\i}a de Maeztu'' MDM-2014-0369 and the "la Caixa" Foundation (fellowship LCF/BQ/PI18/11630012), by the Croatian Science Foundation (HrZZ) Project IP-2016-06-9782 and the University of Rijeka Project 13.12.1.3.02, by the DFG Collaborative Research Centers SFB823/C4 and SFB876/C3, the Polish National Research Centre grant UMO-2016/22/M/ST9/00382 and by the Brazilian MCTIC, CNPq and FAPERJ.\\
The \textit{Fermi}-LAT Collaboration acknowledges generous ongoing support from a number of agencies and institutes that have supported both the development and the operation of the LAT as well as scientific data analysis. These include the National Aeronautics and Space Administration and the Department of Energy in the United States, the Commissariat \`a l'Energie Atomique and the Centre National de la Recherche Scientifique / Institut National de Physique Nucl\'eaire et de Physique des Particules in France, the Agenzia Spaziale Italiana and the Istituto Nazionale di Fisica Nucleare in Italy, the Ministry of Education, Culture, Sports, Science and Technology (MEXT), High Energy Accelerator Research Organization (KEK) and Japan Aerospace Exploration Agency (JAXA) in Japan, and the K.~A.~Wallenberg Foundation, the Swedish Research Council and the Swedish National Space Board in Sweden. Additional support for science analysis during the operations phase is gratefully acknowledged from the Istituto Nazionale di Astrofisica in Italy and the Centre National d'\'Etudes Spatiales in France. This work performed in part under DOE Contract DE-AC02-76SF00515.\\
Part of this work is based on observations made with the Nordic Optical Telescope, operated by the Nordic Optical Telescope Scientific Association at the Observatorio del Roque de los Muchachos, La Palma, Spain, of the Instituto de Astrofisica de Canarias.\\
We acknowledge the support from the Bulgarian NSF through grants DN 18-1220 13/2017, DN 18-10/2017, KP-06-H28/3 (2018) and KP-06-PN38/1 (2019).\\
This research has made use of data from the OVRO 40-m monitoring program which is supported in part by NASA grants NNX08AW31G, NNX11A043G, and NNX14AQ89G and NSF grants AST-0808050 and AST-1109911.\\
This research has made use of data from the MOJAVE database that is maintained by the MOJAVE team \citep{2019ApJ...874...43L}.\\
Part of this work is based on archival data, software, or online services provided by the Space Science Data Centre - ASI.\\
This research has made use of data and/or software provided by the High Energy Astrophysics Science Archive Research Center (HEASARC), which is a service of the Astrophysics Science Division at NASA/GSFC and the High Energy Astrophysics Division of the Smithsonian Astrophysical Observatory.
\end{acknowledgements}

\bibliographystyle{aa}
\bibliography{ref.bib}

\begin{appendix}
\section{Examples of online data}
The examples of online data are presented in the Tables~\ref{tab:tuorla} and \ref{tab:xray} following the analysis procedures described in Sections~\ref{sec:opt_radio} and \ref{sec:xray}, respectively.

\input{tables/tab_tuorla.tex}

\input{tables/tab_xray.tex}

\section{\label{app:lccf}Light curves correlations}
\begin{figure}[ht!]
\includegraphics{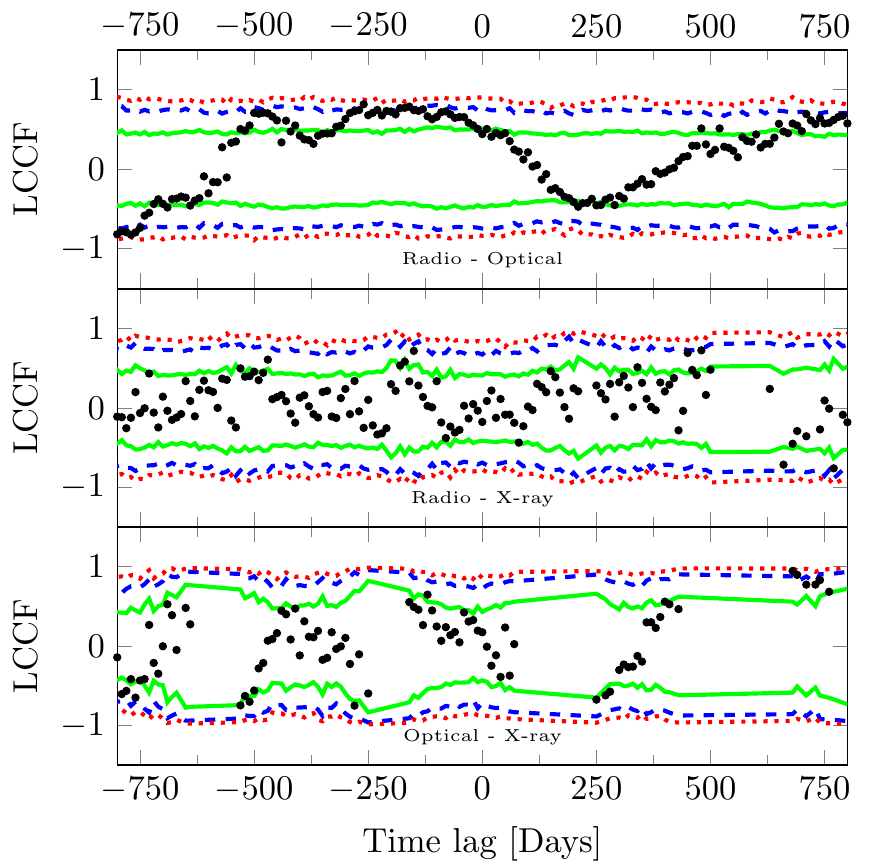}
\caption{\label{0521lccffig}The results of the LCCF study of VER~J0521+211: \textit{(Top)} between radio (15\,GHz) and optical (R-band), \textit{(middle)} between radio (15\,GHz) and X-ray (2-10\,keV), and \textit{(Bottom)} between optical and X-ray (2-10\,keV). We show $1\sigma$ (green solid line), $2\sigma$ (blue dashed line) and $3\sigma$ (red doted line) significance limits. Positive significant lags show that the flare at lower frequency is preceding the other band.}
\end{figure}

\begin{figure}[ht!]
\includegraphics{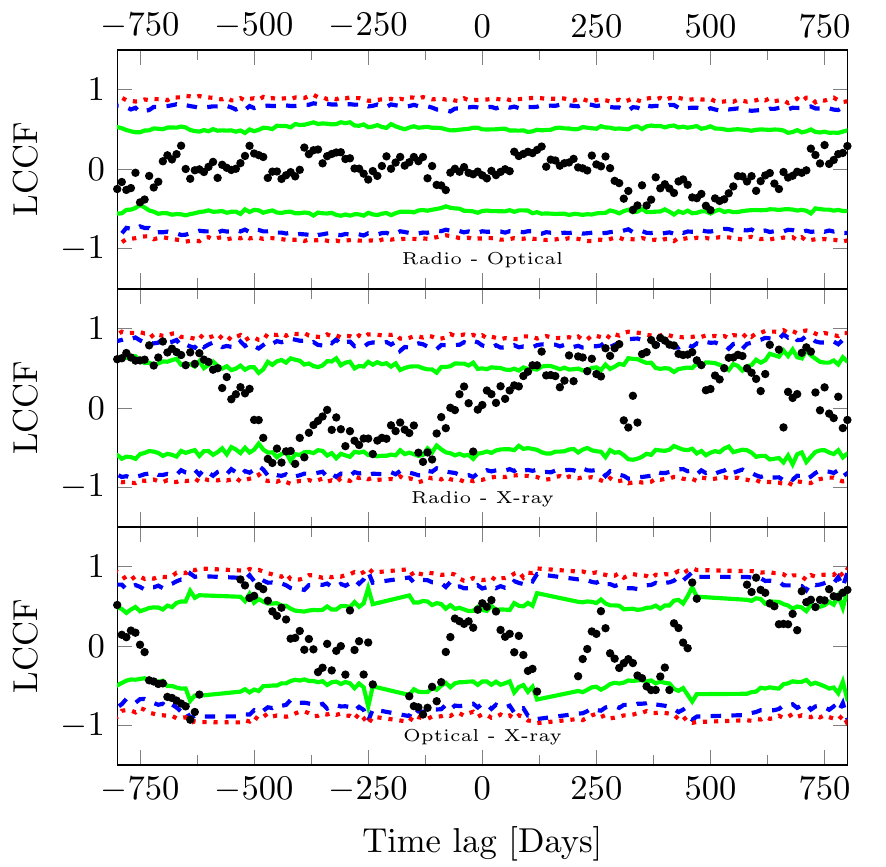}
\caption{\label{1424lccffig}{Same description as in Figure~\ref{0521lccffig} for PKS~1424+240.}}
\end{figure}

\begin{figure}[ht!]
\includegraphics{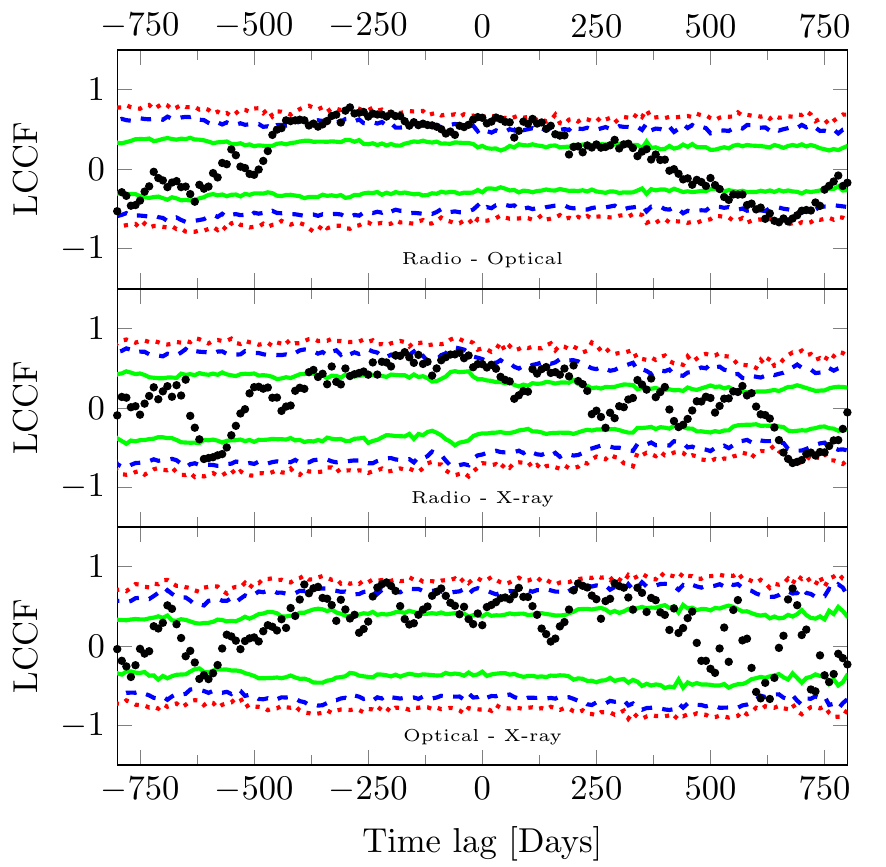}
\caption{\label{1727lccffig}{Same description as in Figure~\ref{0521lccffig} for 1ES~1727+502.}}
\end{figure}

\begin{figure}[ht!]
\includegraphics{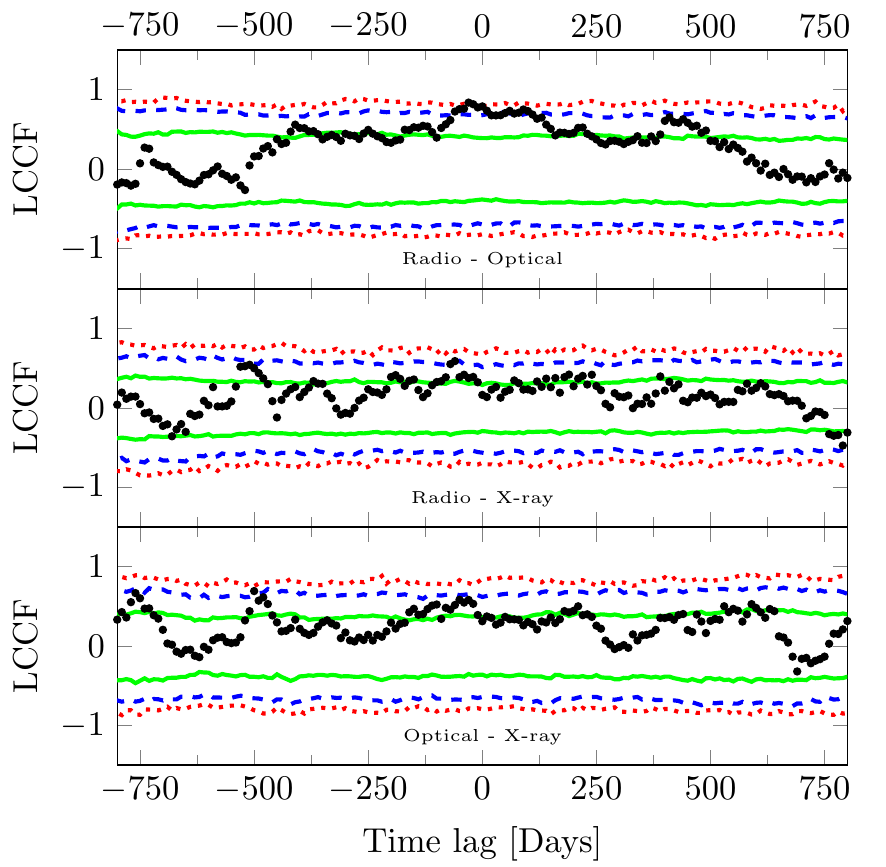}
\caption{\label{1959lccffig}{Same description as in Figure~\ref{0521lccffig} for 1ES~1959+650.}}
\end{figure}

\begin{figure}[ht!]
\includegraphics{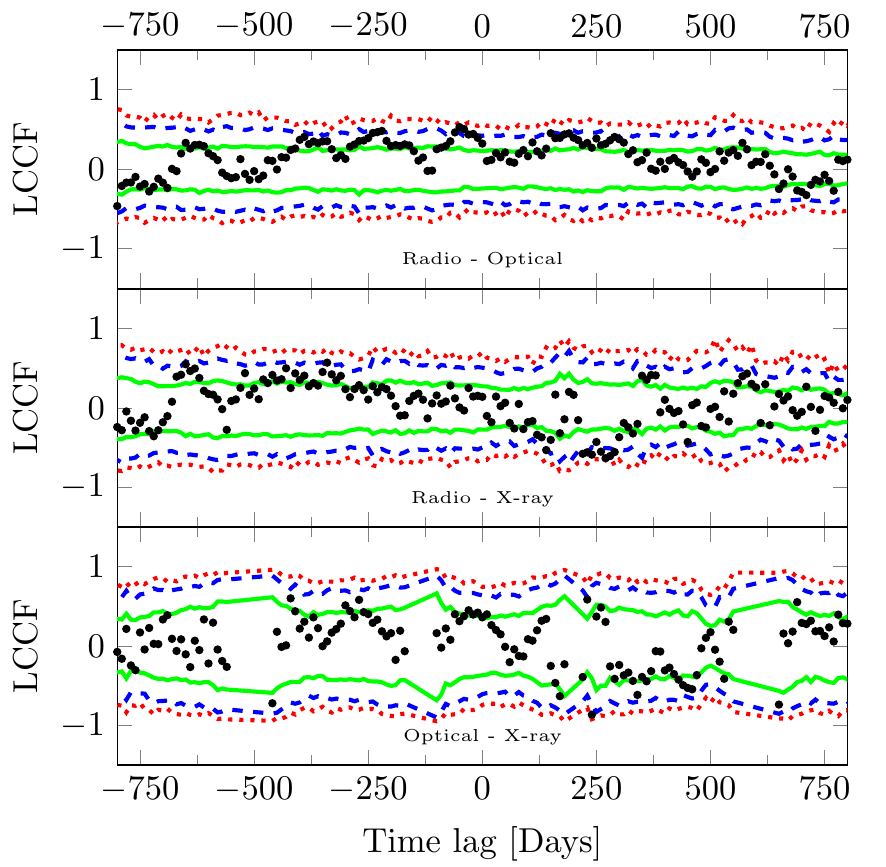}
\caption{\label{2344lccffig}{Same description as in Figure~\ref{0521lccffig} for 1ES~2344+514.}}
\end{figure}

In this section the results of the cross-correlation analysis (Sec.~\ref{sec:lccf}) are shown in Figures~\ref{0521lccffig} to \ref{2344lccffig} for each source. 

\section{\label{app:pol}Details of the polarisation modelling}
Here we provide a detailed description of the model parameters and fitting procedure we used in Section~\ref{sec:pol}. As discussed there, we modelled the variable polarisation component as a homogeneous cylindrical emission region in a jet with a helical magnetic field and computed the Stokes parameters using the formulae in \citet[]{2005MNRAS.360..869L}.

The relativistic outflow with speed $\beta$ is passing through this cylindrical region, with length  $l$ and radius $r$ whose symmetry axis is parallel to the velocity vector $\overline{v} = v z$. This vector forms an angle $\theta$ with the line of sight to the observer and a sky projected angle $\varphi_{\text{0}}$ with the vector pointing to the North in the sky. The cylinder has a uniform electron density $K_{\text{e}}$ and a power-law energy distribution of electrons with a slope $p$. The magnetic field in the cylinder has two components, one parallel to $\overline{v}$, and another one perpendicular to it, i.e.
\begin{equation}
\label{lutikaava}
\overline{B'} = (B'_x, B'_y, B'_z) = B_0[\sin{\psi'}(-\sin{\phi},0,\cos{\phi}) + \cos{\psi'}(0,0,1)],
\end{equation}
where $\psi'$ is the magnetic field pitch angle, $\phi$ is the azimuthal angle in the plane perpendicular to $\overline{v}$ (z-axis) and we follow the notation in \cite{2005MNRAS.360..869L} by marking quantities B and $\psi$ in the co-moving plane as primed. The model for the variable component has now 9 parameters, $\beta$, $\theta$, $\varphi_{\text{0}}$, $l$, $r$, $K_{\text{e}}$, $p$, $B_{\text{0}}$, and $\psi'$. The total number of parameters to model the variable and constant components is 12 the constant component has three free parameters: $I_C$, $Q_C$ and $U_C$.

Most of the parameters in the model cannot be constrained with monochromatic observational data due to a high degree of degeneracy in the model. Since the variable (blob) emission region is assumed to be homogeneous and unresolved, {Equation~(2)} of \cite{2005MNRAS.360..869L} can be reformulated as follow and the Stokes parameter $I$ becomes
\begin{equation}
I = \text{const} \times \frac{p + 7/3}{p + 1} 
\frac{\delta^{{2+(p-1)/2}}}{D^2(1+z)^{2+(p-1)/2}}
\frac{K_e \pi r^2 l}{\sin{\theta}} \int |B' \sin{\psi'}| dS,
\end{equation}
where $D$ is the luminosity distance of the source and the integration is over the volume of the cylinder. There are infinite combinations of $\beta$, $\theta$, $p$, $K_{\text{e}}$, $r$, $l$ and $B_{\text{0}}$ which produce the same $I$ (and $Q$ and $U$), so there is no way to constrain these parameters.

To deal with the degeneracy, the fit was made {in the $Q-U$ plane following these assumptions/procedure}: we fixed $p$ to 2.1, $r$ to $2.5 \times 10^{15}$ cm, $l$ to $5 \times 10^{15}$\,cm $B_0$ to  0.1\,Gauss and $\beta$ to 0.99. These values are similar to those applied for the SED modelling (Sec.~\ref{sec:sed_2comp}). At each iteration, we then determine the $K_{\text{e}}$ that equals the model $I$ to the observed $I$ value for each data point. We use this value of $K_{\text{e}}$ to compute $Q_{\text{V}}$ and $U_{\text{V}}$ \citep[see {Equation 2} in][]{2005MNRAS.360..869L}. We thus assumed that changes in the $I_{\text{V}}$ were due to changes in $K_{\text{e}}$. This also means that we effectively fit two linearly polarised components with constant $Q$ and $U$, but with $I$ of the other (variable) one changing. It is clear from examining the data that this simple model can {fit} only major trends in the data. There is a lot of fast variability (see Figs.~\ref{0521mwlfig}-~\ref{2344mwlfig} panels 2,3 and 4 from the top), which is probably caused by random turbulence and thus cannot be explained by a deterministic model. We treat this fast flickering as pure noise by adding one more parameter to the model. This parameter, $\sigma$, is added in quadrature to the errors of the observed Stokes parameters $Q$ and $U$ when we compute the likelihood.

There are thus 7 free parameters in our final model: the constant-component Stokes parameters $I_{\text{C}}$, $Q_{\text{C}}$ and $U_{\text{C}}$, viewing angle $\theta$, magnetic field pitch angle $\psi'$, jet position angle $\varphi_0$ and {standard deviation} of the turbulence, $\sigma$. The fit was made using a Monte Carlo Markov Chain (MCMC) ensemble sampler \citep{goodman2010}. The posteriori probability was first sampled with 21 walkers using 10000 steps and best-fit values for the parameters were then obtained from the marginalised distributions. {We found that the fit stabilised quite well after 50 iterations. We thus treated the first 50 iterations as the "burn-in" phase, and discarded them.} Since the model "predicts" $I$ always correctly, the posteriori was computed from the $Q$ and $U$ data only. The priors were set up in the following way: $I_{\text{C}}$: flat from 0 to minimum observed flux (given in column 3 of Table \ref{poltable}), $Q_{\text{C}}$ and $U_{\text{C}}$: Gaussian with $\mu = 0.0$ and $\sigma = 0.3$ mJy, $\theta$: flat 0 to $10\degr$, $\psi'$: flat 0 to $90\degr$, $\varphi_0$: flat 0 to $180\degr$ and $\sigma$: Gaussian with $\mu$ = 0.1 mJy and $\sigma$ = 0.1 mJy.

\begin{figure*}
    \centering
    \includegraphics{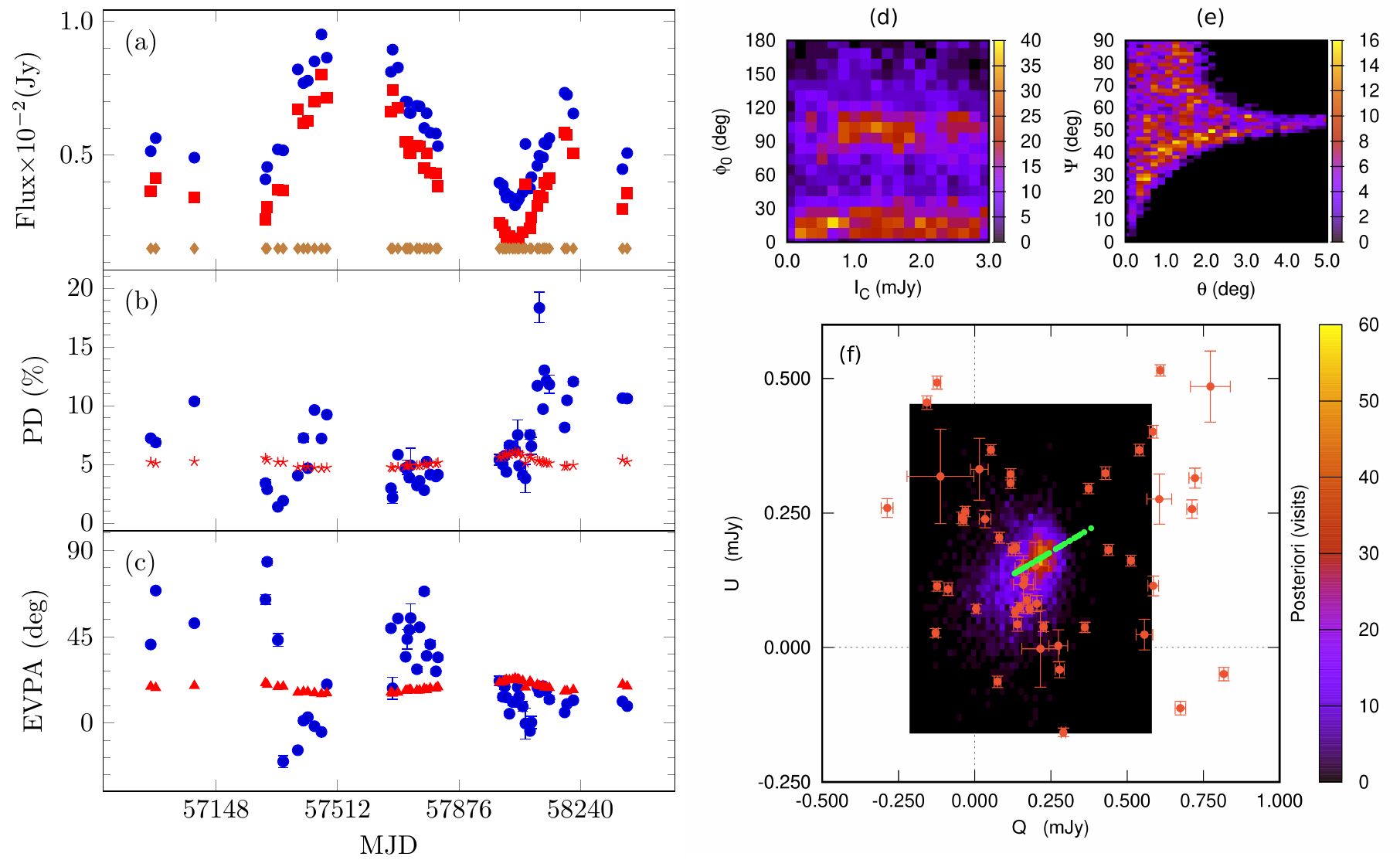}
    \caption{\label{polpost0521}{Results of polarisation analysis for VER~J0521+211. Panels (a): observed optical R-band flux (blue circles), variable component model (red squares), and model constant component (brown diamonds). Panels (b): observed (blue circles) and modelled (red stars) optical polarisation degree. Panels (c): observed (blue circle) and modelled (red triangles) electric vector polarisation angle. Panels (d) to (f): Posteriori distributions of the polarisation fitting. 
    The colour-scale gives the number of visits by the sampler in each cell.
    In addition, panel (f) shows the observations in the $Q-U$ plane (orange) and evolution of the model (light green). }}
\end{figure*}

\begin{figure*}
    \centering
    \includegraphics{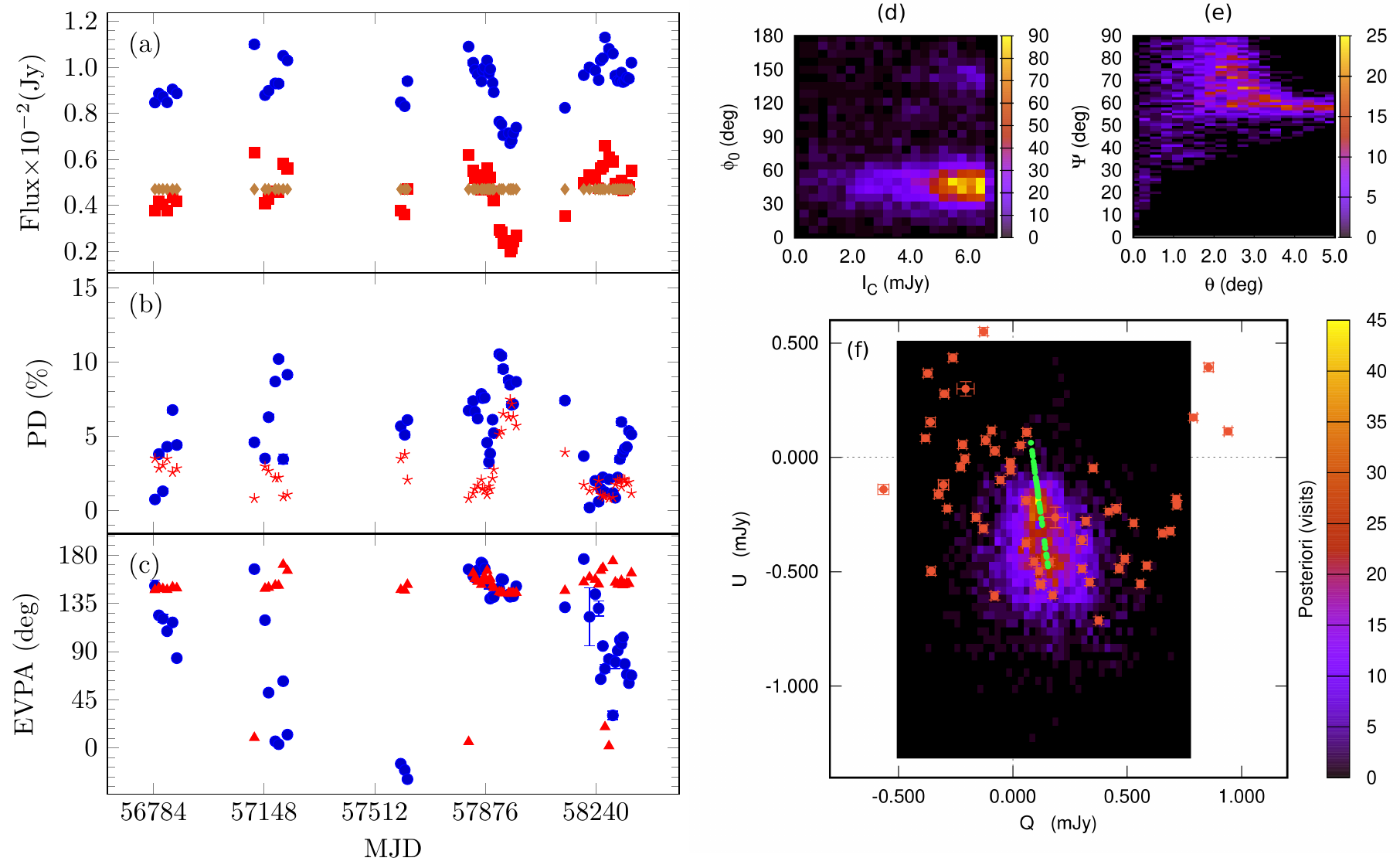}
    \caption{\label{polpost1424}{Same description as in Figure~\ref{polpost0521} for PKS~1424+240.}}
\end{figure*}

\begin{figure*}
    \centering
    \includegraphics{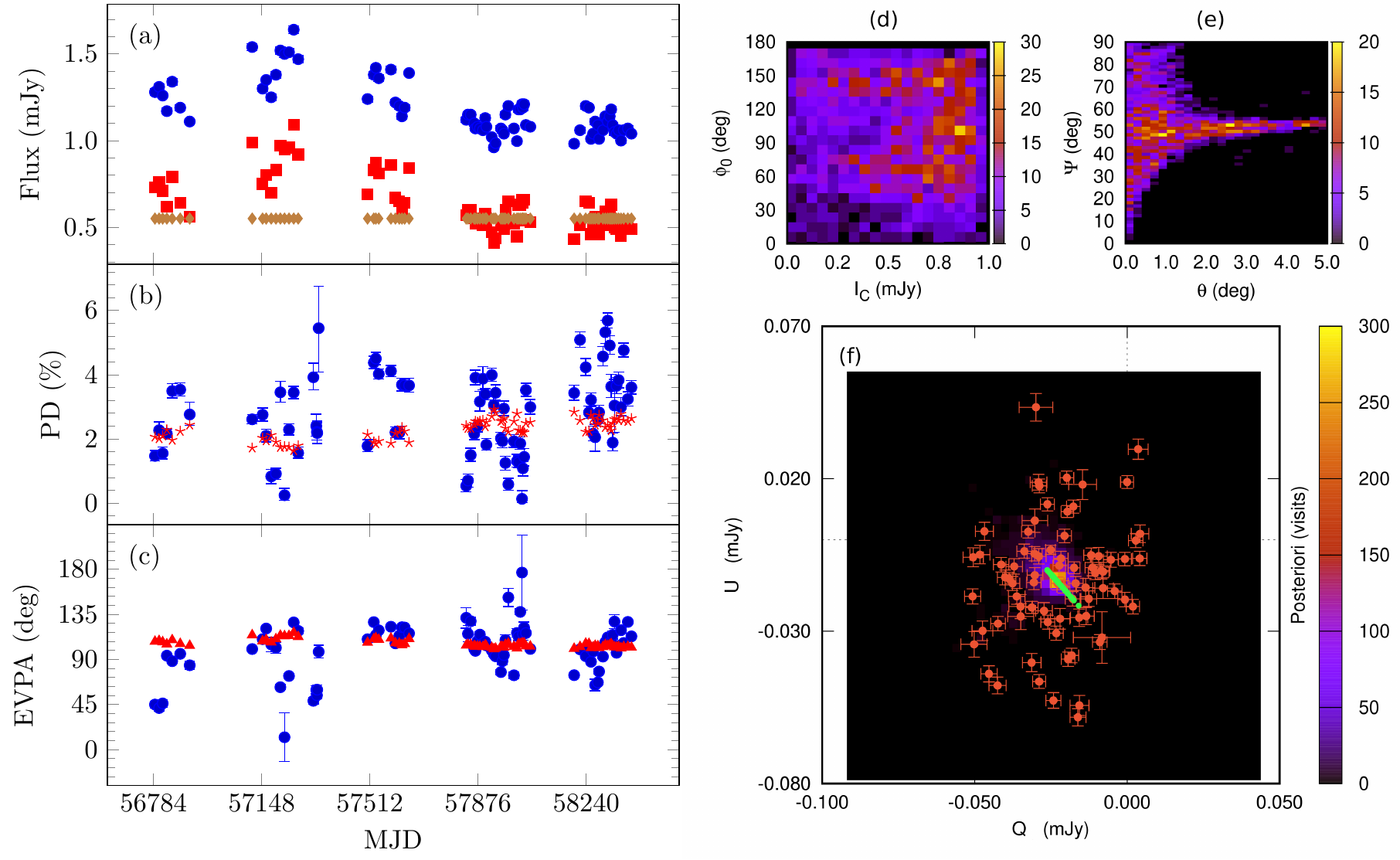}
    \caption{\label{polpost1727}{Same description as in Figure~\ref{polpost0521} for 1ES~1727+502.}}
\end{figure*}

\begin{figure*}
    \centering
    \includegraphics{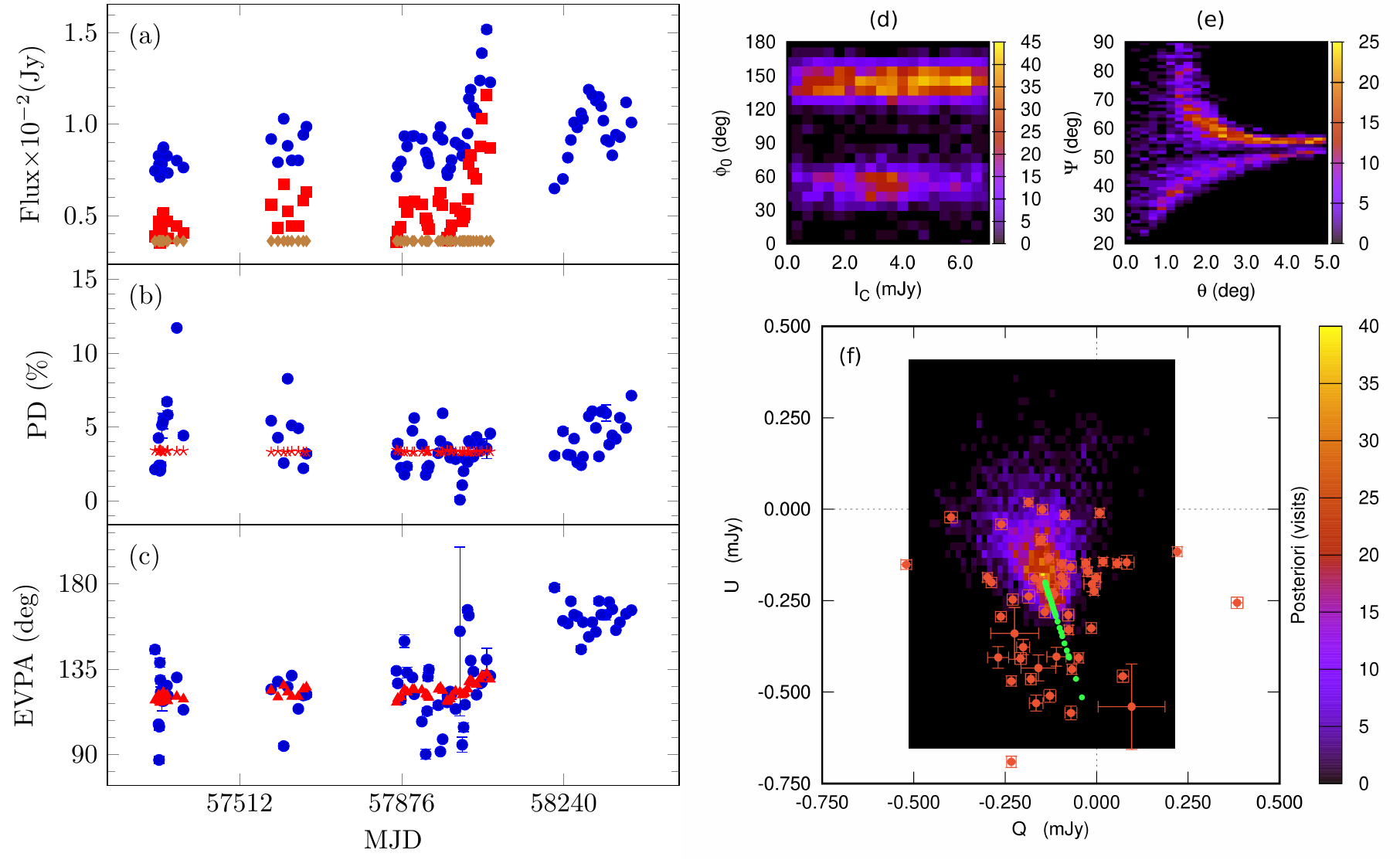}
    \caption{\label{polpost1959}{Same description as in Figure~\ref{polpost0521} for 1ES~1959+650. The last season was excluded from the polarisation fitting (see text).}}
\end{figure*}

\begin{figure*}
    \centering
    \includegraphics{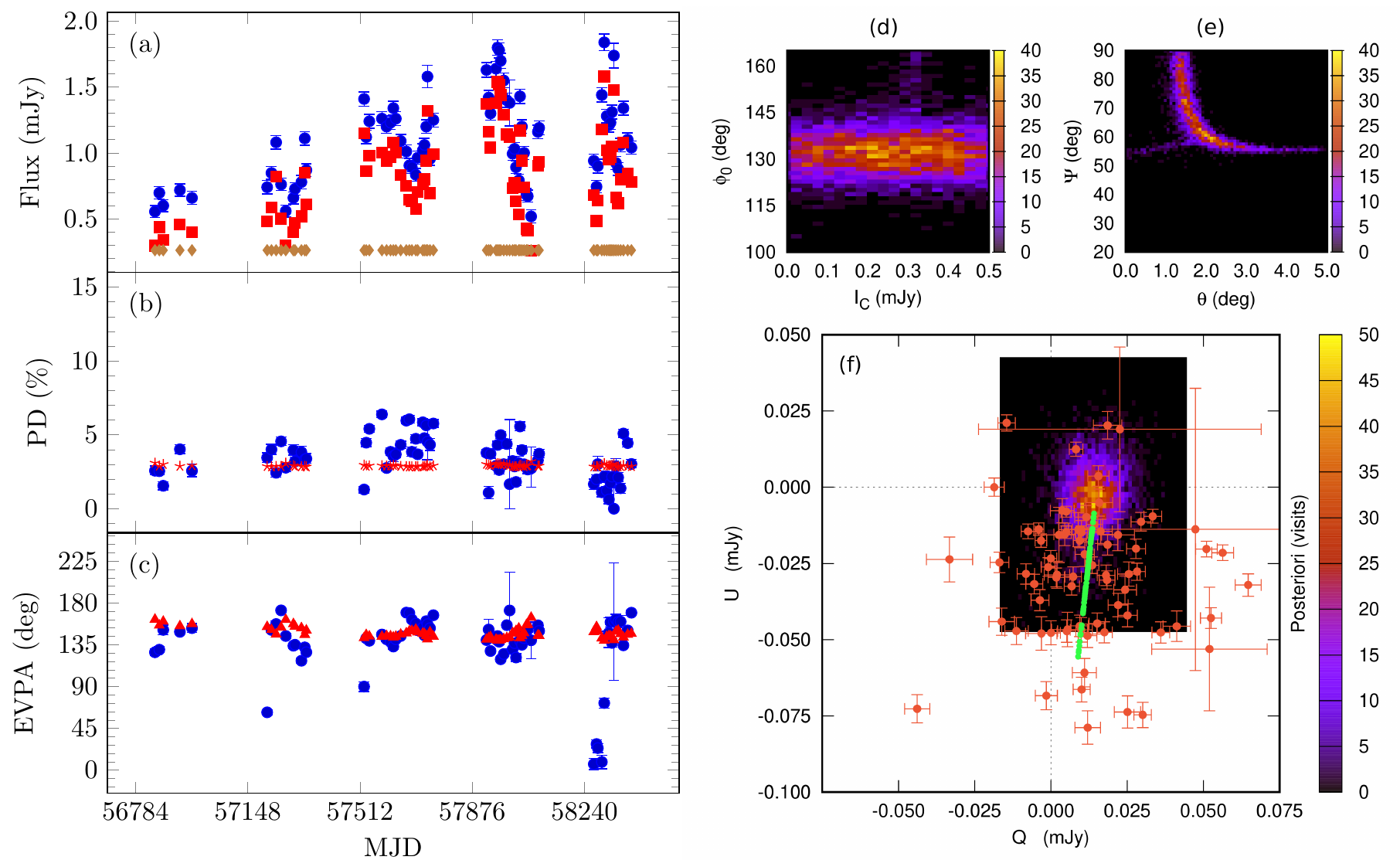}
    \caption{\label{polpost1}{Same description as in Figure~\ref{polpost0521} for 1ES~2344+514.}}
\end{figure*}

The results of the fits are illustrated in Figures~\ref{polpost0521} to \ref{polpost1} and summarised in Table \ref{poltable}. The error bars give the 68\% confidence intervals derived from marginalised distributions. Figure \ref{polpost1} gives an example of typical posteriori distributions. This figure illustrates some common trends, which we will now discuss.


As discussed above, the EVPA of the variable component is always parallel or perpendicular to the projected direction of the jet in the sky. At low pitch angles ($\psi' \lesssim 56\degr$) the EVPA is perpendicular and at higher angles ($\psi' > 56\degr$) parallel to the jet \citep[see Fig. 3 in][]{2017MNRAS.467.3876L}. The degree of polarisation is at maximum when $\psi' = 0\degr$ or $90\degr$ and approaches zero when $\psi'= \psi'_{\text{crit}} \sim 56\degr$. This explains why our fitted $\psi'$ are so similar among our targets; values close to $56\degr$ are the only way to produce the observed low degrees of polarisation. The exact value of $\psi'_{\text{crit}}$ depends on the viewing angle, which explains why the posteriori is confined to such a narrow strip (see upper left panel of Fig. \ref{polpost1}). However, the narrowness of this strip does not mean that we have put strict constrain{t}s on $\psi'$. We have simply forced it to this range by the assumption of a perfectly organised magnetic field. Had we introduced random variations into the magnetic field, the integrated degree of polarisation would have been at the observed level over a wider range of $\psi'$.

The posteriori for the jet angle $\varphi_0$ has typically two branches separated by $90\degr$. In the upper left panel of Figure~\ref{polpost1959} this shows up as two horizontal bands, one of which is much weaker than the other. The weaker branch is visible also in the upper left panel as a small extension towards the lower left corner. This is again a product of the possibility to have two EVPA orientations with respect to the jet axis. Table \ref{poltable} shows a comparison between our findings from this optical polarisation analysis and the VLBI results listed in Section~\ref{sec:VLBA}. The comparison is not straightforward due to the $\pm180\degr$ ambiguity of the EVPA and the 0 or $90\degr$ relative orientation if the jet and the EVPA allowed by our model.

{We also note that our model fails to
fit the EVPA change in 1ES~1959+650 during the last observing season (Fig \ref{polpost1959}). The EVPA changes by $\sim$ 45\degr, which cannot be done by simply
changing the relative contributions of the two components, whose properties are at the same time consistent with the earlier periods. In our formulation the only way is to change $\varphi_0$, i.e. the orientation of the jet.
Therefore, we excluded the last season of data from the polarisation fitting.}

Finally, with our present model, the evolution of $Q$ and $U$ is always along a straight line in the $Q,U$-plane. The observed evolution is much more complex, but we have simplified our treatment by introducing the parameter $\sigma$, the {standard deviation} of the random turbulent variations of $Q$ and $U$. {This
parameter adjusts itself during the fit according to how well the rest of the parameters fit the data. If the predictive power of the latter is weak, $\sigma$ increases to accommodate the increased residuals. This happens automatically since increasing $\sigma$ increases the likelihood up to a point, after which the likelihood begins to decrease with increasing sigma. As a result, $\sigma$ adjusts itself to a value roughly equal to a value the standard deviation of the residuals.} In the case of Figure~\ref{polpost1}, $\sigma = 0.018$ mJy, so adding the turbulent component would mean adding a Gaussian random variable with this $\sigma$ to every point in the evolution of the model in Figure~\ref{polpost1}, which would scatter the model points similarly to the observed points. {As discussed in Section~\ref{sec:pol}, the addition of parameter $\sigma$ makes the model very smooth and thus it does not reproduce the observed complex variability very well. The MCMC loop finds a good fit simply by increasing the errors, implying that the model we selected for the variable component is probably too simple. If our model had more explanatory power, $\sigma$ would tend to lower values and the model would follow the data points more closely. However, the approach introduced in this work, i.e. the Bayesian fitting of physical model to optical polarisation data, will be further investigated in future work.}

\section{\label{app:sedepoch}Details of the SED modelling}
Table~\ref{sedepochs} summarises the detailed description of the selection procedure used for MWL observational data sets used for SED modelling (Sec.~\ref{sec:sed}). Here we discuss the simultaneity of the modelled data and some details of the model parameters for each source. In general for each case of SED modelling, the radio data point at 15\,GHz was selected based on the shortest time lag between observation performed by OVRO and other instruments. All of the UV and optical data points were corrected Galactic extinction and the contribution of host-galaxy flux. 

\subsection*{VER~J0521+211}
The VHE gamma-ray spectrum was constructed by combining all of MAGIC observations between MJD 56580.18 and 56627.95 (4.7\,h distributed in 4 nights). \textit{Fermi}-LAT data obtained between MJD 56580 and 56627 were used for building the HE gamma-ray spectrum. The selected observations performed by \textit{Swift}-XRT and \textit{Swift}-UVOT instruments were carried out on MJD 56625.30, which is the most simultaneous observation to one of the four MAGIC observation windows (MJD 56625.04-56625.12). Similarly, we used optical R-band data on MJD 56625.15 obtained by the KVA telescope. {For this source,} the contribution of the host-galaxy flux {in the UV and optical bands} was neglected based on the uncertainty of the redshift and the reported redshift lower limit. In this period the source was clearly in high state in the optical, X-ray and VHE gamma-ray band compared to the previous observations \citet{2013ApJ...776...69A}. 

\input{tables/tab_sedepochs.tex}

Panel (a) in Figure \ref{fig:seds} illustrates the broadband SED of VER~J0521+211. The two-component model can reproduce the observed quasi-simultaneous data using the parameters sets for the two emission regions reported in Table \ref{sedparams}. The core Doppler factor was calculated from the apparent motion seen in the VLBI data assuming jet viewing angle of $5^{\circ}$ and a redshift of 0.18. The size of the blob emission region was set to be smaller than the shortest variability timescale in the data set (24\,h detected in X-rays band) as the sampling of the light curves would limit our capability to detect shorter timescales of variability. However, due to degeneracy between different parameters, we could achieve equally good representation of the data by increasing $R$ and decreasing $K$.

The comparison of the SED parameters of the blob in Table~\ref{sedparams} with those obtained from single-zone SSC model tested by \citet{2013ApJ...776...69A}, shows that the main differences are: the size of emission region ($4.0 \times 10^{17}$\,cm for one-zone model vs. $1.3 \times 10^{16}$\,cm for two-component model); magnetic field strength (0.0025 vs. 0.1 G) which affect the jet equipartition; the Doppler factor (30 vs. 12); and the maximum electron Lorentz factor ($2.0 \times 10^{6}$ for one-zone model vs. $4.0 \times 10^{5}$ for two-component model). These differences are in line with the general trends discussed in Section~\ref{sec:twoone}.

\subsection*{PKS~1424+240}

As discussed in Section \ref{magic}, we attempted to model the SED of PKS~1424+240 using the data obtained from observations during the two campaigns. Neither of the campaigns were performed during particularly high flux states in lower (optical, X-rays) bands.

\textit{Data set of the observation campaign during 2014:}
The VHE gamma-ray spectrum was constructed by combining all of the MAGIC observations in the time span from MJD 56740.06 and 56825.99. A similar time window was used to compute the HE gamma-ray spectrum from the data obtained with \textit{Fermi}-LAT. We searched for variability of spectral parameters ($F_0$, $\Gamma$, and $\beta$) in HE gamma-ray band in the selected time window. These parameters did not show any significant (at $3\sigma$ confidence level) variability at weekly timescale. In the presented data, there is a mismatch between the HE and VHE gamma-ray spectra at energies between 40 and 60\,GeV. This mismatch is mainly due to non-simultaneity of observations. However, considering the systematic uncertainty of both instruments in that range, the mismatch is negligible. At MJD 56800.93, the VHE gamma-ray flux was the most consistent measurement to the average flux of the entire campaign. Therefore, we selected  \textit{Swift}-XRT (MJD 56801.25), \textit{Swift}-UVOT (MJD 56801.26) and optical (MJD 56800.96) observations which were quasi-simultaneous to MJD 56800.93.
Panel (b) in Figure \ref{fig:seds} shows the broadband SED of PKS~1424+240 (compiled from observations of the 2014 campaign). The parameter sets for the two emission regions can reproduce the observed quasi-simultaneous observational data (see Table \ref{sedparams}). The core Doppler factor is calculated from apparent speed observed in VLBI using a jet viewing angle of $3\degr$. The size of the blob emission region is compatible with a variability timescale seen in the optical and X-rays (1 and 2\,days respectively).

\textit{Data set of the observation campaign during 2015:}
The VHE gamma-ray spectrum was constructed by combining all of the MAGIC observations performed between  MJD 57045.05 and 57186.06. Similar time span was used to compute the HE gamma-ray spectrum from the data obtained with \textit{Fermi}-LAT. Following the procedure used for observation campaign during 2014, the spectral parameters did not show any significant (at $3\sigma$ confidence level) variability at weekly timescale. We selected \textit{Swift}-XRT (MJD 57135.26), \textit{Swift}-UVOT (MJD 57135.26) and optical (MJD 57135.11) observations which were simultaneous to one of the MAGIC observation window.
Panel (c) in Figure \ref{fig:seds} presents the broadband SED of PKS~1424+240 during the 2015 campaign. The observed quasi-simultaneous observational data are reproduced by the sets of parameters summarised in Table \ref{sedparams}. For the core, we use parameters similar to those in 2014, with minor changes to reproduce the overall lower state of the synchrotron part of the SED. For the blob, $n_2$ is softer than in 2014 to reproduce the lower state in X-rays. Also the size of the blob emission region is slightly smaller than in 2014, even if no variability was detected during this campaign, again to reproduce the lower X-ray state. 

\textit{SED modelling results in optical band:}
In contrast with the result in Section \ref{sec:component}, the emission from the core dominates the total flux in the optical band ($F_{\text{core}}/F_{\text{total}}=0.93$). We tried to find sets of parameters in which the optical emission would be dominated by the blob component. Only solutions we found that could produce the HE and VHE gamma-ray part of the SED, had very low magnetic field strength and was far from equipartition, while the set of parameters that we present here give $U'_B/U'_e=2.39$. We emphasis that the derived value for $F_{\text{core}}/F_{\text{total}}$ in Section \ref{sec:component} is a minimum value, so there is no contradiction between these results. The optical polarisation method (Sec.~\ref{sec:pol}) suggests that there {are} two components contributing to the optical flux, which further highlights the general conclusion in Section~\ref{sec:twocomp} that these two methods are complementary. 

\subsection*{1ES~1727+502}
The VHE gamma-ray spectrum was computed by combining all of the MAGIC observations performed between MJD 57306.83 and 57327.83. The HE gamma-ray spectrum was built using the data obtained with \textit{Fermi}-LAT in the time span of MJD 57263 and 57353. The \textit{Swift}-XRT and \textit{Swift}-UVOT observations were selected from the night when the VHE gamma-ray flux was consistent with average flux, i.e. on MJD 57307.99. The optical data point obtained from the observation performed with KVA telescope on MJD 57307.83. 

Panel (d) in Figure \ref{fig:seds} shows the broadband SED of 1ES~1727+502 during flaring activity in 2015. The observed quasi-simultaneous data can be reproduced in the framework of a two-component model using the sets of parameters reported in Table \ref{sedparams}. The size of blob emission region is compatible with variability timescale of 6.3\,h which is in agreement with our observational constrain of 24\,h detected in VHE gamma-ray band. 

The comparison of the SED parameters obtained with a one-zone SSC model used by \citet{2015ApJ...808..110A} with those for the blob emission region (Table \ref{sedparams}) shows similar differences as seen in VER~J0521+211. The size of the emission region is ($4.3-7.4 \times 10^{17}$\,cm for one-zone model vs. $7.1 \times 10^{15}$\,cm for two-component model), magnetic field strength (0.0003-0.0006 vs. 0.1\,G), the Doppler factor (30 vs. 11) and the maximum electron Lorentz factor ($5.5-7.0 \times 10^{6}$ for one-zone model vs. $1.3 \times 10^{6}$ for two-component model).

\subsection*{1ES~1959+650}
Following the discussion in Section \ref{magic}, three data sets were used to build the SEDs of this source in different states of VHE gamma-ray flux. 

\textit{Low state:}
The VHE gamma-ray spectrum was calculated using the 1.2\,h of the MAGIC data from observation starting at MJD 57711.82. The time span between MJD 57711.43 and 57712.35 {was} used for building the HE gamma-ray spectrum adopting \textit{Fermi}-LAT data. For X-ray, UV and optical, the observations which were performed on MJD 57711.58 (\textit{Swift}-XRT), 57711.59 (\textit{Swift}-UVOT) and 57711.82 (KVA) were used. 

\textit{Intermediate state:}
The VHE gamma-ray spectrum was constructed using 1.4\,h of MAGIC data from observation starting at MJD 57547.13. The \textit{Fermi}-LAT data obtained between MJD 57545.16 and 57550.19 were used for building the HE gamma-ray spectrum. For X-ray and UV, the observations which were performed on MJD 57547.13 (\textit{Swift}-XRT and \textit{Swift}-UVOT) were used. The optical and UV data from (MJD 57547.14) were {used in SED modelling}.

\textit{High state:}
The VHE gamma-ray spectrum was computed using 2.2\,h of MAGIC data from observations that started at MJD 57553.06. The \textit{Fermi}-LAT data obtained between MJD 57552.00 and 57554.00 were used for calculating the HE gamma-ray spectrum. For X-ray and UV, the observations which were performed at MJD 57553.10 (\textit{Swift}-XRT and \textit{Swift}-UVOT) were used. The optical data point from (MJD 57553.13) {was used in SED modelling}.

It should be noted that even the "low state" SED is somewhat above the archival data from previous "low state" campaigns, as well visible in Figure~\ref{fig:seds}. The comparison of one-zone SSC models are discussed in details by \citet{1959flare}, but the differences follow the general trend we discussed in Section~\ref{sec:twoone}.

\subsection*{1ES~2344+514}
The VHE gamma-ray spectrum was calculated using 0.5\,h of MAGIC data from observations that started at MJD 57612.06 in order to have quasi-simultaneous MWL coverage. Due to the source faintness in the HE gamma-ray band, the \textit{Fermi}-LAT data obtained between MJD 57520.00 and 57704.00 {were} used for building the spectrum. For X-ray and UV, the observations which {were} performed on MJD 57613.52 (\textit{Swift}-XRT and \textit{Swift}-UVOT) are used. The optical data (R-band) from (MJD 57611.02) {was used in SED modelling}. Figure \ref{fig:seds} (h), demonstrates the broadband SED of 1ES~2344+514 during the 2016. As can be seen the source is relatively bright compared to archival observations in X-rays, HE and VHE gamma-rays during this period.

The one-zone SSC can describe the low state SED of the source \citep{2013A&A...556A..67A} using the data sets of the observation campaign during 2008. However, the radio data is not included in that modelling, and is assumed to origin from different component further out. We find that most of the parameters of {the one-zone model reported} by \citet{2013A&A...556A..67A}, are in good agreement with the blob parameters in this work. It is notable that the one-zone model used the Doppler factor ($\delta=20$) and magnetic field strength ($B=0.07$\,G), while these parameters for the blob emission in this work are more physically realistic ($\delta=6$ and $B=0.1$\,G). Moreover, similar quasi-simultaneous data set is used by \citet{2344fact} and the set of parameters describing the broadband SED have low magnetic field strength ($B=0.02$\,G), emission region size ($R=1 \times 10^{16}$\,cm), and high break and and maximum electron Lorentz factor ($\gamma_{\text{b}}=1.8\times 10^{6}$ and $\gamma_{\text{max}}=8.0\times 10^{6}$). 

\end{appendix}

\end{document}

%% file: authors.tex
\author{
MAGIC Collaboration\thanks{\textit{Send offprint requests to} MAGIC Collaboration (\email{contact.magic@mpp.mpg.de}). Corresponding authors are V.~Fallah~Ramazani, E.~Lindfors, and K.~Nilsson.}:
V.~A.~Acciari\inst{\ref{inst1}} \and
S.~Ansoldi\inst{\ref{inst2},\ref{inst24}} \and
L.~A.~Antonelli\inst{\ref{inst3}} \and
A.~Arbet Engels\inst{\ref{inst4}} \and
D.~Baack\inst{\ref{inst5}} \and
A.~Babi\'c\inst{\ref{inst6}} \and
B.~Banerjee\inst{\ref{inst7}} \and
U.~Barres de Almeida\inst{\ref{inst8}} \and
J.~A.~Barrio\inst{\ref{inst9}} \and
J.~Becerra Gonz\'alez\inst{\ref{inst1}} \and
W.~Bednarek\inst{\ref{inst10}} \and
L.~Bellizzi\inst{\ref{inst11}} \and
E.~Bernardini\inst{\ref{inst12},\ref{inst16}} \and
A.~Berti\inst{\ref{inst13}} \and
J.~Besenrieder\inst{\ref{inst14}} \and
W.~Bhattacharyya\inst{\ref{inst12}} \and
C.~Bigongiari\inst{\ref{inst3}} \and
A.~Biland\inst{\ref{inst4}} \and
O.~Blanch\inst{\ref{inst15}} \and
G.~Bonnoli\inst{\ref{inst11}} \and
\v{Z}.~Bo\v{s}njak\inst{\ref{inst6}} \and
G.~Busetto\inst{\ref{inst16}} \and
R.~Carosi\inst{\ref{inst17}} \and
G.~Ceribella\inst{\ref{inst14}} \and
M.~Cerruti\inst{\ref{inst18}} \and
Y.~Chai\inst{\ref{inst14}} \and
A.~Chilingarian\inst{\ref{inst19}} \and
S.~Cikota\inst{\ref{inst6}} \and
S.~M.~Colak\inst{\ref{inst15}} \and
U.~Colin\inst{\ref{inst14}} \and
E.~Colombo\inst{\ref{inst1}} \and
J.~L.~Contreras\inst{\ref{inst9}} \and
J.~Cortina\inst{\ref{inst20}} \and
S.~Covino\inst{\ref{inst3}} \and
G.~D'Amico\inst{\ref{inst14}} \and
V.~D'Elia\inst{\ref{inst3}} \and
P.~Da Vela\inst{\ref{inst17},\ref{inst26}} \and
F.~Dazzi\inst{\ref{inst3}} \and
A.~De Angelis\inst{\ref{inst16}} \and
B.~De Lotto\inst{\ref{inst2}} \and
M.~Delfino\inst{\ref{inst15},\ref{inst27}} \and
J.~Delgado\inst{\ref{inst15},\ref{inst27}} \and
D.~Depaoli\inst{\ref{inst13}} \and
F.~Di Pierro\inst{\ref{inst13}} \and
L.~Di Venere\inst{\ref{inst13}} \and
E.~Do Souto Espi\~neira\inst{\ref{inst15}} \and
D.~Dominis Prester\inst{\ref{inst6}} \and
A.~Donini\inst{\ref{inst2}} \and
D.~Dorner\inst{\ref{inst21}} \and
M.~Doro\inst{\ref{inst16}} \and
D.~Elsaesser\inst{\ref{inst5}} \and
V.~Fallah Ramazani\inst{\ref{inst22}} \and
A.~Fattorini\inst{\ref{inst5}} \and
G.~Ferrara\inst{\ref{inst3}} \and
L.~Foffano\inst{\ref{inst16}} \and
M.~V.~Fonseca\inst{\ref{inst9}} \and
L.~Font\inst{\ref{inst23}} \and
C.~Fruck\inst{\ref{inst14}} \and
S.~Fukami\inst{\ref{inst24}} \and
R.~J.~Garc\'ia L\'opez\inst{\ref{inst1}} \and
M.~Garczarczyk\inst{\ref{inst12}} \and
S.~Gasparyan\inst{\ref{inst19}} \and
M.~Gaug\inst{\ref{inst23}} \and
N.~Giglietto\inst{\ref{inst13}} \and
F.~Giordano\inst{\ref{inst13}} \and
P.~Gliwny\inst{\ref{inst10}} \and
N.~Godinovi\'c\inst{\ref{inst6}} \and
D.~Green\inst{\ref{inst14}} \and
D.~Hadasch\inst{\ref{inst24}} \and
A.~Hahn\inst{\ref{inst14}} \and
J.~Herrera\inst{\ref{inst1}} \and
J.~Hoang\inst{\ref{inst9}} \and
D.~Hrupec\inst{\ref{inst6}} \and
M.~H\"utten\inst{\ref{inst14}} \and
T.~Inada\inst{\ref{inst24}} \and
S.~Inoue\inst{\ref{inst24}} \and
K.~Ishio\inst{\ref{inst14}} \and
Y.~Iwamura\inst{\ref{inst24}} \and
L.~Jouvin\inst{\ref{inst15}} \and
Y.~Kajiwara\inst{\ref{inst24}} \and
M.~Karjalainen\inst{\ref{inst1}} \and
D.~Kerszberg\inst{\ref{inst15}} \and
Y.~Kobayashi\inst{\ref{inst24}} \and
H.~Kubo\inst{\ref{inst24}} \and
J.~Kushida\inst{\ref{inst24}} \and
A.~Lamastra\inst{\ref{inst3}} \and
D.~Lelas\inst{\ref{inst6}} \and
F.~Leone\inst{\ref{inst3}} \and
E.~Lindfors\inst{\ref{inst22}} \and
S.~Lombardi\inst{\ref{inst3}} \and
F.~Longo\inst{\ref{inst2},\ref{inst28}} \and
M.~L\'opez\inst{\ref{inst9}} \and
R.~L\'opez-Coto\inst{\ref{inst16}} \and
A.~L\'opez-Oramas\inst{\ref{inst1}} \and
S.~Loporchio\inst{\ref{inst13}} \and
B.~Machado de Oliveira Fraga\inst{\ref{inst8}} \and
C.~Maggio\inst{\ref{inst23}} \and
P.~Majumdar\inst{\ref{inst7}} \and
M.~Makariev\inst{\ref{inst25}} \and
M.~Mallamaci\inst{\ref{inst16}} \and
G.~Maneva\inst{\ref{inst25}} \and
M.~Manganaro\inst{\ref{inst6}} \and
K.~Mannheim\inst{\ref{inst21}} \and
L.~Maraschi\inst{\ref{inst3}} \and
M.~Mariotti\inst{\ref{inst16}} \and
M.~Mart\'inez\inst{\ref{inst15}} \and
D.~Mazin\inst{\ref{inst14},\ref{inst24}} \and
S.~Mender\inst{\ref{inst5}} \and
S.~Mi\'canovi\'c\inst{\ref{inst6}} \and
D.~Miceli\inst{\ref{inst2}} \and
T.~Miener\inst{\ref{inst9}} \and
M.~Minev\inst{\ref{inst25}} \and
J.~M.~Miranda\inst{\ref{inst11}} \and
R.~Mirzoyan\inst{\ref{inst14}} \and
E.~Molina\inst{\ref{inst18}} \and
A.~Moralejo\inst{\ref{inst15}} \and
D.~Morcuende\inst{\ref{inst9}} \and
V.~Moreno\inst{\ref{inst23}} \and
E.~Moretti\inst{\ref{inst15}} \and
P.~Munar-Adrover\inst{\ref{inst23}} \and
V.~Neustroev\inst{\ref{inst22}} \and
C.~Nigro\inst{\ref{inst15}} \and
K.~Nilsson\inst{\ref{inst22}} \and
D.~Ninci\inst{\ref{inst15}} \and
K.~Nishijima\inst{\ref{inst24}} \and
K.~Noda\inst{\ref{inst24}} \and
L.~Nogu\'es\inst{\ref{inst15}} \and
S.~Nozaki\inst{\ref{inst24}} \and
Y.~Ohtani\inst{\ref{inst24}} \and
T.~Oka\inst{\ref{inst24}} \and
J.~Otero-Santos\inst{\ref{inst1}} \and
M.~Palatiello\inst{\ref{inst2}} \and
D.~Paneque\inst{\ref{inst14}} \and
R.~Paoletti\inst{\ref{inst11}} \and
J.~M.~Paredes\inst{\ref{inst18}} \and
L.~Pavleti\'c\inst{\ref{inst6}} \and
P.~Pe\~nil\inst{\ref{inst9}} \and
M.~Peresano\inst{\ref{inst2}} \and
M.~Persic\inst{\ref{inst2},\ref{inst29}} \and
P.~G.~Prada Moroni\inst{\ref{inst17}} \and
E.~Prandini\inst{\ref{inst16}} \and
I.~Puljak\inst{\ref{inst6}} \and
W.~Rhode\inst{\ref{inst5}} \and
M.~Rib\'o\inst{\ref{inst18}} \and
J.~Rico\inst{\ref{inst15}} \and
C.~Righi\inst{\ref{inst3}} \and
A.~Rugliancich\inst{\ref{inst17}} \and
L.~Saha\inst{\ref{inst9}} \and
N.~Sahakyan\inst{\ref{inst19}} \and
T.~Saito\inst{\ref{inst24}} \and
S.~Sakurai\inst{\ref{inst24}} \and
K.~Satalecka\inst{\ref{inst12}} \and
B.~Schleicher\inst{\ref{inst21}} \and
K.~Schmidt\inst{\ref{inst5}} \and
T.~Schweizer\inst{\ref{inst14}} \and
J.~Sitarek\inst{\ref{inst10}} \and
I.~\v{S}nidari\'c\inst{\ref{inst6}} \and
D.~Sobczynska\inst{\ref{inst10}} \and
A.~Spolon\inst{\ref{inst16}} \and
A.~Stamerra\inst{\ref{inst3}} \and
D.~Strom\inst{\ref{inst14}} \and
M.~Strzys\inst{\ref{inst24}} \and
Y.~Suda\inst{\ref{inst14}} \and
T.~Suri\'c\inst{\ref{inst6}} \and
M.~Takahashi\inst{\ref{inst24}} \and
F.~Tavecchio\inst{\ref{inst3}} \and
P.~Temnikov\inst{\ref{inst25}} \and
T.~Terzi\'c\inst{\ref{inst6}} \and
M.~Teshima\inst{\ref{inst14},\ref{inst24}} \and
N.~Torres-Alb\`a\inst{\ref{inst18}} \and
L.~Tosti\inst{\ref{inst13}} \and
J.~van Scherpenberg\inst{\ref{inst14}} \and
G.~Vanzo\inst{\ref{inst1}} \and
M.~Vazquez Acosta\inst{\ref{inst1}} \and
S.~Ventura\inst{\ref{inst11}} \and
V.~Verguilov\inst{\ref{inst25}} \and
C.~F.~Vigorito\inst{\ref{inst13}} \and
V.~Vitale\inst{\ref{inst13}} \and
I.~Vovk\inst{\ref{inst24}} \and
M.~Will\inst{\ref{inst14}} \and
D.~Zari\'c\inst{\ref{inst6}} \and \\
M.~Nievas-Rosillo \inst{\ref{inst12}} \and
C. Arcaro \inst{\ref{inst88},\ref{inst87}} \and
F. D' Ammando \inst{\ref{inst89}} \and 
F.~de~Palma \inst{\ref{inst86}} \and 
M.~Hodges \inst{\ref{inst98}} \and
T.~Hovatta \inst{\ref{inst99},\ref{inst97}} \and
S.~Kiehlmann \inst{\ref{inst96},\ref{inst95}} \and
W.~Max-Moerbeck \inst{\ref{inst94}} \and
A.~C.~S.~Readhead \inst{\ref{inst98}} \and
R.~Reeves \inst{\ref{inst93}} \and 
L.~Takalo \inst{\ref{inst92}}\fnmsep\thanks{This paper is dedicated to the memory of our colleague and dear friend Leo Takalo 1952--2018, who played a crucial role in starting the Tuorla blazar monitoring program and contributed significantly to the data acquisition.}
R.~Reinthal \inst{\ref{inst92}} \and
J.~Jormanainen \inst{\ref{inst22}} \and
F.~Wierda \inst{\ref{inst92}} \and
S.~M.~Wagner \inst{\ref{inst21}} \and
A.~Berdyugin \inst{\ref{inst92}} \and
A.~Nabizadeh \inst{\ref{inst92}} \and
N.~Talebpour~Sheshvan \inst{\ref{inst92}} \and
A.~Oksanen \inst{\ref{inst85}} \and
R.~Bachev \inst{\ref{inst84}} \and
A.~Strigachev \inst{\ref{inst84}} \and
P.~Kehusmaa \inst{\ref{inst83}}
}
\authorrunning{MAGIC Collaboration}

\institute { 
\label{inst1}Inst. de Astrof\'isica de Canarias, E-38200 La Laguna, and Universidad de La Laguna, Dpto. Astrof\'isica, E-38206 La Laguna, Tenerife, Spain
\and
\label{inst2}Universit\`a di Udine, and INFN Trieste, I-33100 Udine, Italy
\and
\label{inst24}Japanese MAGIC Consortium: ICRR, The University of Tokyo, 277-8582 Chiba, Japan; Department of Physics, Kyoto University, 606-8502 Kyoto, Japan; Tokai University, 259-1292 Kanagawa, Japan; RIKEN, 351-0198 Saitama, Japan
\and
\label{inst3}National Institute for Astrophysics (INAF), I-00136 Rome, Italy
\and
\label{inst4}ETH Zurich, CH-8093 Zurich, Switzerland
\and
\label{inst5}Technische Universit\"at Dortmund, D-44221 Dortmund, Germany
\and
\label{inst6}Croatian Consortium: University of Rijeka, Department of Physics, 51000 Rijeka; University of Split - FESB, 21000 Split; University of Zagreb - FER, 10000 Zagreb; University of Osijek, 31000 Osijek; Rudjer Boskovic Institute, 10000 Zagreb, Croatia
\and
\label{inst7}Saha Institute of Nuclear Physics, HBNI, 1/AF Bidhannagar, Salt Lake, Sector-1, Kolkata 700064, India
\and
\label{inst8}Centro Brasileiro de Pesquisas F\'isicas (CBPF), 22290-180 URCA, Rio de Janeiro (RJ), Brasil
\and
\label{inst9}IPARCOS Institute and EMFTEL Department, Universidad Complutense de Madrid, E-28040 Madrid, Spain
\and
\label{inst10}University of Lodz, Faculty of Physics and Applied Informatics, Department of Astrophysics, 90-236 Lodz, Poland
\and
\label{inst11}Universit\`a di Siena and INFN Pisa, I-53100 Siena, Italy
\and
\label{inst12}Deutsches Elektronen-Synchrotron (DESY), D-15738 Zeuthen, Germany
\and
\label{inst16}Universit\`a di Padova and INFN, I-35131 Padova, Italy
\and
\label{inst13}Istituto Nazionale Fisica Nucleare (INFN), 00044 Frascati (Roma) Italy
\and
\label{inst14}Max-Planck-Institut f\"ur Physik, D-80805 M\"unchen, Germany
\and
\label{inst15}Institut de F\'isica d'Altes Energies (IFAE), The Barcelona Institute of Science and Technology (BIST), E-08193 Bellaterra (Barcelona), Spain
\and
\label{inst17}Universit\`a di Pisa, and INFN Pisa, I-56126 Pisa, Italy
\and
\label{inst18}Universitat de Barcelona, ICCUB, IEEC-UB, E-08028 Barcelona, Spain
\and
\label{inst19}The Armenian Consortium: ICRANet-Armenia at NAS RA, A. Alikhanyan National Laboratory
\and
\label{inst20}Centro de Investigaciones Energ\'eticas, Medioambientales y Tecnol\'ogicas, E-28040 Madrid, Spain
\and
\label{inst21}Universit\"at W\"urzburg, D-97074 W\"urzburg, Germany
\and
\label{inst22}Finnish MAGIC Consortium: Finnish Centre of Astronomy with ESO (FINCA), University of Turku, FI-20014 Turku, Finland; Astronomy Research Unit, University of Oulu, FI-90014 Oulu, Finland
\and
\label{inst23}Departament de F\'isica, and CERES-IEEC, Universitat Aut\`onoma de Barcelona, E-08193 Bellaterra, Spain
\and
\label{inst25}Inst. for Nucl. Research and Nucl. Energy, Bulgarian Academy of Sciences, BG-1784 Sofia, Bulgaria
\and
\label{inst26}now at University of Innsbruck
\and
\label{inst27}also at Port d'Informaci\'o Cient\'ifica (PIC) E-08193 Bellaterra (Barcelona) Spain
\and
\label{inst28}also at Dipartimento di Fisica, Universit\`a di Trieste, I-34127 Trieste, Italy
\and
\label{inst29}also at INAF-Trieste and Dept. of Physics \& Astronomy, University of Bologna
\and
\label{inst88}Centre for Space Research, North-West University, Potchefstroom 2520, South Africa 
\and
\label{inst87}also at INAF-Osservatorio Astronomico di Padova, Vicolo dell'Osservatorio 5, I-35122, Padova, Italy 
\and
\label{inst89}{Istituto di RadioAstronomia, I-40129, Bologna, Italy}
\and
\label{inst86}{Istituto Nazionale di Fisica Nucleare, Sezione di Torino, I-10125 Torino, Italy}
\and
\label{inst98}Owens Valley Radio Observatory, California Institute of Technology, Pasadena, CA 91125, USA
\and
\label{inst99}Finnish Center for Astronomy with ESO (FINCA), University of Turku, FI-20014, Turku, Finland
\and
\label{inst97}Aalto University Metsähovi Radio Observatory, Metsähovintie 114, 02540 Kylmälä, Finland
\and
\label{inst96}Institute of Astrophysics, Foundation for Research and Technology-Hellas, GR-71110 Heraklion,Greece, 
\and
\label{inst95}Department of Physics, Univ. of Crete, GR-70013 Heraklion, Greece
\and
\label{inst94}Departamento de Astronomía, Universidad de Chile, Camino El Observatorio 1515, Las Condes, Santiago, Chile
\and
\label{inst93}Departamento de Astronom\'ia, Universidad de Concepci\'on, Concepci\'on, Chile
\and
\label{inst92}Department of Physics and Astronomy, University of Turku, FI-20014, Turku, Finland
\and
\label{inst85}Hankasalmi Observatory, Murtoistentie 116-124, FI-41500, Hankasalmi, Finland.
\and
\label{inst84}Institute of Astronomy and NAO, Bulgarian Academy of Sciences, 1784 Sofia, Bulgaria
\and
\label{inst83}Harlingten New Mexico Observatory, USA
}

%% file: tables/tab_general.tex
\begin{table*}
\caption{\label{tab_general}General properties of the selected TeV BL Lacs and the correction coefficients used in optical, UV and X-ray data analysis.}          
\centering  
\setlength{\tabcolsep}{0.51em}
\begin{tabular}{cccccccccc}          
\hline 
(1) & (2) & (3) & (4) & (5) & (6) & (7) & (8) & (9) & (10)\\
\multirow{2}{*}{Source name}  &RA   &Dec 	&\multirow{2}{*}{z}	&$A_{\text{R}}$ & $N_{\text{H}}$ & $r_{\text{ap}}$(phot)&$r_{\text{ap}}$(pol) & $F_{\text{host,phot}}$ & $F_{\text{host,pol}}$\\ 
& J2000 & J2000 &  & (Mag) & ($\times 10^{21}\,\text{cm}^{-2}$) & (arcsecond) & (arcsecond) & (mJy) & (mJy)\\
\hline  
\object{VER J0521+211} &05 21 45.9  &+21 12 51    &0.180\tablefootmark{a} & 1.481 & 2.94 & 5.0 & 1.5 & 0.0\tablefootmark{b} & 0.0\tablefootmark{b}\\
\object{PKS 1424+240} &14 27 00.4  &+23 48 00	&0.604  &0.123&0.28 & 7.5 & 1.5 & 0.0\tablefootmark{c} & 0.0\tablefootmark{c}\\
\object{1ES 1727+502} &17 28 18.6  &+50 13 10    &0.055  &0.064&0.24 & 7.5 & 1.5 & 1.25\tablefootmark{d} & 0.45\tablefootmark{d}\\
\object{1ES 1959+650} &19 59 59.8  &+65 08 55    &0.047	&0.375&1.00 & 7.5 & 1.5 & 1.73\tablefootmark{d} & 0.38\tablefootmark{d}\\
\object{1ES 2344+514} &23 47 04.8  &+51 42 18    &0.044  &0.458&1.50 & 7.5 & 4.0 & 3.71\tablefootmark{d} & 2.57\tablefootmark{d}\\

\hline
\end{tabular}
\tablefoot{Columns: (1) source name. (2) right ascension. (3) declination.  (4) redshift. (5) R-band Galactic extinctions reported by \citet{2011ApJ...737..103S} used for correcting the optical observations. (6) equivalent Galactic hydrogen column density reported by \citet{2005A&A...440..775K} used for correcting UV and X-ray observations. (7) and (8) aperture radius in arcsecond for optical photometry and polarisation observations. (9) and (10) contribution of the host-galaxy flux (R-band) within the aperture for optical photometry and polarisation observations.
\\
\tablefoottext{a}{Lower limit based on spectroscopy \citep{2017ApJ...837..144P};}
\tablefoottext{b}{Assumed to be zero based on the uncertainty of the redshift and the reported redshift lower limit;}
\tablefoottext{c}{Reported by \citet{2000ApJ...532..740S};}
\tablefoottext{d}{Reported by \citet{2007A&A...475..199N};}
}
\end{table*}

%% file: tables/tab_magic_flux.tex
\begin{table*}
\centering
\caption{\label{tab_magic_flux}Observed VHE gamma-ray integral flux of the sample.}
\begin{tabular}{cccccc} 
\hline
(1) & (2) & (3) & (4) & (5) \\
Source name & Epoch  & $E_{\mathrm{thr}}$& F$_{>{E}_{\mathrm{thr}}}$ & Prob\tablefootmark{a}\\
&  (MJD)       &  (GeV) &($\times 10^{-11}$cm$^{-2}$s$^{-1}$)& \%\\
                  	
\hline
VER J0521+211      & 56580-56627    &  200 & $5.8  \pm 0.6$ & 1.1 \\
\multirow{2}{*}{PKS 1424+240\Big\{}      & 56740-56826    &  150  & $1.1  \pm 0.2$ & 69.5  \\
                  & 57045-57187    &  150  & $0.6  \pm 0.2$ & 62.4  \\
1ES 1727+502      & 57307-57327    &  300 & $1.8  \pm 0.3$ & 0.08  \\
\multirow{3}{*}{1ES 1959+650\Bigg\{}     & 57547          &  300 & $18.9 \pm 1.1$ & 12.8  \\
                  & 57553          &  300 & $32.8 \pm 1.3$ & 12.7  \\
                  & 57711          &  300 & $3.8  \pm 0.3$ & 55.8  \\
\multirow{3}{*}{1ES 2344+514\Bigg\{}     & 57611-57612    &  300 & $3.8  \pm 0.4$ & $8.3\times10^{-7}$  \\
                  & 57611          &  300 & $6.9  \pm 0.9$ & 4.0  \\
                  & 57612          &  300 & $2.2  \pm 0.5$ & 82.0 \\

\hline
\end{tabular}
\tablefoot{Columns: (1) source name. (2) observation epoch. (3) energy threshold. (4) observed integral flux above energy threshold. (5) probability for a fit of the flux with a constant. \tablefoottext{a}{The constant-flux hypothesis (daily timescale) is rejected at a 3-$\sigma$ confidence level if the fit probability is less than $0.27\%$.}}
\end{table*}

%% file: tables/tab_magic_spec.tex
\begin{table*}
\centering
\caption{\label{tab_magic_spec} Results of the VHE gamma-ray spectral analysis of the sample.}
\begin{tabular}{cccccccccc} 
\hline
(1) & (2) & (3) & (4) & (5) & (6) & (7) \\
Source name & Epoch & Model & $E_{0}$ & $F_{0}$ & \multirow{2}{*}{$\Gamma$} &  \multirow{2}{*}{$\beta$}  \\

&  (MJD)       & & (GeV)       & ($\times 10^{-11}$cm$^{-2}$s$^{-1}$) &  & \\
                  	
\hline
VER J0521+211  & 56580-56627 & LP  &  300 &  $27.43 \pm 0.51$  & $2.69 \pm 0.02$ & $0.47 \pm 0.07$  \\
\multirow{2}{*}{PKS 1424+240\tablefootmark{a}\Big\{} & 56740-56826 & PL &  111 &  $98.9 \pm 6.5$  & $2.77 \pm 0.16$ &  \\
              & 57045-57187 & LP  &  104 &  $82 \pm 15$ & $2.19 \pm 0.52$ & $1.93 \pm 0.87$  \\
1ES 1727+502\tablefootmark{a}  & 57307-57327 & PL &  585 &  $2.08 \pm 0.15$   & $2.21 \pm 0.08$ &    \\
\multirow{3}{*}{1ES 1959+650\Bigg\{}  & 57547       & LP  &  261 &  $133.8 \pm 4.4$ & $2.04 \pm 0.05$ & $0.23 \pm 0.07$   \\
              & 57553       & LP  &  307 &  $153.2 \pm 3.9$ & $1.81 \pm 0.04$ & $0.37 \pm 0.04$  \\
              & 57711       & LP  &  293 &  $26.0  \pm 2.1$ & $2.30 \pm 0.15$ & $0.34 \pm 0.22$ \\
\multirow{3}{*}{1ES 2344+514\Bigg\{}   & 57611-57612 & PL &  487 &  $7.27 \pm 0.97 $  & $2.07 \pm 0.22$ &  \\
              & 57611       & PL &  465 &  $13.4 \pm 1.5 $ & $2.07 \pm 0.13$ &  \\
              & 57612       & PL &  396 &  $5.7 \pm 1.1 $  & $2.11 \pm 0.21$ &  \\

\hline
\end{tabular}
\tablefoot{Columns: (1) source name. (2) observation epoch. (3) best-fitted model, log parabola (LP) and power law (PL). (4) normalisation {(decorrelation)} energy of spectrum. (5) flux at normalisation energy. (6) and (7) spectral index and the curvature parameter{. All of the spectral parameters are calculated after taking into account the effect of} EBL absorption using the model described by \citet{2011MNRAS.410.2556D}. \tablefoottext{a}{The VHE gamma-ray spectral points are extracted from \citet{2019arXiv190400134M}.}}
\end{table*}

%% file: tables/tab_radio_opt.tex
\begin{table*}
\caption{\label{radio-opt}Analysis results of the long-term radio and optical light curves}
\centering  
\begin{tabular}{cccccccccc}
\hline
(1) & (2) & (3) & (4) & (5) & (6) & (7) & (8) & (9) & (10)    \\
\multirow{2}{*}{Source name} & \multirow{2}{*}{$N_{\text{rad}}$} & $F_{\text{ave, rad}}$ & \multirow{2}{*}{$r_{\text{s, rad}}$} & \multirow{2}{*}{p-value$_{\text{rad}}$} & \multirow{2}{*}{$N_{\text{opt}}$} & $F_{\text{ave, opt}}$ & \multirow{2}{*}{$r_{\text{s, opt}}$} & \multirow{2}{*}{p-value$_{\text{opt}}$}&\multirow{2}{*}{Fraction}\\ 
&&(Jy)&&&&(mJy)&&&\\
\hline
VER J0521+211& 191&  0.444&-0.489&$<2.2\times 10^{-16}$ &109&7.35&-0.754&$< 2.2\times 10^{-16}$&0.36\\
PKS~1424+240& 178&  0.498&0.627&$<2.2\times 10^{-16}$ &201&9.31&0.015&0.831&0\\ 
1ES~1727+502& 196&  0.144&-0.494&$<2.2\times 10^{-16}$ &209&1.27&-0.591&$< 2.2\times 10^{-16}$&0.26\\
1ES~1959+650& 222& 0.264&0.510&$<2.2\times 10^{-16}$ &330&7.06&0.785&$<2.2\times 10^{-16}$&0.52\\
1ES~2344+514& 243& 0.184&0.356&$1.451\times 10^{-8}$ &140&1.00&0.574&$1.28\times 10^{-13}$&0.06\\
\hline
\end{tabular}
\tablefoot{Columns: (1) source name. (2) number of observational data points in the radio light curve (15\,GHz). (3) average radio flux at 15\,GHz. (4) Spearman's rank correlation coefficient of the linear trend in the radio light curve. (5) null-hypothesis probability of the linear fit of the radio light curve.  (6)  number of observational data points in the optical light curve (R-band). (7)  average optical (R-band) flux. (8) Spearman's rank correlation coefficient of the linear trend in the optical light curve. (9) null-hypothesis probability of the linear fit of the optical light curve. (10) fractional contribution of the slowly varying radio component to the total optical flux density.
}
\end{table*}

%% file: tables/tab_pol.tex
\begin{table*}
\caption{\label{poltable} The results of the model fitting to the optical polarisation data and the jet orientation parameters for comparison.}
\centering
\renewcommand{\arraystretch}{1.5}

\begin{tabular}{ccccccccccc}
\hline
\vspace{-0.15cm}
(1) & (2) & (3) & (4) & (5) & (6) & (7) & (8) & (9) & (10) & (11)\\
\vspace{-0.15cm}
\multirow{2}{*}{Source name} & $I_C (max)$ & $I_C$ & $Q_C$ & $U_C$ & $\theta$ & $\psi'$ & $\varphi_0$ & $\sigma$ & EVPA$_{\text{core}}$  & PA$_{\text{core}}$\\
            &          ($\mu$Jy) & ($\mu$Jy) & ($\mu$Jy) & ($\mu$Jy) & ($\degr$) & ($\degr$) & ($\degr$) & ($\mu$Jy) & ($\degr$) & ($\degr$)\\
\hline            
VER J0521+211
 & 3000 & 1500$^{+1000}_{-1000}$ & 170$^{+80}_{-80}$ & 140$^{+60}_{-50}$ &
1.8$^{+0.9}_{-1.4}$ & 56$^{+18}_{-14}$  & 13$^{+20}_{-20}$ & 220 & 200 & 250\\
{PKS 1424+240} 
 & 6700 & 4700$^{+1400}_{-1800}$ & 140$^{+120}_{-110}$ & -400$^{+180}_{-180}$ &
4.0$^{+3.5}_{-2.4}$ & 62$^{+13}_{-8}$  & 47$^{+13}_{-13}$ & 320 & 145 & 140\\
{1ES 1727+502} 
 & 960 & 550$^{+300}_{-330}$ & -25$^{+4}_{-4}$ & -11$^{+4}_{-3}$ &
1.8$^{+1.4}_{-1.5}$ & 53$^{+14}_{-12}$ & 140$^{+11}_{-11}$ & 17 & - & 270\\
{1ES 1959+650} 
 & 7100 & 3600$^{+2200}_{-2200}$ & -159$^{+67}_{-62}$ & -108$^{+90}_{-98}$ &
2.9$^{+2.9}_{-1.8}$ & 55$^{+10}_{-11}$ & 140$^{+11}_{-11}$ & 160 & 152 &150\\
{1ES 2344+514} 
 & 520 & 260$^{+170}_{-160}$ & 14$^{+6}_{-6}$ & -2$^{+9}_{-7}$ &
1.8$^{+0.5}_{-0.5}$ & 67$^{+12}_{-10}$ & 132$^{+6}_{-6}$ & 18 & - & 137\\

\hline
\end{tabular}
\tablefoot{Columns: (1) source name. (2) upper limit for the constant component prior. (3),(4) and (5)  constant-component Stokes parameters. (6) viewing angle. (7) magnetic-field pitch angle. (8) jet position angle. (9) RMS of the turbulence. (10) radio-core EVPA. (11) VLBI jet position angle.}
\end{table*}

%% file: tables/tab_sedparams.tex
\begin{table*}
\caption{\label{sedparams}SED modelling parameters {for one-zone SSC and two-component models.}}
\centering  
\setlength{\tabcolsep}{0.4em}
\renewcommand{\arraystretch}{1.2}

\begin{tabular}{c:c:cccccccccc}
\hline
  (1) & (2) & (3) & (4) & (5) & (6) & (7) & (8) & (9) & (10) & (11) & (12)  \\
  \multirow{2}{*}{Source name} & Campaign/ & Model & $\gamma_{\text{min}}$ & $\gamma_{\text{b}}$ & $\gamma_{\text{max}}$ & \multirow{2}{*}{$n_1$} & \multirow{2}{*}{$n_2$} & B & K & R & \multirow{2}{*}{$\delta$} \\
  
  & state & (region) & ($\times 10^{3}$) &($\times 10^{4}$) &($\times 10^{5}$) &  &  & (G) & ($\times 10^{3}$\,cm$^{-3}$) & ($\times 10^{15}$\,cm) & \\
\hline
\multirow{3}{*}{VER J0521+211}  & \multirow{3}{*}{2013} 
      & one-zone & 5.5 & 1.4 & 9.0 & 2.1 & 3.7 & 0.04 & 85 & 13.5 & 36 \\
      & & 2-comp (blob) & 1.0 & 3.0 & 4.0 & 1.95 & 3.1 & 0.1 & 31.5 & 13 & 12 \\
      & & 2-comp (core) & 0.35 & 0.11 & 0.16 & 1.64 & 2.77 & 0.1 & 0.012 & 370 & 11 \\
\hdashline

\multirow{6}{*}{PKS 1424+240}  & \multirow{3}{*}{2014} 
      & one-zone & 3.6 & 2.3 & 8.9 & 1.9 & 4.3 & 0.017 & 0.4 & 55 & 80 \\
         & & 2-comp (blob) & 9.0 & 3.2 & 3.0 & 1.98 & 3.35 & 0.1 & 17 & 19 & 20\\
      &  & 2-comp (core) & 0.35 & 0.3 & 0.28 & 1.69 & 3.0 & 0.1 & 0.007 & 1020 & 10\\
\cdashline{2-12}

  & \multirow{3}{*}{2015} 
      & one-zone & 6.0 & 2.8 & 6.0 & 2.0 & 4.8 & 0.015 & 1.4 & 56 & 75 \\
         & & 2-comp (blob) & 6.0 & 4.5 & 3.3 & 1.98 & 3.85 & 0.1 & 32 & 13.1 & 18 \\
      &  & 2-comp (core) & 0.33 & 0.3 & 0.3 & 1.68 & 3.0 & 0.1 & 0.007 & 1020 & 10 \\
\hdashline

\multirow{3}{*}{1ES~1727+502}  & \multirow{3}{*}{2015}
      & one-zone & 2.5 & 1.3 & 18 & 2.0 & 2.7 & 0.03 & 8.8 & 7.0 & 29 \\
         & & 2-comp (blob) &5.0 & 5.0 & 13 & 1.95 & 2.45 & 0.1 & 7.0 & 7.1  & 11\\
      &  & 2-comp (core) & 0.16 & 0.3 & 0.8 & 1.95 & 2.7 & 0.1 & 0.15 & 154 & 4 \\
\hdashline

\multirow{9}{*}{1ES~1959+650}  & \multirow{3}{1.75cm}{\centering 2016/ low}     
 & one-zone  & 0.4 & 0.7 & 4.5 & 1.98 & 2.7 & 0.06 & 5.0 & 7.2 & 41\\
 &  & 2-comp (blob) & 3.0  & 7.0 & 6.5  & 1.97 & 3.35 & 0.2 & 0.9  & 7.1 & 27 \\ 
 &  & 2-comp (core) & 0.29 & 0.2 & 0.45 & 1.68 & 2.90 & 0.2 & 0.08 & 126 & 4  \\
\cdashline{2-12}

 & \multirow{3}{1.75cm}{\centering 2016/ intermediate}      
 & one-zone & 0.5 & 6.0 & 8.0 & 2.0 & 2.85 & 0.06 & 14 & 5.5 & 30\\
 &  & 2-comp (blob)  & 3.8  & 9.5  & 6.54 & 1.98 & 2.5  & 0.1 & 7.5  & 5.5 & 23 \\
 &  & 2-comp (core)  & 0.33 & 0.26 & 0.57 & 1.67 & 2.85 & 0.1 & 0.13 & 126 & 4\\
\cdashline{2-12}

 & \multirow{3}{1.75cm}{\centering 2016/ high}        
 & one-zone & 1.0 & 6.0 & 15 & 1.95 & 2.8 & 0.07 & 13 & 4.3 & 31\\
 &  & 2-comp (blob)  & 7.0  & 6.0 & 13.0 & 1.95 & 2.72 & 0.1 & 10.5 & 4.3 & 27 \\
 &  & 2-comp (core)  & 0.33 & 0.3 & 0.35 & 1.67 & 3.0  & 0.1 & 0.13 & 126 & 4 \\ 
\hdashline

\multirow{3}{*}{1ES~2344+514} &  \multirow{3}{*}{2016} 
       & one-zone & 1.0 & 5.0 & 30 & 2.0 & 2.93 & 0.02 & 5 & 12.2 & 20\\
        & & 2-comp (blob) & 5.8 & 5.4 & 28 & 2.0 & 2.65 & 0.1 & 19 & 10.7 & 6\\
      &  & 2-comp (core) & 0.26 & 0.2 & 1.3 & 1.8 & 2.95 & 0.1 & 0.06 & 160 & 4\\
\hline
\end{tabular}
\tablefoot{Columns: (1) source name. (2) observation campaign/state. (3) {model (emission region).} (4), (5) and (6) minimum, break and maximum electron Lorentz factor. (7) and (8) slopes of electron distribution below and above $\gamma_{b}$.  (9) magnetic field strength. (10) electron density.  (11) emission-region size. (12) Doppler factor.}
\end{table*}

%% file: tables/tab_sedresults.tex
\begin{table*}
\caption{\label{sedresults}Results obtained from the {one-zone SSC and} two-component SED modelling}
\centering  
\renewcommand{\arraystretch}{1.3}
\begin{tabular}{c:c:c:ccccccc}
\hline
  (1) & (2) & (3) & (4) & (5) & (6) & (7) & (8) & (9)  & (10)  \\
  \multirow{2}{*}{Source name} & Campaign/ & \multirow{2}{*}{Model} & $L_\text{B}$ & $L_\text{e}$ & $L_\text{p}$  & \multirow{2}{*}{$\dfrac{U'_\text{B}}{U'_\text{e}}$} & \multirow{2}{*}{CD}  & $\log \nu_{\text{sync}}$ & $\log \nu_{\text{IC}}$ \\
\cline{4-6}  
\cline{9-10}
  &  state & & \multicolumn{3}{c}{($\times 10^{43}$ erg/s)} &  &  & \multicolumn{2}{c}{(Hz)}\\
\hline

\multirow{2}{*}{VER J0521+211} & \multirow{2}{*}{2013} 
& one-zone & 0.19 & 129 & 13.6 & 0.001 & 2.05 & 15.12 & 24.17\\
& & 2-comp & 82.8 & 40.9 & 46.8 & 0.9 & 1.71  & 14.38 & 24.56 \\

\hdashline

\multirow{4}{*}{PKS 1424+240} & \multirow{2}{*}{2014}   
& one-zone & 2.79 & 624 & 68.1 & 0.004 & 0.41 & 15.23 & 24.02 \\
& & 2-comp & 520 & 198 & 176 & 2.4  & 0.55 & 14.70 & 24.82 \\
    
\cdashline{2-10}
   & \multirow{2}{*}{2015}                
& one-zone & 1.98 & 697 & 52.3 & 0.003 & 0.58 & 15.08 & 24.02 \\
& & 2-comp & 520 & 219 & 199 & 1.1  & 0.43 & 14.67 & 24.63 \\

\hdashline
   
\multirow{2}{*}{1ES~1727+502} & \multirow{2}{*}{2015}                     
& one-zone & 0.02 & 5.5 & 1.0 & 0.003 & 0.33 & 18.41 & 25.70 \\
& & 2-comp & 1.9  & 2.1  & 3.8  & 4.5  & 0.33 & 18.28 & 25.87 \\

\hdashline

\multirow{6}{*}{1ES~1959+650} 
   & \multirow{2}{*}{2016/ low}                          
& one-zone & 0.16 & 12.8 & 12.4 & 0.01 & 0.21 & 17.59 & 25.36\\
& & 2-comp & 5.1  & 5.1  & 4.6  & 25.6 & 0.24 & 17.60 & 25.93 \\ 
\cdashline{2-10}

   & \multirow{2}{1.75cm}{\centering 2016/ intermediate}                  
& one-zone & 0.05 & 17.4 & 9.8 & 0.003 & 0.68 & 17.92 & 25.59 \\
& & 2-comp & 1.3  & 9.8  & 7.5  & 16.7 & 0.59 & 18.04 & 25.71 \\
\cdashline{2-10}

   & \multirow{2}{*}{2016/ high}                         
& one-zone & 0.04 & 1.4 & 4.2 & 0.003 & 0.65 & 18.62 & 25.82 \\
& & 2-comp & 1.3  & 10.9 & 2.3  & 3.1  & 0.64 & 18.55 & 25.86 \\

\hdashline

\multirow{2}{*}{1ES~2344+514} & \multirow{2}{*}{2016}                      
& one-zone & 0.01 & 11.7 & 3.6 & 0.001 & 0.72 & 18.22 & 25.40 \\
& & 2-comp & 2.1  & 2.5  & 2.8  & 3.4  & 0.65 & 18.66 & 25.48 \\
\hline
\end{tabular}
\tablefoot{Columns: (1) source name. (2) observation campaign/state. (3) model. (4), (5) and (6) kinetic energy of the magnetic field, electrons, and cold protons of the {emission region (core in the case of two-component model)}, respectively. (7) ratio between the magnetic field and relativistic electron energy densities. (8) the luminosity ratio between the IC and Synchrotron peak (Compton Dominance parameter). (9) and (10) {observed} synchrotron and IC peak frequencies.}
\end{table*}

%% file: tables/tab_core_blob.tex
\begin{table*}
\centering  
\caption{\label{corblob}The fraction of emission at optical (R-band) which originated from the core emission.}

\begin{tabular}{ccccc}
\hline
  (1) & (2) & (3) & (4) & (5)  \\
  Source name & Campaign/ state & Radio -- optical long-term & Polarisation & SED Modelling\\
\hline
VER J0521+211
   & 2013 & $>0.36$ & [0.05 - 0.22]   & 0.87\\
         
\multirow{2}{*}{PKS 1424+240\,\Big\{} 
   & 2014 & $>0.0$ & [0.31 - 0.72] &   0.94 \\
   & 2015 & $>0.0$ & [0.28 - 0.67] &   0.96 \\

1ES~1727+502
   & 2015 & $>0.26$ & [0.18 - 0.70] &   0.63 \\

\multirow{3}{*}{1ES~1959+650\,\Bigg\{} 
   & 2016/ low  & $>0.52$ & [0.14 - 0.56] &  0.44 \\ 
   & 2016/ intermediate & $>0.52$ & [0.21 - 0.87] &  0.49 \\
   & 2016/ high & $>0.52$ & [0.21 - 0.89] & 0.41 \\

1ES~2344+514 
   & 2016       & $>0.06$ & [0.08 - 0.35] &  0.63 \\
\hline
\end{tabular}

\tablefoot{Columns: (1) source name. (2) observation campaign/state. (3), (4) and (5) ratio between core flux and total flux obtained from radio-optical long-term light curve, polarisation analysis, and {two-component} SED modelling. The fractions are calculated using the methods described in Section ~\ref{sec:component} and \ref{sec:pol}. The results from the SED modelling are also presented for comparison.}
\end{table*}

%% file: tables/tab_tuorla.tex
\begin{table}[ht!]
\centering  
\caption{\label{tab:tuorla}An example of the optical (R-band) light curve data available at the CDS for 1ES~1959+650.}
\begin{tabular}{ccc}
\hline
(1) & (2) & (3)  \\
Date & Flux & Flux error \\
(MJD) & (mJy) & (mJy)\\
\hline
56206.88	&	6.63	&	0.13	\\
56210.89	&	6.17	&	0.13	\\
56213.85	&	6.71	&	0.13	\\
56220.83	&	6.91	&	0.14	\\
56222.82	&	7.02	&	0.14	\\
56242.84	&	8.07	&	0.14	\\
56255.83	&	9.22	&	0.17	\\
56401.38	&	4.53	&	0.07	\\
56402.42	&	4.53	&	0.07	\\
56403.46	&	4.50	&	0.07	\\
\hline
\end{tabular}

\tablefoot{Columns: (1) observations date. (2) and (3) optical (R-band) flux and its error. The flux is corrected for the host-galaxy contribution (when applicable) and galactic extinction (see Sec.~\ref{sec:opt_radio} for details). Only the first ten lines of the table are shown. Data are available for all five targets.}
\end{table}

%% file: tables/tab_xray.tex
\begin{table*}
\caption{\label{tab:xray}Example of the {\it Swift}-XRT results for 1ES~1959+650.}
\centering
\setlength{\tabcolsep}{0.37em}
\begin{tabular}{ccccccccccc} 
\hline
  (1) & (2) & (3) & (4) & (5) & (6) & (7) & (8) & (9) & (10) & (11)   \\
Date    & \multirow{2}{*}{OBS ID}  &  Exp. & \multirow{2}{*}{$\Gamma_{\text{PL}}$} & \multirow{2}{*}{$\chi ^2$/d.o.f.}   & \multirow{2}{*}{$\Gamma_{\text{LP}}$} & \multirow{2}{*}{$\beta_{\text{LP}}$} & \multirow{2}{*}{$\chi ^2$/d.o.f.} & Prob.$^\star$ & F$_{2-10\mathrm{\,keV}}$    &   F$_{0.3-10\mathrm{\,keV}}$ \\ 
\cline{10-11}

(MJD)  &  &  (s)     &   &      &   &   &     & (\%)  &  \multicolumn{2}{c}{($\times10^{-11}$ erg cm$^{-2}$s$^{-1}$)}         \\
\hline                                 
56572.10 & 00035025108 & 1064 &      &    &$2.08 \pm 0.03$&$0.45 \pm 0.06$& 307.8/233 & $10^{-4}$ &$11.8 \pm 0.4$&$25.9 \pm 0.4$\\
56779.02 & 00035025109 & 1041 &      &    &$2.04 \pm 0.05$&$0.49 \pm 0.10$& 107.4/117 & $10^{-3}$ &$4.3 \pm 0.2$&$9.3 \pm 0.2$\\
56786.88 & 00035025110 & 648 &$2.33 \pm 0.04$& 84.6/70 &      &      &  & 1.00 &$3.6 \pm 0.3$&$7.9 \pm 0.3$\\
56793.73 & 00035025111 & 754 &      &    &$2.16 \pm 0.06$&$0.50 \pm 0.14$& 76.1/80 & 0.02 &$3.1 \pm 0.3$&$7.5 \pm 0.3$\\
56807.95 & 00035025112 & 962 &      &    &$2.19 \pm 0.05$&$0.39 \pm 0.11$& 59.8/87 & $10^{-3}$ &$2.9 \pm 0.2$&$6.8 \pm 0.2$\\
56821.13 & 00035025114 & 993 &$2.10 \pm 0.05$& 77.0/58 &      &      &  & 3.21 &$6.3 \pm 0.4$&$11.6 \pm 0.4$\\
56842.19 & 00035025116 & 994 &$2.08 \pm 0.03$& 133.3/132 &      &      &  & 2.65 &$7.9 \pm 0.3$&$14.2 \pm 0.3$\\
56848.52 & 00035025117 & 910 &      &    &$2.06 \pm 0.05$&$0.43 \pm 0.12$& 96.3/82 & 0.08 &$4.3 \pm 0.3$&$9.2 \pm 0.3$\\
56856.67 & 00035025118 & 1260 &      &    &$2.05 \pm 0.03$&$0.46 \pm 0.07$& 182.8/174 & $10^{-4}$ &$5.4 \pm 0.3$&$11.6 \pm 0.3$\\
56863.53 & 00035025119 & 885 &      &    &$2.01 \pm 0.03$&$0.38 \pm 0.07$& 166.8/164 & $10^{-3}$ &$7.7 \pm 0.3$&$15.6 \pm 0.3$\\

\hline
\end{tabular}
\tablefoot{Columns: (1) observation date. (2) observation ID. (3) exposure time. (4) spectral index of the PL model. (5) $\chi^2$/d.o.f. of the fitted PL model. (6) spectral index of the LP model. (7) curvature parameter of the fitted LP model. (8) $\chi^2$/d.o.f. of the LP model. (9)  Null-hypotheses probability of the F-test. (10) X-ray flux in the range of 2-10 keV. (11) X-ray flux in the range of 0.3-10 keV. 

$^\star$ The LP model is preferred over the PL model at $3\sigma$ confidence level if the F-test probability value is less than $0.27\%$. Only the first ten lines of the table are shown. Data are available for all five targets.}
\end{table*}

%% file: tables/tab_sedepochs.tex
\begin{table*}
\caption{\label{sedepochs}The observation date/epochs of the MWL data used for SED modelling}
\centering  
\setlength{\tabcolsep}{0.45em}
\begin{tabular}{cccccccc}
\hline
  (1) & (2) & (3) & (4) & (5) & (6) & (7) & (8)\\
  \multirow{2}{*}{Source name} & Campaign/ &VHE gamma rays & HE gamma rays & X-rays & UV  & Optical & Radio \\

  & state & (MJD) & (MJD) & (MJD) & (MJD) & (MJD) & (MJD)    \\
\hline
VER J0521+211   & 2013 & 56580.18-56627.95 & 56580.00-56627.00 & 56625.30 & 56625.30 & 56625.15 & 56625.45 \\
         
\multirow{2}{*}{PKS 1424+240\,\Big\{} 
   & 2014 & 56740.06-56825.99 & 56740.00-56826.00 & 56801.25 & 56801.26 & 56800.96 & 56781.21  \\
   & 2015 & 57045.05-57186.06 & 57045.00-57187.00 & 57135.26 & 57135.26 & 57135.11 & 57125.48  \\

1ES~1727+502
   & 2015 & 57306.83-57327.83 & 57263.00-57353.00 & 57307.99 & 57307.99 &  57307.83 & 57304.84 \\

\multirow{3}{*}{1ES~1959+650\,\Bigg\{} 
   & 2016/ low & 57711.82-57711.87  & 57711.43-57712.35	 & 57711.58  & 57711.59 & 57711.82 & 57714.79  \\ 
   & 2016/ intermediate & 57547.13-57547.19  & 57545.16-57550.19 & 57547.13 & 57547.13 & 57547.14 & 57548.38  \\
   & 2016/ high & 57553.06-57553.14 & 57552.00-57554.00 & 57553.10 & 57553.10 & 57553.13 & 57556.31  \\

1ES~2344+514 
   & 2016 & 57612.06-57612.08  & 57520.00-57704.00 & 57613.52 & 57613.52 & 57611.02 & 57613.32  \\
\hline
\end{tabular}
\tablefoot{Columns: (1) source name. (2) observation campaign/state. (3) and (4) start-end time of VHE gamma-ray and HE gamma-ray observations. (5), (6), (7), and (8) start time of X-ray, UV, optical (R-band) and radio (15 GHz) observations.}
\end{table*}

%% file: Monster_BL_LacV0.bbl
\begin{thebibliography}{121}
\expandafter\ifx\csname natexlab\endcsname\relax\def\natexlab#1{#1}\fi

\bibitem[{Abdo {et~al.}(2010)Abdo, Ackermann, Agudo, Ajello, Aller, Aller,
  Angelakis, Arkharov, Axelsson, Bach, Baldini, Ballet, Barbiellini, Bastieri,
  Baughman, Bechtol, \& Bellazzini}]{0004-637X-716-1-30}
Abdo, A.~A., Ackermann, M., Agudo, I., {et~al.} 2010, \apj, 716, 30

\bibitem[{{Acciari} {et~al.}(2011){Acciari}, {Aliu}, {Arlen}, {Aune},
  {Beilicke}, {Benbow}, {Boltuch}, {Bradbury}, {Buckley}, {Bugaev}, {Byrum},
  {Cannon}, {Cesarini}, {Ciupik}, {Cui}, {Dickherber}, {Duke}, {Falcone},
  {Finley}, {Finnegan}, {Fortson}, {Furniss}, {Galante}, {Gall}, {Gillanders},
  {Godambe}, {Grube}, {Guenette}, {Gyuk}, {Hanna}, {Holder}, {Hui}, {Humensky},
  {Imran}, {Kaaret}, {Karlsson}, {Kertzman}, {Kieda}, {Konopelko},
  {Krawczynski}, {Krennrich}, {Lang}, {Maier}, {McArthur}, {McCutcheon},
  {Moriarty}, {Ong}, {Otte}, {Ouellette}, {Pandel}, {Perkins}, {Pichel},
  {Pohl}, {Quinn}, {Ragan}, {Reyes}, {Reynolds}, {Roache}, {Rose}, {Rovero},
  {Schroedter}, {Sembroski}, {Senturk}, {Steele}, {Swordy}, {Theiling},
  {Thibadeau}, {Varlotta}, {Vassiliev}, {Vincent}, {Wagner}, {Wakely}, {Ward},
  {Weekes}, {Weinstein}, {Weisgarber}, {Williams}, {Wissel}, {Wood}, {Zitzer},
  {Garson}, {Lee}, {Sadun}, {Carini}, {Barnaby}, {Cook}, {Maune}, {Pease},
  {Smith}, {Walters}, {Berdyugin}, {Lindfors}, {Nilsson}, {Pasanen}, {Sainio},
  {Sillanpaa}, {Takalo}, {Villforth}, {Montaruli}, {Baker}, {Lahteenmaki},
  {Tornikoski}, {Hovatta}, {Nieppola}, {Aller}, \&
  {Aller}}]{2011ApJ...738...25A}
{Acciari}, V.~A., {Aliu}, E., {Arlen}, T., {et~al.} 2011, \apj, 738, 25

\bibitem[{{Acciari} {et~al.}(2019){Acciari}, {Ansoldi}, {Antonelli}, {Arbet
  Engels}, {Baack}, {Babi{\'c}}, {}, {Banerjee}, {Barres de Almeida}, {Barrio},
  {Becerra Gonz{\'a}lez}, {Bednarek}, {Bellizzi}, {Bernardini}, {Berti},
  {Besenrieder}, {Bhattacharyya}, {Bigongiari}, {Biland}, {Blanch}, {Bonnoli},
  {Busetto}, {Carosi}, {Ceribella}, {Chai}, {Cikota}, {Colak}, {Colin},
  {Colombo}, {Contreras}, {Cortina}, {Covino}, {D'Elia}, {Da Vela}, {Dazzi},
  {De Angelis}, {De Lotto}, {Delfino}, {Delgado}, {Di Pierro}, {Do Souto
  Espi{\~n}eira}, {Dom{\'\i}nguez}, {Dominis Prester}, {Dorner}, {Doro},
  {Elsaesser}, {Fallah Ramazani}, {Fattorini}, {Fern{\'a}ndez-Barral},
  {Ferrara}, {Fidalgo}, {Foffano}, {Fonseca}, {Font}, {Fruck}, {Galindo},
  {Gallozzi}, {Garc{\'\i}a L{\'o}pez}, {Garczarczyk}, {Gasparyan}, {Gaug},
  {Godinovi{\'c}}, {}, {Green}, {Guberman}, {Hadasch}, {Hahn}, {Hassan},
  {Herrera}, {Hoang}, {Hrupec}, {Inoue}, {Ishio}, {Iwamura}, {Kubo}, {Kushida},
  {Lamastra}, {Lelas}, {Leone}, {Lindfors}, {Lombardi}, {Longo}, {L{\'o}pez},
  {L{\'o}pez-Coto}, {L{\'o}pez-Oramas}, {Machado de Oliveira Fraga}, {Maggio},
  {Majumdar}, {Makariev}, {Mallamaci}, {Maneva}, {Manganaro}, {Mannheim},
  {Maraschi}, {Mariotti}, {Mart{\'\i}nez}, {Masuda}, {Mazin}, {Mi{\'c}},
  {anovi{\'c}}, {}, {Miceli}, {Minev}, {Miranda}, {Mirzoyan}, {Molina},
  {Moralejo}, {Morcuende}, {Moreno}, {Moretti}, {Munar-Adrover}, {Neustroev},
  {Niedzwiecki}, {Nievas Rosillo}, {Nigro}, {Nilsson}, {Ninci}, {Nishijima},
  {Noda}, {Nogu{\'e}s}, {N{\"o}the}, {Paiano}, {Palacio}, {Palatiello},
  {Paneque}, {Paoletti}, {Paredes}, {Pe{\~n}il}, {Peresano}, {Persic}, {Prada
  Moroni}, {Prandini}, {Puljak}, {Rhode}, {Rib{\'o}}, {Rico}, {Righi},
  {Rugliancich}, {Saha}, {Sahakyan}, {Saito}, {Satalecka}, {Schweizer},
  {Sitarek}, {{\v{S}}nidari{\'c}}, {}, {Sobczynska}, {Somero}, {Stamerra},
  {Strom}, {Strzys}, {Suri{\'c}}, {}, {Tavecchio}, {Temnikov}, {Terzi{\'c}},
  {}, {Teshima}, {Torres-Alb{\`a}}, {Tsujimoto}, {van Scherpenberg}, {Vanzo},
  {V{\'a}zquez Acosta}, {Vovk}, {Will}, {Zari{\'c}}, \&
  {}}]{2019arXiv190400134M}
{Acciari}, V.~A., {Ansoldi}, S., {Antonelli}, L.~A., {et~al.} 2019, \mnras,
  486, 4233

\bibitem[{{Acero} {et~al.}(2015){Acero}, {Ackermann}, {Ajello}, {Albert},
  {Atwood}, {Axelsson}, {Baldini}, {Ballet}, {Barbiellini}, {Bastieri},
  {Belfiore}, {Bellazzini}, {Bissaldi}, {Blandford}, {Bloom}, {Bogart},
  {Bonino}, {Bottacini}, {Bregeon}, {Britto}, {Bruel}, {Buehler}, {Burnett},
  {Buson}, {Caliandro}, {Cameron}, {Caputo}, {Caragiulo}, {Caraveo},
  {Casandjian}, {Cavazzuti}, {Charles}, {Chaves}, {Chekhtman}, {Cheung},
  {Chiang}, {Chiaro}, {Ciprini}, {Claus}, {Cohen-Tanugi}, {Cominsky}, {Conrad},
  {Cutini}, {D'Ammando}, {de Angelis}, {DeKlotz}, {de Palma}, {Desiante},
  {Digel}, {Di Venere}, {Drell}, {Dubois}, {Dumora}, {Favuzzi}, {Fegan},
  {Ferrara}, {Finke}, {Franckowiak}, {Fukazawa}, {Funk}, {Fusco}, {Gargano},
  {Gasparrini}, {Giebels}, {Giglietto}, {Giommi}, {Giordano}, {Giroletti},
  {Glanzman}, {Godfrey}, {Grenier}, {Grondin}, {Grove}, {Guillemot}, {Guiriec},
  {Hadasch}, {Harding}, {Hays}, {Hewitt}, {Hill}, {Horan}, {Iafrate}, {Jogler},
  {J{\'o}hannesson}, {Johnson}, {Johnson}, {Johnson}, {Johnson}, {Kamae},
  {Kataoka}, {Katsuta}, {Kuss}, {La Mura}, {Landriu}, {Larsson}, {Latronico},
  {Lemoine-Goumard}, {Li}, {Li}, {Longo}, {Loparco}, {Lott}, {Lovellette},
  {Lubrano}, {Madejski}, {Massaro}, {Mayer}, {Mazziotta}, {McEnery},
  {Michelson}, {Mirabal}, {Mizuno}, {Moiseev}, {Mongelli}, {Monzani},
  {Morselli}, {Moskalenko}, {Murgia}, {Nuss}, {Ohno}, {Ohsugi}, {Omodei},
  {Orienti}, {Orlando}, {Ormes}, {Paneque}, {Panetta}, {Perkins},
  {Pesce-Rollins}, {Piron}, {Pivato}, {Porter}, {Racusin}, {Rando}, {Razzano},
  {Razzaque}, {Reimer}, {Reimer}, {Reposeur}, {Rochester}, {Romani},
  {Salvetti}, {S{\'a}nchez-Conde}, {Saz Parkinson}, {Schulz}, {Siskind},
  {Smith}, {Spada}, {Spandre}, {Spinelli}, {Stephens}, {Strong}, {Suson},
  {Takahashi}, {Takahashi}, {Tanaka}, {Thayer}, {Thayer}, {Thompson},
  {Tibaldo}, {Tibolla}, {Torres}, {Torresi}, {Tosti}, {Troja}, {Van Klaveren},
  {Vianello}, {Winer}, {Wood}, {Wood}, {Zimmer}, \& {Fermi-LAT
  Collaboration}}]{2015ApJS..218...23A}
{Acero}, F., {Ackermann}, M., {Ajello}, M., {et~al.} 2015, \apjs, 218, 23

\bibitem[{{Ackermann} {et~al.}(2011){Ackermann}, {Ajello}, {Allafort},
  {Antolini}, {Atwood}, {Axelsson}, {Baldini}, {Ballet}, {Barbiellini},
  {Bastieri}, {Bechtol}, {Bellazzini}, {Berenji}, {Blandford}, {Bloom},
  {Bonamente}, {Borgland}, {Bottacini}, {Bouvier}, {Bregeon}, {Brigida},
  {Bruel}, {Buehler}, {Burnett}, {Buson}, {Caliandro}, {Cameron}, {Caraveo},
  {Casandjian}, {Cavazzuti}, {Cecchi}, {Charles}, {Cheung}, {Chiang},
  {Ciprini}, {Claus}, {Cohen-Tanugi}, {Conrad}, {Costamante}, {Cutini}, {de
  Angelis}, {de Palma}, {Dermer}, {Digel}, {Silva}, {Drell}, {Dubois},
  {Escande}, {Favuzzi}, {Fegan}, {Ferrara}, {Finke}, {Focke}, {Fortin},
  {Frailis}, {Fukazawa}, {Funk}, {Fusco}, {Gargano}, {Gasparrini}, {Gehrels},
  {Germani}, {Giebels}, {Giglietto}, {Giommi}, {Giordano}, {Giroletti},
  {Glanzman}, {Godfrey}, {Grenier}, {Grove}, {Guiriec}, {Gustafsson},
  {Hadasch}, {Hayashida}, {Hays}, {Healey}, {Horan}, {Hou}, {Hughes},
  {Iafrate}, {J{\'o}hannesson}, {Johnson}, {Johnson}, {Kamae}, {Katagiri},
  {Kataoka}, {Kn{\"o}dlseder}, {Kuss}, {Lande}, {Larsson}, {Latronico},
  {Longo}, {Loparco}, {Lott}, {Lovellette}, {Lubrano}, {Madejski}, {Mazziotta},
  {McConville}, {McEnery}, {Michelson}, {Mitthumsiri}, {Mizuno}, {Moiseev},
  {Monte}, {Monzani}, {Moretti}, {Morselli}, {Moskalenko}, {Murgia},
  {Nakamori}, {Naumann-Godo}, {Nolan}, {Norris}, {Nuss}, {Ohno}, {Ohsugi},
  {Okumura}, {Omodei}, {Orienti}, {Orlando}, {Ormes}, {Ozaki}, {Paneque},
  {Parent}, {Pesce-Rollins}, {Pierbattista}, {Piranomonte}, {Piron}, {Pivato},
  {Porter}, {Rain{\`o}}, {Rando}, {Razzano}, {Razzaque}, {Reimer}, {Reimer},
  {Ritz}, {Rochester}, {Romani}, {Roth}, {Sanchez}, {Sbarra}, {Scargle},
  {Schalk}, {Sgr{\`o}}, {Shaw}, {Siskind}, {Spandre}, {Spinelli}, {Strong},
  {Suson}, {Tajima}, {Takahashi}, {Takahashi}, {Tanaka}, {Thayer}, {Thayer},
  {Thompson}, {Tibaldo}, {Tinivella}, {Torres}, {Tosti}, {Troja}, {Uchiyama},
  {Vandenbroucke}, {Vasileiou}, {Vianello}, {Vitale}, {Waite}, {Wallace},
  {Wang}, {Winer}, {Wood}, {Wood}, \& {Zimmer}}]{2011ApJ...743..171A}
{Ackermann}, M., {Ajello}, M., {Allafort}, A., {et~al.} 2011, \apj, 743, 171

\bibitem[{{Ahnen} {et~al.}(2017{\natexlab{a}}){Ahnen}, {Ansoldi}, {Antonelli},
  {Antoranz}, {Arcaro}, {Babic}, {Banerjee}, {Bangale}, {de Almeida}, {Barrio},
  {Bednarek}, {Bernardini}, {Berti}, {Biasuzzi}, {Biland}, {Blanch},
  {Bonnefoy}, {Bonnoli}, {Borracci}, {Bretz}, {Buson}, {Carosi}, {Chatterjee},
  {Clavero}, {Colin}, {Colombo}, {Contreras}, {Cortina}, {Covino}, {Da Vela},
  {Dazzi}, {De Angelis}, {De Lotto}, {Wilhelmi}, {Pierro}, {Doert},
  {Dom{\'{\i}}nguez}, {Prester}, {Dorner}, {Doro}, {Einecke}, {Glawion},
  {Elsaesser}, {Engelkemeier}, {Ramazani}, {Fern{\'a}ndez-Barral}, {Fidalgo},
  {Fonseca}, {Font}, {Frantzen}, {Fruck}, {Galindo}, {L{\'o}pez},
  {Garczarczyk}, {Terrats}, {Gaug}, {Giammaria}, {Godinovi{\'c}}, {Gora},
  {Guberman}, {Hadasch}, {Hahn}, {Hayashida}, {Herrera}, {Hose}, {Hrupec},
  {Hughes}, {Idec}, {Kodani}, {Konno}, {Kubo}, {Kushida}, {Barbera}, {Lelas},
  {Lindfors}, {Lombardi}, {Longo}, {L{\'o}pez}, {L{\'o}pez-Coto}, {Majumdar},
  {Makariev}, {Mallot}, {Maneva}, {Manganaro}, {Mankuzhiyil}, {Mannheim},
  {Maraschi}, {Marcote}, {Mariotti}, {Mart{\'{\i}}nez}, {Mazin}, {Menzel},
  {Miranda}, {Mirzoyan}, {Moralejo}, {Moretti}, {Nakajima}, {Neustroev},
  {Niedzwiecki}, {Rosillo}, {Nilsson}, {Nishijima}, {Noda}, {Nogu{\'e}s},
  {Paiano}, {Palacio}, {Palatiello}, {Paneque}, {Paoletti}, {Paredes},
  {Paredes-Fortuny}, {Pedaletti}, {Peresano}, {Perri}, {Persic}, {Poutanen},
  {Moroni}, {Prandini}, {Puljak}, {Garcia}, {Reichardt}, {Rhode}, {Rib{\'o}},
  {Rico}, {Saito}, {Satalecka}, {Schroeder}, {Schweizer}, {Shore},
  {Sillanp{\"a}{\"a}}, {Sitarek}, {Snidaric}, {Sobczynska}, {Stamerra},
  {Strzys}, {Suri{\'c}}, {Takalo}, {Takami}, {Tavecchio}, {Temnikov},
  {Terzi{\'c}}, {Tescaro}, {Teshima}, {Torres}, {Toyama}, {Treves}, {Vanzo},
  {Verguilov}, {Vovk}, {Ward}, {Will}, {Wu}, {Zanin}, {Becerra Gonz{\'a}lez},
  {Rani}, {Krauss}, {Perri}, {Verrecchia}, \& {Reinthal}}]{2017MNRAS.468.1534A}
{Ahnen}, M.~L., {Ansoldi}, S., {Antonelli}, L.~A., {et~al.} 2017{\natexlab{a}},
  \mnras, 468, 1534

\bibitem[{{Ahnen} {et~al.}(2016){Ahnen}, {Ansoldi}, {Antonelli}, {Antoranz},
  {Babic}, {Banerjee}, {Bangale}, {Barres de Almeida}, {Barrio}, {Becerra
  Gonz{\'a}lez}, {Bednarek}, {Bernardini}, {Biasuzzi}, {Biland}, {Blanch},
  {Bonnefoy}, {Bonnoli}, {Borracci}, {Bretz}, {Carmona}, {Carosi},
  {Chatterjee}, {Clavero}, {Colin}, {Colombo}, {Contreras}, {Cortina},
  {Covino}, {Da Vela}, {Dazzi}, {De Angelis}, {De Caneva}, {De Lotto}, {de
  O{\~n}a Wilhelmi}, {Delgado Mendez}, {Di Pierro}, {Dominis Prester},
  {Dorner}, {Doro}, {Einecke}, {Elsaesser}, {Fern{\'a}ndez-Barral}, {Fidalgo},
  {Fonseca}, {Font}, {Frantzen}, {Fruck}, {Galindo}, {Garc{\'\i}a L{\'o}pez},
  {Garczarczyk}, {Garrido Terrats}, {Gaug}, {Giammaria}, {Eisenacher Glawion},
  {Godinovi{\'c}}, {Gonz{\'a}lez Mu{\~n}oz}, {Guberman}, {Hanabata},
  {Hayashida}, {Herrera}, {Hose}, {Hrupec}, {Hughes}, {Idec}, {Kodani},
  {Konno}, {Kubo}, {Kushida}, {La Barbera}, {Lelas}, {Lindfors}, {Lombardi},
  {Longo}, {L{\'o}pez}, {L{\'o}pez-Coto}, {L{\'o}pez-Oramas}, {Lorenz},
  {Majumdar}, {Makariev}, {Mallot}, {Maneva}, {Manganaro}, {Mannheim},
  {Maraschi}, {Marcote}, {Mariotti}, {Mart{\'\i}nez}, {Mazin}, {Menzel},
  {Miranda}, {Mirzoyan}, {Moralejo}, {Nakajima}, {Neustroev}, {Niedzwiecki},
  {Nievas Rosillo}, {Nilsson}, {Nishijima}, {Noda}, {Orito}, {Overkemping},
  {Paiano}, {Palacio}, {Palatiello}, {Paneque}, {Paoletti}, {Paredes},
  {Paredes-Fortuny}, {Persic}, {Poutanen}, {Prada Moroni}, {Prand ini},
  {Puljak}, {Reinthal}, {Rhode}, {Rib{\'o}}, {Rico}, {Rodriguez Garcia},
  {R{\"u}gamer}, {Saito}, {Satalecka}, {Scapin}, {Schultz}, {Schweizer},
  {Shore}, {Sillanp{\"a}{\"a}}, {Sitarek}, {Snidaric}, {Sobczynska},
  {Stamerra}, {Steinbring}, {Strzys}, {Takalo}, {Takami}, {Tavecchio},
  {Temnikov}, {Terzi{\'c}}, {Tescaro}, {Teshima}, {Thaele}, {Torres}, {Toyama},
  {Treves}, {Verguilov}, {Vovk}, {Ward}, {Will}, {Wu}, {Zanin}, {Lucarelli},
  {Pittori}, {Vercellone}, {Berdyugin}, {Carini}, {L{\"a}hteenm{\"a}ki},
  {Pasanen}, {Pease}, {Sainio}, {Tornikoski}, \&
  {Walters}}]{2016MNRAS.459.2286A}
{Ahnen}, M.~L., {Ansoldi}, S., {Antonelli}, L.~A., {et~al.} 2016, \mnras, 459,
  2286

\bibitem[{{Ahnen} {et~al.}(2017{\natexlab{b}}){Ahnen}, {Ansoldi}, {Antonelli},
  {Antoranz}, {Babic}, {Banerjee}, {Bangale}, {Barres de Almeida}, {Barrio},
  {Becerra Gonz{\'a}lez}, \& et~al.}]{2017A&A...603A..31A}
{Ahnen}, M.~L., {Ansoldi}, S., {Antonelli}, L.~A., {et~al.} 2017{\natexlab{b}},
  \aap, 603, A31

\bibitem[{{Aleksi{\'c}} {et~al.}(2016){Aleksi{\'c}}, {Ansoldi}, {Antonelli},
  {Antoranz}, {Babic}, {Bangale}, {Barcel{\'o}}, {Barrio}, {Becerra
  Gonz{\'a}lez}, {Bednarek}, {Bernardini}, {Biasuzzi}, {Biland}, {Bitossi},
  {Blanch}, {Bonnefoy}, {Bonnoli}, {Borracci}, {Bretz}, {Carmona}, {Carosi},
  {Cecchi}, {Colin}, {Colombo}, {Contreras}, {Corti}, {Cortina}, {Covino}, {Da
  Vela}, {Dazzi}, {De Angelis}, {De Caneva}, {De Lotto}, {de O{\~n}a Wilhelmi},
  {Delgado Mendez}, {Dettlaff}, {Dominis Prester}, {Dorner}, {Doro}, {Einecke},
  {Eisenacher}, {Elsaesser}, {Fidalgo}, {Fink}, {Fonseca}, {Font}, {Frantzen},
  {Fruck}, {Galindo}, {Garc{\'{\i}}a L{\'o}pez}, {Garczarczyk}, {Garrido
  Terrats}, {Gaug}, {Giavitto}, {Godinovi{\'c}}, {Gonz{\'a}lez Mu{\~n}oz},
  {Gozzini}, {Haberer}, {Hadasch}, {Hanabata}, {Hayashida}, {Herrera},
  {Hildebrand}, {Hose}, {Hrupec}, {Idec}, {Illa}, {Kadenius}, {Kellermann},
  {Knoetig}, {Kodani}, {Konno}, {Krause}, {Kubo}, {Kushida}, {La Barbera},
  {Lelas}, {Lemus}, {Lewandowska}, {Lindfors}, {Lombardi}, {Longo},
  {L{\'o}pez}, {L{\'o}pez-Coto}, {L{\'o}pez-Oramas}, {Lorca}, {Lorenz},
  {Lozano}, {Makariev}, {Mallot}, {Maneva}, {Mankuzhiyil}, {Mannheim},
  {Maraschi}, {Marcote}, {Mariotti}, {Mart{\'{\i}}nez}, {Mazin}, {Menzel},
  {Miranda}, {Mirzoyan}, {Moralejo}, {Munar-Adrover}, {Nakajima}, {Negrello},
  {Neustroev}, {Niedzwiecki}, {Nilsson}, {Nishijima}, {Noda}, {Orito},
  {Overkemping}, {Paiano}, {Palatiello}, {Paneque}, {Paoletti}, {Paredes},
  {Paredes-Fortuny}, {Persic}, {Poutanen}, {Prada Moroni}, {Prandini},
  {Puljak}, {Reinthal}, {Rhode}, {Rib{\'o}}, {Rico}, {Rodriguez Garcia},
  {R{\"u}gamer}, {Saito}, {Saito}, {Satalecka}, {Scalzotto}, {Scapin},
  {Schultz}, {Schlammer}, {Schmidl}, {Schweizer}, {Shore}, {Sillanp{\"a}{\"a}},
  {Sitarek}, {Snidaric}, {Sobczynska}, {Spanier}, {Stamerra}, {Steinbring},
  {Storz}, {Strzys}, {Takalo}, {Takami}, {Tavecchio}, {Tejedor}, {Temnikov},
  {Terzi{\'c}}, {Tescaro}, {Teshima}, {Thaele}, {Tibolla}, {Torres}, {Toyama},
  {Treves}, {Vogler}, {Wetteskind}, {Will}, \& {Zanin}}]{2016APh....72...76A}
{Aleksi{\'c}}, J., {Ansoldi}, S., {Antonelli}, L.~A., {et~al.} 2016,
  Astroparticle Physics, 72, 76

\bibitem[{{Aleksi{\'c}} {et~al.}(2014){Aleksi{\'c}}, {Ansoldi}, {Antonelli},
  {Antoranz}, {Babic}, {Bangale}, {Barres de Almeida}, {Barrio}, {Becerra
  Gonz{\'a}lez}, {Bednarek}, {Berger}, {Bernardini}, {Biland}, {Blanch},
  {Bock}, {Bonnefoy}, {Bonnoli}, {Borracci}, {Bretz}, {Carmona}, {Carosi},
  {Carreto Fidalgo}, {Colin}, {Colombo}, {Contreras}, {Cortina}, {Covino}, {Da
  Vela}, {Dazzi}, {De Angelis}, {De Caneva}, {De Lotto}, {Delgado Mendez},
  {Doert}, {Dom{\'{\i}}nguez}, {Dominis Prester}, {Dorner}, {Doro}, {Einecke},
  {Eisenacher}, {Elsaesser}, {Farina}, {Ferenc}, {Fonseca}, {Font}, {Frantzen},
  {Fruck}, {Garc{\'{\i}}a L{\'o}pez}, {Garczarczyk}, {Garrido Terrats}, {Gaug},
  {Giavitto}, {Godinovi{\'c}}, {Gonz{\'a}lez Mu{\~n}oz}, {Gozzini}, {Hadasch},
  {Hayashida}, {Herrero}, {Hildebrand}, {Hose}, {Hrupec}, {Idec}, {Kadenius},
  {Kellermann}, {Kodani}, {Konno}, {Krause}, {Kubo}, {Kushida}, {La Barbera},
  {Lelas}, {Lewandowska}, {Lindfors}, {Lombardi}, {L{\'o}pez},
  {L{\'o}pez-Coto}, {L{\'o}pez-Oramas}, {Lorenz}, {Lozano}, {Makariev},
  {Mallot}, {Maneva}, {Mankuzhiyil}, {Mannheim}, {Maraschi}, {Marcote},
  {Mariotti}, {Mart{\'{\i}}nez}, {Mazin}, {Menzel}, {Meucci}, {Miranda},
  {Mirzoyan}, {Moralejo}, {Munar-Adrover}, {Nakajima}, {Niedzwiecki},
  {Nilsson}, {Nishijima}, {Nowak}, {Orito}, {Overkemping}, {Paiano},
  {Palatiello}, {Paneque}, {Paoletti}, {Paredes}, {Paredes-Fortuny}, {Partini},
  {Persic}, {Prada}, {Prada Moroni}, {Prandini}, {Preziuso}, {Puljak},
  {Reinthal}, {Rhode}, {Rib{\'o}}, {Rico}, {Rodriguez Garcia}, {R{\"u}gamer},
  {Saggion}, {Saito}, {Saito}, {Salvati}, {Satalecka}, {Scalzotto}, {Scapin},
  {Schultz}, {Schweizer}, {Shore}, {Sillanp{\"a}{\"a}}, {Sitarek}, {Snidaric},
  {Sobczynska}, {Spanier}, {Stamatescu}, {Stamerra}, {Steinbring}, {Storz},
  {Sun}, {Suri{\'c}}, {Takalo}, {Takami}, {Tavecchio}, {Temnikov},
  {Terzi{\'c}}, {Tescaro}, {Teshima}, {Thaele}, {Tibolla}, {Torres}, {Toyama},
  {Treves}, {Uellenbeck}, {Vogler}, {Wagner}, {Zandanel}, {Zanin}, {MAGIC
  Collaboration}, {Cutini}, {Gasparrini}, {Furniss}, {Hovatta}, {Kangas},
  {Kankare}, {Kotilainen}, {Lister}, {L{\"a}hteenm{\"a}ki}, {Max-Moerbeck},
  {Pavlidou}, {Readhead}, \& {Richards}}]{2014A&A...567A.135A}
{Aleksi{\'c}}, J., {Ansoldi}, S., {Antonelli}, L.~A., {et~al.} 2014, \aap, 567,
  A135

\bibitem[{{Aleksi{\'c}} {et~al.}(2015){Aleksi{\'c}}, {Ansoldi}, {Antonelli},
  {Antoranz}, {Babic}, {Bangale}, {Barres de Almeida}, {Barrio}, {Becerra
  Gonz{\'a}lez}, {Bednarek}, {Berger}, {Bernardini}, {Biland}, {Blanch},
  {Bock}, {Bonnefoy}, {Bonnoli}, {Borracci}, {Bretz}, {Carmona}, {Carosi},
  {Carreto Fidalgo}, {Colin}, {Colombo}, {Contreras}, {Cortina}, {Covino}, {Da
  Vela}, {Dazzi}, {De Angelis}, {De Caneva}, {De Lotto}, {Delgado Mendez},
  {Doert}, {Dom{\'\i}nguez}, {Dominis Prester}, {Dorner}, {Doro}, {Einecke},
  {Eisenacher}, {Elsaesser}, {Farina}, {Ferenc}, {Fonseca}, {Font}, {Frantzen},
  {Fruck}, {Garc{\'\i}a L{\'o}pez}, {Garczarczyk}, {Garrido Terrats}, {Gaug},
  {Giavitto}, {Godinovi{\'c}}, {Gonz{\'a}lez Mu{\~n}oz}, {Gozzini}, {Hadamek},
  {Hadasch}, {Herrero}, {Hildebrand}, {Hose}, {Hrupec}, {Idec}, {Kadenius},
  {Kellermann}, {Knoetig}, {Krause}, {Kushida}, {La Barbera}, {Lelas},
  {Lewandowska}, {Lindfors}, {Longo}, {Lombardi}, {L{\'o}pez},
  {L{\'o}pez-Coto}, {L{\'o}pez-Oramas}, {Lorenz}, {Lozano}, {Makariev},
  {Mallot}, {Maneva}, {Mankuzhiyil}, {Mannheim}, {Maraschi}, {Marcote},
  {Mariotti}, {Mart{\'\i}nez}, {Mazin}, {Menzel}, {Meucci}, {Miranda},
  {Mirzoyan}, {Moralejo}, {Munar-Adrover}, {Nakajima}, {Niedzwiecki},
  {Nilsson}, {Nowak}, {Orito}, {Overkemping}, {Paiano}, {Palatiello},
  {Paneque}, {Paoletti}, {Paredes}, {Paredes-Fortuny}, {Partini}, {Persic},
  {Prada}, {Prada Moroni}, {Prand ini}, {Preziuso}, {Puljak}, {Reinthal},
  {Rhode}, {Rib{\'o}}, {Rico}, {RodriguezGarcia}, {R{\"u}gamer}, {Saggion},
  {Saito}, {Salvati}, {Satalecka}, {Scalzotto}, {Scapin}, {Schultz},
  {Schweizer}, {Shore}, {Sillanp{\"a}{\"a}}, {Sitarek}, {Snidaric},
  {Sobczynska}, {Spanier}, {Stamatescu}, {Stamerra}, {Steinbring}, {Storz},
  {Sun}, {Suri{\'c}}, {Takalo}, {Tavecchio}, {Temnikov}, {Terzi{\'c}},
  {Tescaro}, {Teshima}, {Thaele}, {Tibolla}, {Torres}, {Toyama}, {Treves},
  {Uellenbeck}, {Vogler}, {Wagner}, {Zandanel}, {Zanin}, {MAGIC Collaboration},
  {Archambault}, {Behera}, {Beilicke}, {Benbow}, {Bird}, {Buckley}, {Bugaev},
  {Cerruti}, {Chen}, {Ciupik}, {Collins-Hughes}, {Cui}, {Dumm}, {Eisch},
  {Falcone}, {Federici}, {Feng}, {Finley}, {Fleischhack}, {Fortin}, {Fortson},
  {Furniss}, {Griffin}, {Griffiths}, {Grube}, {Gyuk}, {Hanna}, {Holder},
  {Hughes}, {Humensky}, {Johnson}, {Kaaret}, {Kertzman}, {Khassen}, {Kieda},
  {Krawczynski}, {Krennrich}, {Kumar}, {Lang}, {Maier}, {McArthur}, {Meagher},
  {Moriarty}, {Mukherjee}, {Ong}, {Otte}, {Park}, {Pichel}, {Pohl}, {Popkow},
  {Prokoph}, {Quinn}, {Ragan}, {Rajotte}, {Reynolds}, {Richards}, {Roache},
  {Rovero}, {Sembroski}, {Shahinyan}, {Staszak}, {Telezhinsky}, {Theiling},
  {Tucci}, {Tyler}, {Varlotta}, {Wakely}, {Weekes}, {Weinstein}, {Welsing},
  {Wilhelm}, {Williams}, {Zitzer}, {VERITAS Collaboration}, {Villata},
  {Raiteri}, {Aller}, {Aller}, {Chen}, {Jordan}, {Koptelova}, {Kurtanidze},
  {L{\"a}hteenm{\"a}ki}, {McBreen}, {Larionov}, {Lin}, {Nikolashvili},
  {Angelakis}, {Capalbi}, {Carrami{\~n}ana}, {Carrasco}, {Cassaro}, {Cesarini},
  {Fuhrmann}, {Giroletti}, {Hovatta}, {Krichbaum}, {Krimm}, {Max-Moerbeck},
  {Moody}, {Maccaferri}, {Mori}, {Nestoras}, {Orlati}, {Pace}, {Pearson},
  {Perri}, {Readhead}, {Richards}, {Sadun}, {Sakamoto}, {Tammi}, {Tornikoski},
  {Yatsu}, \& {Zook}}]{2015A&A...576A.126A}
{Aleksi{\'c}}, J., {Ansoldi}, S., {Antonelli}, L.~A., {et~al.} 2015, \aap, 576,
  A126

\bibitem[{{Aleksi{\'c}} {et~al.}(2013){Aleksi{\'c}}, {Antonelli}, {Antoranz},
  {Asensio}, {Backes}, {Barres de Almeida}, {Barrio}, {Bednarek}, {Berger},
  {Bernardini}, {Biland}, {Blanch}, {Bock}, {Boller}, {Bonnefoy}, {Bonnoli},
  {Borla Tridon}, {Bretz}, {Carmona}, {Carosi}, {Carreto Fidalgo}, {Colin},
  {Colombo}, {Contreras}, {Cortina}, {Cossio}, {Covino}, {Da Vela}, {Dazzi},
  {De Angelis}, {De Caneva}, {De Lotto}, {Delgado Mendez}, {Doert},
  {Dom{\'{\i}}nguez}, {Dominis Prester}, {Dorner}, {Doro}, {Eisenacher},
  {Elsaesser}, {Ferenc}, {Fonseca}, {Font}, {Fruck}, {Garc{\'{\i}}a L{\'o}pez},
  {Garczarczyk}, {Garrido Terrats}, {Gaug}, {Giavitto}, {Godinovi{\'c}},
  {Gonz{\'a}lez Mu{\~n}oz}, {Gozzini}, {Hadamek}, {Hadasch}, {Herrero}, {Hose},
  {Hrupec}, {Jankowski}, {Kadenius}, {Klepser}, {Knoetig},
  {Kr{\"a}henb{\"u}hl}, {Krause}, {Kushida}, {La Barbera}, {Lelas}, {Leonardo},
  {Lewandowska}, {Lindfors}, {Lombardi}, {L{\'o}pez}, {L{\'o}pez-Coto},
  {L{\'o}pez-Oramas}, {Lorenz}, {Lozano}, {Makariev}, {Mallot}, {Maneva},
  {Mankuzhiyil}, {Mannheim}, {Maraschi}, {Marcote}, {Mariotti},
  {Mart{\'{\i}}nez}, {Masbou}, {Mazin}, {Meucci}, {Miranda}, {Mirzoyan},
  {Mold{\'o}n}, {Moralejo}, {Munar-Adrover}, {Nakajima}, {Niedzwiecki},
  {Nieto}, {Nilsson}, {Nowak}, {Orito}, {Paiano}, {Palatiello}, {Paneque},
  {Paoletti}, {Paredes}, {Partini}, {Persic}, {Pilia}, {Prada}, {Prada Moroni},
  {Prandini}, {Puljak}, {Reichardt}, {Reinthal}, {Rhode}, {Rib{\'o}}, {Rico},
  {R{\"u}gamer}, {Saggion}, {Saito}, {Saito}, {Salvati}, {Satalecka},
  {Scalzotto}, {Scapin}, {Schultz}, {Schweizer}, {Shore}, {Sillanp{\"a}{\"a}},
  {Sitarek}, {Snidaric}, {Sobczynska}, {Spanier}, {Spiro}, {Stamatescu},
  {Stamerra}, {Steinke}, {Storz}, {Sun}, {Suri{\'c}}, {Takalo}, {Takami},
  {Tavecchio}, {Temnikov}, {Terzi{\'c}}, {Tescaro}, {Teshima}, {Tibolla},
  {Torres}, {Toyama}, {Treves}, {Uellenbeck}, {Vogler}, {Wagner}, {Weitzel},
  {Zandanel}, {Zanin}, {MAGIC Collaboration}, {Longo}, {Lucarelli}, {Pittori},
  {Vercellone}, {AGILE Team}, {Bastieri}, {Sbarra}, {Fermi-LAT Collaboration},
  {Angelakis}, {Fuhrmann}, {Nestoras}, {Krichbaum}, {Sievers}, {Zensus},
  {F-GAMMA program}, {Antonyuk}, {Baumgartner}, {Berduygin}, {Carini}, {Cook},
  {Gehrels}, {Kadler}, {Kovalev}, {Kovalev}, {Krauss}, {Krimm},
  {L{\"a}hteenm{\"a}ki}, {Lister}, {Max-Moerbeck}, {Pasanen}, {Pushkarev},
  {Readhead}, {Richards}, {Sainio}, {Shakhovskoy}, {Sokolovsky}, {Tornikoski},
  {Tueller}, {Weidinger}, \& {Wilms}}]{2013A&A...556A..67A}
{Aleksi{\'c}}, J., {Antonelli}, L.~A., {Antoranz}, P., {et~al.} 2013, \aap,
  556, A67

\bibitem[{{Aliu} {et~al.}(2013){Aliu}, {Archambault}, {Arlen}, {Aune},
  {Beilicke}, {Benbow}, {Bird}, {B{\"o}ttcher}, {Bouvier}, {Bugaev}, {Byrum},
  {Cesarini}, {Ciupik}, {Collins-Hughes}, {Connolly}, {Cui}, {Dickherber},
  {Duke}, {Dumm}, {Errand o}, {Falcone}, {Federici}, {Feng}, {Finley},
  {Finnegan}, {Fortson}, {Furniss}, {Galante}, {Gall}, {Gillanders}, {Griffin},
  {Grube}, {Gyuk}, {Hanna}, {Holder}, {Hughes}, {Humensky}, {Kaaret},
  {Kertzman}, {Khassen}, {Kieda}, {Krawczynski}, {Krennrich}, {Lang},
  {Madhavan}, {Maier}, {Majumdar}, {McArthur}, {McCann}, {Moriarty},
  {Mukherjee}, {Nelson}, {O'Faol{\'a}in de Bhr{\'o}ithe}, {Ong}, {Orr}, {Otte},
  {Park}, {Perkins}, {Pichel}, {Pohl}, {Popkow}, {Prokoph}, {Quinn}, {Ragan},
  {Reyes}, {Reynolds}, {Roache}, {Saxon}, {Schroedter}, {Sembroski}, {Skole},
  {Smith}, {Staszak}, {Telezhinsky}, {Theiling}, {Tyler}, {Varlotta},
  {Vassiliev}, {Wakely}, {Weekes}, {Weinstein}, {Welsing}, {Williams}, \&
  {Zitzer}}]{2013ApJ...775....3A}
{Aliu}, E., {Archambault}, S., {Arlen}, T., {et~al.} 2013, \apj, 775, 3

\bibitem[{{Angel} {et~al.}(1978){Angel}, {Boroson}, {Adams}, {Duerr},
  {Giampapa}, {Gresham}, {Gural}, {Hubbard}, {Kopriva}, {Moore}, {Peterson},
  {Schmidt}, {Turnshek}, {Wilkerson}, {Zotov}, {Maza}, \&
  {Kinman}}]{1978bllo.conf..117A}
{Angel}, J.~R.~P., {Boroson}, T.~A., {Adams}, M.~T., {et~al.} 1978, in BL Lac
  Objects, ed. A.~M. {Wolfe}, 117--146

\bibitem[{{Archambault} {et~al.}(2015){Archambault}, {Archer}, {Beilicke},
  {Benbow}, {Bird}, {Biteau}, {Bouvier}, {Bugaev}, {Cardenzana}, {Cerruti},
  {Chen}, {Ciupik}, {Connolly}, {Cui}, {Dickinson}, {Dumm}, {Eisch}, {Errando},
  {Falcone}, {Feng}, {Finley}, {Fleischhack}, {Fortin}, {Fortson}, {Furniss},
  {Gillanders}, {Griffin}, {Griffiths}, {Grube}, {Gyuk}, {H{\aa}kansson},
  {Hanna}, {Holder}, {Humensky}, {Johnson}, {Kaaret}, {Kar}, {Kertzman},
  {Khassen}, {Kieda}, {Krause}, {Krennrich}, {Kumar}, {Lang}, {Maier},
  {McArthur}, {McCann}, {Meagher}, {Millis}, {Moriarty}, {Mukherjee}, {Nieto},
  {O'Faol{\'a}in de Bhr{\'o}ithe}, {Ong}, {Otte}, {Park}, {Pohl}, {Popkow},
  {Prokoph}, {Pueschel}, {Quinn}, {Ragan}, {Reyes}, {Reynolds}, {Richards},
  {Roache}, {Santander}, {Sembroski}, {Shahinyan}, {Smith}, {Staszak},
  {Telezhinsky}, {Tucci}, {Tyler}, {Varlotta}, {Vincent}, {Wakely},
  {Weinstein}, {Welsing}, {Wilhelm}, {Williams}, {Zitzer}, {Veritas
  Collaboration}, \& {Hughes}}]{2015ApJ...808..110A}
{Archambault}, S., {Archer}, A., {Beilicke}, M., {et~al.} 2015, \apj, 808, 110

\bibitem[{{Archambault} {et~al.}(2013){Archambault}, {Arlen}, {Aune}, {Behera},
  {Beilicke}, {Benbow}, {Bird}, {Bouvier}, {Buckley}, {Bugaev}, {Byrum},
  {Cesarini}, {Ciupik}, {Connolly}, {Cui}, {Errando}, {Falcone}, {Federici},
  {Feng}, {Finley}, {Fortson}, {Furniss}, {Galante}, {Gall}, {Gillanders},
  {Griffin}, {Grube}, {Gyuk}, {Hanna}, {Holder}, {Hughes}, {Humensky},
  {Kaaret}, {Kertzman}, {Khassen}, {Kieda}, {Krawczynski}, {Krennrich},
  {Kumar}, {Lang}, {Madhavan}, {Maier}, {Majumdar}, {McArthur}, {McCann},
  {Millis}, {Moriarty}, {Mukherjee}, {O'Faol{\'a}in de Bhr{\'o}ithe}, {Ong},
  {Otte}, {Park}, {Perkins}, {Pohl}, {Popkow}, {Prokoph}, {Quinn}, {Ragan},
  {Reyes}, {Reynolds}, {Richards}, {Roache}, {Saxon}, {Sembroski}, {Smith},
  {Staszak}, {Telezhinsky}, {Theiling}, {Varlotta}, {Vassiliev}, {Vincent},
  {Wakely}, {Weekes}, {Weinstein}, {Welsing}, {Williams}, {Zitzer}, {VERITAS
  Collaboration}, {B{\"o}ttcher}, {Fegan}, {Fortin}, {Halpern}, {Kovalev},
  {Lister}, {Liu}, {Pushkarev}, \& {Smith}}]{2013ApJ...776...69A}
{Archambault}, S., {Arlen}, T., {Aune}, T., {et~al.} 2013, \apj, 776, 69

\bibitem[{{Archambault} {et~al.}(2014){Archambault}, {Aune}, {Behera},
  {Beilicke}, {Benbow}, {Berger}, {Bird}, {Biteau}, {Bugaev}, {Byrum},
  {Cardenzana}, {Cerruti}, {Chen}, {Ciupik}, {Connolly}, {Cui}, {Dumm},
  {Errando}, {Falcone}, {Federici}, {Feng}, {Finley}, {Fleischhack}, {Fortson},
  {Furniss}, {Galante}, {Gillanders}, {Griffin}, {Griffiths}, {Grube}, {Gyuk},
  {Hanna}, {Holder}, {Hughes}, {Humensky}, {Johnson}, {Kaaret}, {Kertzman},
  {Khassen}, {Kieda}, {Krawczynski}, {Krennrich}, {Kumar}, {Lang}, {Madhavan},
  {Maier}, {McCann}, {Meagher}, {Moriarty}, {Mukherjee}, {Nieto},
  {O'Faol{\'a}in de Bhr{\'o}ithe}, {Ong}, {Otte}, {Park}, {Pohl}, {Popkow},
  {Prokoph}, {Quinn}, {Ragan}, {Rajotte}, {Reyes}, {Reynolds}, {Richards},
  {Roache}, {Sembroski}, {Shahinyan}, {Staszak}, {Telezhinsky}, {Tucci},
  {Tyler}, {Varlotta}, {Vassiliev}, {Vincent}, {Wakely}, {Weinstein},
  {Welsing}, {Wilhelm}, {Williams}, {VERITAS Collaboration}, {Ackermann},
  {Ajello}, {Albert}, {Baldini}, {Bastieri}, {Bellazzini}, {Bissaldi},
  {Bregeon}, {Buehler}, {Buson}, {Caliandro}, {Cameron}, {Caraveo},
  {Cavazzuti}, {Charles}, {Chiang}, {Ciprini}, {Claus}, {Cutini}, {D'Ammando},
  {de Angelis}, {de Palma}, {Dermer}, {Digel}, {Di Venere}, {Drell}, {Favuzzi},
  {Franckowiak}, {Fusco}, {Gargano}, {Gasparrini}, {Giglietto}, {Giordano},
  {Giroletti}, {Grenier}, {Guiriec}, {Jogler}, {Kuss}, {Larsson}, {Latronico},
  {Longo}, {Loparco}, {Lubrano}, {Madejski}, {Mayer}, {Mazziotta}, {Michelson},
  {Mizuno}, {Monzani}, {Morselli}, {Murgia}, {Nuss}, {Ohsugi}, {Ormes},
  {Paneque}, {Perkins}, {Piron}, {Pivato}, {Rain{\`o}}, {Razzano}, {Reimer},
  {Reimer}, {Ritz}, {Schaal}, {Sgr{\`o}}, {Siskind}, {Spinelli}, {Takahashi},
  {Tibaldo}, {Tinivella}, {Troja}, {Vianello}, {Werner}, {Wood}, \& {Fermi LAT
  Collaboration}}]{2014ApJ...785L..16A}
{Archambault}, S., {Aune}, T., {Behera}, B., {et~al.} 2014, \apjl, 785, L16

\bibitem[{{Attridge} {et~al.}(1999){Attridge}, {Roberts}, \&
  {Wardle}}]{attridge99}
{Attridge}, J.~M., {Roberts}, D.~H., \& {Wardle}, J.~F.~C. 1999, \apjl, 518,
  L87

\bibitem[{{Atwood} {et~al.}(2009){Atwood}, {Abdo}, {Ackermann}, {Althouse},
  {Anderson}, {Axelsson}, {Baldini}, {Ballet}, {Band}, {Barbiellini}, \&
  et~al.}]{2009ApJ...697.1071A}
{Atwood}, W.~B., {Abdo}, A.~A., {Ackermann}, M., {et~al.} 2009, \apj, 697, 1071

\bibitem[{{Barres de Almeida} {et~al.}(2014){Barres de Almeida}, {Tavecchio},
  \& {Mankuzhiyil}}]{2014MNRAS.441.2885B}
{Barres de Almeida}, U., {Tavecchio}, F., \& {Mankuzhiyil}, N. 2014, \mnras,
  441, 2885

\bibitem[{{Barres de Almeida} {et~al.}(2010){Barres de Almeida}, {Ward},
  {Dominici}, {Abraham}, {Franco}, {Daniel}, {Chadwick}, \&
  {Boisson}}]{2010MNRAS.408.1778B}
{Barres de Almeida}, U., {Ward}, M.~J., {Dominici}, T.~P., {et~al.} 2010,
  \mnras, 408, 1778

\bibitem[{{Bhatta} {et~al.}(2018){Bhatta}, {Mohorian}, \&
  {Bilinsky}}]{2018A&A...619A..93B}
{Bhatta}, G., {Mohorian}, M., \& {Bilinsky}, I. 2018, \aap, 619, A93

\bibitem[{{Blandford} \& {K{\"o}nigl}(1979)}]{1979ApJ...232...34B}
{Blandford}, R.~D. \& {K{\"o}nigl}, A. 1979, \apj, 232, 34

\bibitem[{{Bloom} \& {Marscher}(1996)}]{1996ApJ...461..657B}
{Bloom}, S.~D. \& {Marscher}, A.~P. 1996, \apj, 461, 657

\bibitem[{{Bottacini} {et~al.}(2010){Bottacini}, {B{\"o}ttcher}, {Schady},
  {Rau}, {Zhang}, {Ajello}, {Fendt}, \& {Greiner}}]{2010ApJ...719L.162B}
{Bottacini}, E., {B{\"o}ttcher}, M., {Schady}, P., {et~al.} 2010, \apjl, 719,
  L162

\bibitem[{{Breeveld} {et~al.}(2010){Breeveld}, {Curran}, {Hoversten}, {Koch},
  {Landsman}, {Marshall}, {Page}, {Poole}, {Roming}, {Smith}, {Still},
  {Yershov}, {Blustin}, {Brown}, {Gronwall}, {Holland}, {Kuin}, {McGowan},
  {Rosen}, {Boyd}, {Broos}, {Carter}, {Chester}, {Hancock}, {Huckle}, {Immler},
  {Ivanushkina}, {Kennedy}, {Mason}, {Morgan}, {Oates}, {de Pasquale},
  {Schady}, {Siegel}, \& {vanden Berk}}]{2010MNRAS.406.1687B}
{Breeveld}, A.~A., {Curran}, P.~A., {Hoversten}, E.~A., {et~al.} 2010, \mnras,
  406, 1687

\bibitem[{{Burrows} {et~al.}(2004){Burrows}, {Hill}, {Nousek}, {Wells},
  {Chincarini}, {Abbey}, {Beardmore}, {Bosworth}, {Br{\"a}uninger}, {Burkert},
  {Campana}, {Capalbi}, {Chang}, {Citterio}, {Freyberg}, {Giommi}, {Hartner},
  {Killough}, {Kittle}, {Klar}, {Mangels}, {McMeekin}, {Miles}, {Moretti},
  {Mori}, {Morris}, {Mukerjee}, {Osborne}, {Short}, {Tagliaferri},
  {Tamburelli}, {Watson}, {Willingale}, \& {Zugger}}]{2004SPIE.5165..201B}
{Burrows}, D.~N., {Hill}, J.~E., {Nousek}, J.~A., {et~al.} 2004, in \procspie,
  Vol. 5165, X-Ray and Gamma-Ray Instrumentation for Astronomy XIII, ed. K.~A.
  {Flanagan} \& O.~H.~W. {Siegmund}, 201--216

\bibitem[{{Cardelli} {et~al.}(1989){Cardelli}, {Clayton}, \&
  {Mathis}}]{1989ApJ...345..245C}
{Cardelli}, J.~A., {Clayton}, G.~C., \& {Mathis}, J.~S. 1989, \apj, 345, 245

\bibitem[{{Celotti} \& {Ghisellini}(2008)}]{2008MNRAS.385..283C}
{Celotti}, A. \& {Ghisellini}, G. 2008, \mnras, 385, 283

\bibitem[{{Cerruti} {et~al.}(2017){Cerruti}, {Benbow}, {Chen}, {Dumm},
  {Fortson}, \& {Shahinyan}}]{2017A&A...606A..68C}
{Cerruti}, M., {Benbow}, W., {Chen}, X., {et~al.} 2017, \aap, 606, A68

\bibitem[{{Dermer} \& {Schlickeiser}(1993)}]{1993ApJ...416..458D}
{Dermer}, C.~D. \& {Schlickeiser}, R. 1993, \apj, 416, 458

\bibitem[{{Dom{\'{\i}}nguez} {et~al.}(2011){Dom{\'{\i}}nguez}, {Primack},
  {Rosario}, {Prada}, {Gilmore}, {Faber}, {Koo}, {Somerville},
  {P{\'e}rez-Torres}, {P{\'e}rez-Gonz{\'a}lez}, {Huang}, {Davis},
  {Guhathakurta}, {Barmby}, {Conselice}, {Lozano}, {Newman}, \&
  {Cooper}}]{2011MNRAS.410.2556D}
{Dom{\'{\i}}nguez}, A., {Primack}, J.~R., {Rosario}, D.~J., {et~al.} 2011,
  \mnras, 410, 2556

\bibitem[{{Edelson} \& {Krolik}(1988)}]{1988ApJ...333..646E}
{Edelson}, R.~A. \& {Krolik}, J.~H. 1988, \apj, 333, 646

\bibitem[{{Emmanoulopoulos} {et~al.}(2013){Emmanoulopoulos}, {McHardy}, \&
  {Papadakis}}]{2013MNRAS.433..907E}
{Emmanoulopoulos}, D., {McHardy}, I.~M., \& {Papadakis}, I.~E. 2013, \mnras,
  433, 907

\bibitem[{{Evans} {et~al.}(2009){Evans}, {Beardmore}, {Page}, {Osborne},
  {O'Brien}, {Willingale}, {Starling}, {Burrows}, {Godet}, {Vetere}, {Racusin},
  {Goad}, {Wiersema}, {Angelini}, {Capalbi}, {Chincarini}, {Gehrels}, {Kennea},
  {Margutti}, {Morris}, {Mountford}, {Pagani}, {Perri}, {Romano}, \&
  {Tanvir}}]{2009MNRAS.397.1177E}
{Evans}, P.~A., {Beardmore}, A.~P., {Page}, K.~L., {et~al.} 2009, \mnras, 397,
  1177

\bibitem[{{Fallah Ramazani} {et~al.}(2017){Fallah Ramazani}, {Lindfors}, \&
  {Nilsson}}]{2017A&A...608A..68F}
{Fallah Ramazani}, V., {Lindfors}, E., \& {Nilsson}, K. 2017, \aap, 608, A68

\bibitem[{{Fermi-LAT collaboration}(2019)}]{4LAC}
{Fermi-LAT collaboration}. 2019, arXiv e-prints, arXiv:1905.10771

\bibitem[{{Fukugita} {et~al.}(1995){Fukugita}, {Shimasaku}, \&
  {Ichikawa}}]{1995PASP..107..945F}
{Fukugita}, M., {Shimasaku}, K., \& {Ichikawa}, T. 1995, \pasp, 107, 945

\bibitem[{{Fumagalli} {et~al.}(2012){Fumagalli}, {Dessauges-Zavadsky},
  {Furniss}, {Prochaska}, {Williams}, {Kaplan}, \&
  {Hogan}}]{2012MNRAS.424.2276F}
{Fumagalli}, M., {Dessauges-Zavadsky}, M., {Furniss}, A., {et~al.} 2012,
  \mnras, 424, 2276

\bibitem[{{Furniss} {et~al.}(2013){Furniss}, {Williams}, {Danforth},
  {Fumagalli}, {Prochaska}, {Primack}, {Urry}, {Stocke}, {Filippenko}, \&
  {Neely}}]{2013ApJ...768L..31F}
{Furniss}, A., {Williams}, D.~A., {Danforth}, C., {et~al.} 2013, \apjl, 768,
  L31

\bibitem[{{Gabuzda} {et~al.}(2014){Gabuzda}, {Reichstein}, \&
  {O'Neill}}]{gabuzda14}
{Gabuzda}, D.~C., {Reichstein}, A.~R., \& {O'Neill}, E.~L. 2014, \mnras, 444,
  172

\bibitem[{{Ghisellini} {et~al.}(2005){Ghisellini}, {Tavecchio}, \&
  {Chiaberge}}]{2005A&A...432..401G}
{Ghisellini}, G., {Tavecchio}, F., \& {Chiaberge}, M. 2005, \aap, 432, 401

\bibitem[{{Giroletti} {et~al.}(2004){Giroletti}, {Giovannini}, {Feretti},
  {Cotton}, {Edwards}, {Lara}, {Marscher}, {Mattox}, {Piner}, \&
  {Venturi}}]{giroletti04}
{Giroletti}, M., {Giovannini}, G., {Feretti}, L., {et~al.} 2004, \apj, 600, 127

\bibitem[{Goodman \& Weare(2010)}]{goodman2010}
Goodman, J. \& Weare, J. 2010, Commun. Appl. Math. Comput. Sci., 5, 65

\bibitem[{{Gutierrez} {et~al.}(2006){Gutierrez}, {Badran}, {Bradbury},
  {Buckley}, {Celik}, {Chow}, {Cogan}, {Cui}, {Daniel}, {Falcone}, {Fegan},
  {Finley}, {Gillanders}, {Grube}, {Holder}, {Horan}, {Hughes}, {Jung},
  {Kieda}, {Kosack}, {Krawczynski}, {Krennrich}, {Lang}, {Le Bohec}, {Maier},
  {Moriarty}, {Perkins}, {Pohl}, {Quinn}, {Rebillot}, {Rose}, {Schroedter},
  {Sembroski}, {Wakely}, {Weekes}, {White}, {VERITAS Collaboration}, {Aller},
  {Aller}, {Charlot}, \& {Le Campion}}]{2006ApJ...644..742G}
{Gutierrez}, K., {Badran}, H.~M., {Bradbury}, S.~M., {et~al.} 2006, \apj, 644,
  742

\bibitem[{{H.E.S.S.~Collaboration} {et~al.}(2013){H.E.S.S.~Collaboration},
  {Abramowski}, {Acero}, {Aharonian}, {Ait Benkhali}, {Akhperjanian},
  {Ang{\"u}ner}, {Anton}, {Balenderan}, {Balzer}, \&
  et~al.}]{2013A&A...559A.136H}
{H.E.S.S.~Collaboration}, {Abramowski}, A., {Acero}, F., {et~al.} 2013, \aap,
  559, A136

\bibitem[{{Hodge} {et~al.}(2018){Hodge}, {Lister}, {Aller}, {Aller}, {Kovalev},
  {Pushkarev}, \& {Savolainen}}]{2018ApJ...862..151H}
{Hodge}, M.~A., {Lister}, M.~L., {Aller}, M.~F., {et~al.} 2018, \apj, 862, 151

\bibitem[{{Hovatta} {et~al.}(2016){Hovatta}, {Lindfors}, {Blinov}, {Pavlidou},
  {Nilsson}, {Kiehlmann}, {Angelakis}, {Fallah Ramazani}, {Liodakis},
  {Myserlis}, {Panopoulou}, \& {Pursimo}}]{2016A&A...596A..78H}
{Hovatta}, T., {Lindfors}, E., {Blinov}, D., {et~al.} 2016, \aap, 596, A78

\bibitem[{{Hughes} {et~al.}(1989){Hughes}, {Aller}, \&
  {Aller}}]{1989ApJ...341...68H}
{Hughes}, P.~A., {Aller}, H.~D., \& {Aller}, M.~F. 1989, \apj, 341, 68

\bibitem[{{Jannuzi} {et~al.}(1994){Jannuzi}, {Smith}, \&
  {Elston}}]{1994ApJ...428..130J}
{Jannuzi}, B.~T., {Smith}, P.~S., \& {Elston}, R. 1994, \apj, 428, 130

\bibitem[{{Jorstad} {et~al.}(2007){Jorstad}, {Marscher}, {Stevens}, {Smith},
  {Forster}, {Gear}, {Cawthorne}, {Lister}, {Stirling}, {G{\'o}mez}, {Greaves},
  \& {Robson}}]{2007AJ....134..799J}
{Jorstad}, S.~G., {Marscher}, A.~P., {Stevens}, J.~A., {et~al.} 2007, \aj, 134,
  799

\bibitem[{{Kalberla} {et~al.}(2005){Kalberla}, {Burton}, {Hartmann}, {Arnal},
  {Bajaja}, {Morras}, \& {P{\"o}ppel}}]{2005A&A...440..775K}
{Kalberla}, P.~M.~W., {Burton}, W.~B., {Hartmann}, D., {et~al.} 2005, \aap,
  440, 775

\bibitem[{{Kang} {et~al.}(2016){Kang}, {Zheng}, {Wu}, \&
  {Chen}}]{2016MNRAS.461.1862K}
{Kang}, S.-J., {Zheng}, Y.-G., {Wu}, Q., \& {Chen}, L. 2016, \mnras, 461, 1862

\bibitem[{{Konigl}(1981)}]{1981ApJ...243..700K}
{Konigl}, A. 1981, \apj, 243, 700

\bibitem[{{Krawczynski} {et~al.}(2004){Krawczynski}, {Hughes}, {Horan},
  {Aharonian}, {Aller}, {Aller}, {Boltwood}, {Buckley}, {Coppi}, {Fossati},
  {G{\"o}tting}, {Holder}, {Horns}, {Kurtanidze}, {Marscher}, {Nikolashvili},
  {Remillard}, {Sadun}, \& {Schr{\"o}der}}]{2004ApJ...601..151K}
{Krawczynski}, H., {Hughes}, S.~B., {Horan}, D., {et~al.} 2004, \apj, 601, 151

\bibitem[{{Lindfors} {et~al.}(2016){Lindfors}, {Hovatta}, {Nilsson},
  {Reinthal}, {Fallah Ramazani}, {Pavlidou}, {Max-Moerbeck}, {Richards},
  {Berdyugin}, {Takalo}, {Sillanp{\"a}{\"a}}, \&
  {Readhead}}]{2016A&A...593A..98L}
{Lindfors}, E.~J., {Hovatta}, T., {Nilsson}, K., {et~al.} 2016, \aap, 593, A98

\bibitem[{{Liodakis} {et~al.}(2018){Liodakis}, {Romani}, {Filippenko},
  {Kiehlmann}, {Max-Moerbeck}, {Readhead}, \& {Zheng}}]{2018MNRAS.480.5517L}
{Liodakis}, I., {Romani}, R.~W., {Filippenko}, A.~V., {et~al.} 2018, \mnras,
  480, 5517

\bibitem[{{Lister} {et~al.}(2009){Lister}, {Aller}, {Aller}, {Cohen}, {Homan},
  {Kadler}, {Kellermann}, {Kovalev}, {Ros}, {Savolainen}, {Zensus}, \&
  {Vermeulen}}]{2009AJ....137.3718L}
{Lister}, M.~L., {Aller}, H.~D., {Aller}, M.~F., {et~al.} 2009, \aj, 137, 3718

\bibitem[{{Lister} {et~al.}(2016){Lister}, {Aller}, {Aller}, {Homan},
  {Kellermann}, {Kovalev}, {Pushkarev}, {Richards}, {Ros}, \&
  {Savolainen}}]{2016AJ....152...12L}
{Lister}, M.~L., {Aller}, M.~F., {Aller}, H.~D., {et~al.} 2016, \aj, 152, 12

\bibitem[{{Lister} \& {Homan}(2005)}]{2005AJ....130.1389L}
{Lister}, M.~L. \& {Homan}, D.~C. 2005, \aj, 130, 1389

\bibitem[{{Lister} {et~al.}(2019){Lister}, {Homan}, {Hovatta}, {Kellermann},
  {Kiehlmann}, {Kovalev}, {Max-Moerbeck}, {Pushkarev}, {Readhead}, {Ros}, \&
  {Savolainen}}]{2019ApJ...874...43L}
{Lister}, M.~L., {Homan}, D.~C., {Hovatta}, T., {et~al.} 2019, \apj, 874, 43

\bibitem[{{Lyutikov} \& {Kravchenko}(2017)}]{2017MNRAS.467.3876L}
{Lyutikov}, M. \& {Kravchenko}, E.~V. 2017, \mnras, 467, 3876

\bibitem[{{Lyutikov} {et~al.}(2005){Lyutikov}, {Pariev}, \&
  {Gabuzda}}]{2005MNRAS.360..869L}
{Lyutikov}, M., {Pariev}, V.~I., \& {Gabuzda}, D.~C. 2005, \mnras, 360, 869

\bibitem[{{MAGIC Collaboration} {et~al.}(Submitted){MAGIC Collaboration},
  {Acciari}, {Ansoldi}, {Antonelli}, {Arbet Engels}, {Baack}, {Babi{\'c}},
  {Banerjee}, {Bangale}, {Barres de Almeida}, {Barrio}, {Becerra Gonz{\'a}lez},
  {Bednarek}, {Bernardini}, {Berti}, {Besenrieder}, {Bhattacharyya},
  {Bigongiari}, {Biland}, {Blanch}, {Bonnoli}, {Carosi}, {Ceribella}, {Cikota},
  {Colak}, {Colin}, {Colombo}, {Contreras}, \& {Cortina}}]{2344fact}
{MAGIC Collaboration}, {Acciari}, V.~A., {Ansoldi}, S., {et~al.} Submitted, to
  MNRAS

\bibitem[{{MAGIC Collaboration} {et~al.}(2019){MAGIC Collaboration}, {Acciari},
  {Ansoldi}, {Antonelli}, {Arbet Engels}, {Baack}, {Babi{\'c}}, {Banerjee},
  {Bangale}, {Barres de Almeida}, {Barrio}, {Becerra Gonz{\'a}lez}, {Bednarek},
  {Bernardini}, {Berti}, {Besenrieder}, {Bhattacharyya}, {Bigongiari},
  {Biland}, {Blanch}, {Bonnoli}, {Carosi}, {Ceribella}, {Cikota}, {Colak},
  {Colin}, {Colombo}, {Contreras}, {Cortina}, {Covino}, {D'Elia}, {da Vela},
  {Dazzi}, {de Angelis}, {de Lotto}, {Delfino}, {Delgado}, {di Pierro}, {Do
  Souto Espi{\~n}era}, {Dom{\'\i}nguez}, {Dominis Prester}, {Dorner}, {Doro},
  {Einecke}, {Elsaesser}, {Fallah Ramazani}, {Fattorini},
  {Fern{\'a}ndez-Barral}, {Ferrara}, {Fidalgo}, {Foffano}, {Fonseca}, {Font},
  {Fruck}, {Galindo}, {Gallozzi}, {Garc{\'\i}a L{\'o}pez}, {Garczarczyk},
  {Gaug}, {Giammaria}, {Godinovi{\'c}}, {Guberman}, {Hadasch}, {Hahn},
  {Hassan}, {Herrera}, {Hoang}, {Hrupec}, {Inoue}, {Ishio}, {Iwamura}, {Kubo},
  {Kushida}, {Kuve{\v{z}}di{\'c}}, {Lamastra}, {Lelas}, {Leone}, {Lindfors},
  {Lombardi}, {Longo}, {L{\'o}pez}, {L{\'o}pez-Oramas}, {Maggio}, {Majumdar},
  {Makariev}, {Maneva}, {Manganaro}, {Mannheim}, {Maraschi}, {Mariotti},
  {Mart{\'\i}nez}, {Masuda}, {Mazin}, {Minev}, {Miranda}, {Mirzoyan}, {Molina},
  {Moralejo}, {Moreno}, {Moretti}, {Munar-Adrover}, {Neustroev}, {Niedzwiecki},
  {Nievas Rosillo}, {Nigro}, {Nilsson}, {Ninci}, {Nishijima}, {Noda},
  {Nogu{\'e}s}, {N{\"o}the}, {Paiano}, {Palacio}, {Paneque}, {Paoletti},
  {Paredes}, {Pedaletti}, {Pe{\~n}il}, {Peresano}, {Persic}, {Prada Moroni},
  {Prand ini}, {Puljak}, {Garcia}, {Rhode}, {Rib{\'o}}, {Rico}, {Righi},
  {Rugliancich}, {Saha}, {Saito}, {Satalecka}, {Schweizer}, {Sitarek},
  {{\v{S}}nidari{\'c}}, {Sobczynska}, {Somero}, {Stamerra}, {Strzys},
  {Suri{\'c}}, {Tavecchio}, {Temnikov}, {Terzi{\'c}}, {Teshima},
  {Torres-Alb{\`a}}, {Tsujimoto}, {van Scherpenberg}, {Vanzo}, {Vazquez
  Acosta}, {Vovk}, {Will}, {Zari{\'c}}, {D'Ammando}, {Hada}, {Jorstad},
  {Marscher}, {Mobeen}, {Hovatta}, {Larionov}, {Borman}, {Grishina},
  {Kopatskaya}, {Morozova}, {Nikiforova}, {L{\"a}hteenm{\"a}ki}, {Tornikoski},
  \& {Agudo}}]{2019A&A...623A.175M}
{MAGIC Collaboration}, {Acciari}, V.~A., {Ansoldi}, S., {et~al.} 2019, \aap,
  623, A175

\bibitem[{{MAGIC Collaboration} {et~al.}(2020){MAGIC Collaboration}, {Acciari},
  {Ansoldi}, {Antonelli}, {Arbet Engels}, {Baack}, {Babi{\'c}}, {Banerjee},
  {Barres de Almeida}, {Barrio}, {Becerra Gonz{\'a}lez}, {Bednarek},
  {Bellizzi}, {Bernardini}, {Berti}, {Besenrieder}, {Bhattacharyya},
  {Bigongiari}, {Biland }, {Blanch}, {Bonnoli}, {Bosnjak}, {Busetto}, {Carosi},
  {Ceribella}, {Chai}, {Cikota}, {Colak}, {Colin}, {Colombo}, {Contreras},
  {Cortina}, {Covino}, {D'Elia}, {Da Vela}, {Dazzi}, {De Angelis}, {De Lotto},
  {Delfino}, {Delgado}, {Di Pierro}, {Souto Do Espi{\~n}eira}, {Dominis
  Prester}, {Donini}, {Dorner}, {Doro}, {Elsaesser}, {Fallah Ramazani},
  {Fattorini}, {Fern{\'a}ndez-Barral}, {Ferrara}, {Fidalgo}, {Foffano},
  {Fonseca}, {Font}, {Fruck}, {Fukami}, {Gallozzi}, {Garc{\'\i}a L{\'o}pez},
  {Garczarczyk}, {Gasparyan}, {Gaug}, {Godinovi{\'c}}, {Green}, {Guberman},
  {Hadasch}, {Hahn}, {Herrera}, {Hoang}, {Hrupec}, {Inada}, {Inoue}, {Ishio},
  {Iwamura}, {Jouvin}, {Kubo}, {Kushida}, {Lamastra}, {Lelas}, {Leone},
  {Lindfors}, {Lombardi}, {Longo}, {L{\'o}pez}, {L{\'o}pez-Coto},
  {L{\'o}pez-Oramas}, {Oliveira de Machado Fraga}, {Maggio}, {Majumdar},
  {Makariev}, {Mallamaci}, {Maneva}, {Manganaro}, {Mannheim}, {Maraschi},
  {Mariotti}, {Mart{\'\i}nez}, {Masuda}, {Mazin}, {Mi{\'c}anovi{\'c}},
  {Miceli}, {Minev}, {Miranda}, {Mirzoyan}, {Molina}, {Moralejo}, {Morcuende},
  {Moreno}, {Moretti}, {Munar-Adrover}, {Neustroev}, {Niedzwiecki}, {Nigro},
  {Nilsson}, {Ninci}, {Nishijima}, {Noda}, {Nogu{\'e}s}, {N{\"o}the}, {Nozaki},
  {Paiano}, {Palacio}, {Palatiello}, {Paneque}, {Paoletti}, {Paredes},
  {Pe{\~n}il}, {Peresano}, {Persic}, {Prada Moroni}, {Prand ini}, {Puljak},
  {Rhode}, {Rib{\'o}}, {Rico}, {Righi}, {Rugliancich}, {Saha}, {Sahakyan},
  {Saito}, {Sakurai}, {Satalecka}, {Schweizer}, {Sitarek},
  {{\v{S}}nidari{\'c}}, {Sobczynska}, {Somero}, {Stamerra}, {Strom}, {Strzys},
  {Suri{\'c}}, {Takahashi}, {Tavecchio}, {Temnikov}, {Terzi{\'c}}, {Teshima},
  {Torres-Alb{\`a}}, {Tsujimoto}, {van Scherpenberg}, {Vanzo}, {Vazquez
  Acosta}, {Vovk}, {Will}, {Zari{\'c}}, {Fermi-LAT Collaboration}, {:}, \&
  {Hayashida}}]{1959flare}
{MAGIC Collaboration}, {Acciari}, V.~A., {Ansoldi}, S., {et~al.} 2020, arXiv
  e-prints, arXiv:2002.00129

\bibitem[{{MAGIC Collaboration} {et~al.}(2018{\natexlab{a}}){MAGIC
  Collaboration}, {Ahnen}, {Ansoldi}, {Antonelli}, {Arcaro}, {Baack},
  {Babi{\'c}}, {Banerjee}, {Bangale}, {Barres de Almeida}, {Barrio}, {Becerra
  Gonz{\'a}lez}, {Bednarek}, {Bernardini}, {Ch Berse}, {Berti},
  {Bhattacharyya}, {Biland}, {Blanch}, {Bonnoli}, {Carosi}, {Carosi},
  {Ceribella}, {Chatterjee}, {Colak}, {Colin}, {Colombo}, {Contreras},
  {Cortina}, {Covino}, {Cumani}, {da Vela}, {Dazzi}, {de Angelis}, {de Lotto},
  {Delfino}, {Delgado}, {di Pierro}, {Dom{\'{\i}}nguez}, {Dominis Prester},
  {Dorner}, {Doro}, {Einecke}, {Elsaesser}, {Fallah Ramazani},
  {Fern{\'a}ndez-Barral}, {Fidalgo}, {Fonseca}, {Font}, {Fruck}, {Galindo},
  {Gallozzi}, {Garc{\'{\i}}a L{\'o}pez}, {Garczarczyk}, {Gaug}, {Giammaria},
  {Godinovi{\'c}}, {Gora}, {Guberman}, {Hadasch}, {Hahn}, {Hassan},
  {Hayashida}, {Herrera}, {Hose}, {Hrupec}, {Ishio}, {Konno}, {Kubo},
  {Kushida}, {Kuve{\v z}di{\'c}}, {Lelas}, {Lindfors}, {Lombardi}, {Longo},
  {L{\'o}pez}, {Maggio}, {Majumdar}, {Makariev}, {Maneva}, {Manganaro},
  {Mannheim}, {Maraschi}, {Mariotti}, {Mart{\'{\i}}nez}, {Masuda}, {Mazin},
  {Mielke}, {Minev}, {Miranda}, {Mirzoyan}, {Moralejo}, {Moreno}, {Moretti},
  {Nagayoshi}, {Neustroev}, {Niedzwiecki}, {Nievas Rosillo}, {Nigro},
  {Nilsson}, {Ninci}, {Nishijima}, {Noda}, {Nogu{\'e}s}, {Paiano}, {Palacio},
  {Paneque}, {Paoletti}, {Paredes}, {Pedaletti}, {Peresano}, {Persic}, {Prada
  Moroni}, {Prandini}, {Puljak}, {Garcia}, {Reichardt}, {Rhode}, {Rib{\'o}},
  {Rico}, {Righi}, {Rugliancich}, {Saito}, {Satalecka}, {Schweizer}, {Sitarek},
  {{\v S}nidari{\'c}}, {Sobczynska}, {Stamerra}, {Strzys}, {Suri{\'c}},
  {Takahashi}, {Takalo}, {Tavecchio}, {Temnikov}, {Terzi{\'c}}, {Teshima},
  {Torres-Alb{\`a}}, {Treves}, {Tsujimoto}, {Vanzo}, {Vazquez Acosta}, {Vovk},
  {Ward}, {Will}, {Zari{\'c}}, {Fermi-Lat Collaboration}, {Bastieri},
  {Gasparrini}, {Lott}, {Rani}, {Thompson}, {MWL Collaborators}, {Agudo},
  {Angelakis}, {Borman}, {Casadio}, {Grishina}, {Gurwell}, {Hovatta}, {Itoh},
  {J{\"a}rvel{\"a}}, {Jermak}, {Jorstad}, {Kopatskaya}, {Kraus}, {Krichbaum},
  {Kuin}, {L{\"a}hteenm{\"a}ki}, {Larionov}, {Larionova}, {Lien}, {Madejski},
  {Marscher}, {Myserlis}, {Max-Moerbeck}, {Molina}, {Morozova}, {Nalewajko},
  {Pearson}, {Ramakrishnan}, {Readhead}, {Reeves}, {Savchenko}, {Steele},
  {Tornikoski}, {Troitskaya}, {Troitsky}, {Vasilyev}, \&
  {Zensus}}]{2018A&A...619A..45M}
{MAGIC Collaboration}, {Ahnen}, M.~L., {Ansoldi}, S., {et~al.}
  2018{\natexlab{a}}, \aap, 619, A45

\bibitem[{{MAGIC Collaboration} {et~al.}(2018{\natexlab{b}}){MAGIC
  Collaboration}, {Ansoldi}, {Antonelli}, {Arcaro}, {Baack}, {Babi{\'c}},
  {Banerjee}, {Bangale}, {Barres de Almeida}, {Barrio}, {Becerra Gonz{\'a}lez},
  {Bednarek}, {Bernardini}, {Berse}, {Berti}, {Besenrieder}, {Bhattacharyya},
  {Bigongiari}, {Biland}, {Blanch}, {Bonnoli}, {Carosi}, {Ceribella},
  {Chatterjee}, {Colak}, {Colin}, {Colombo}, {Contreras}, {Cortina}, {Covino},
  {Cumani}, {D'Elia}, {da Vela}, {Dazzi}, {de Angelis}, {de Lotto}, {Delfino},
  {Delgado}, {di Pierro}, {Dom{\'{\i}}nguez}, {Dominis Prester}, {Dorner},
  {Doro}, {Einecke}, {Elsaesser}, {Fallah Ramazani}, {Fattorini},
  {Fern{\'a}ndez-Barral}, {Ferrara}, {Fidalgo}, {Foffano}, {Fonseca}, {Font},
  {Fruck}, {Gallozzi}, {Garc{\'{\i}}a L{\'o}pez}, {Garczarczyk}, {Gaug},
  {Giammaria}, {Godinovi{\'c}}, {Guberman}, {Hadasch}, {Hahn}, {Hassan},
  {Hayashida}, {Herrera}, {Hoang}, {Hrupec}, {Inoue}, {Ishio}, {Iwamura},
  {Konno}, {Kubo}, {Kushida}, {Lamastra}, {Lelas}, {Leone}, {Lindfors},
  {Lombardi}, {Longo}, {L{\'o}pez}, {Maggio}, {Majumdar}, {Makariev}, {Maneva},
  {Manganaro}, {Mannheim}, {Maraschi}, {Mariotti}, {Mart{\'{\i}}nez}, {Masuda},
  {Mazin}, {Mielke}, {Minev}, {Miranda}, {Mirzoyan}, {Moralejo}, {Moreno},
  {Moretti}, {Neustroev}, {Niedzwiecki}, {Nievas Rosillo}, {Nigro}, {Nilsson},
  {Ninci}, {Nishijima}, {Noda}, {Nogu{\'e}s}, {Paiano}, {Palacio}, {Paneque},
  {Paoletti}, {Paredes}, {Pedaletti}, {Pe{\~n}il}, {Peresano}, {Persic},
  {Pfrang}, {Prada Moroni}, {Prandini}, {Puljak}, {Garcia}, {Rhode},
  {Rib{\'o}}, {Rico}, {Righi}, {Rugliancich}, {Saha}, {Saito}, {Satalecka},
  {Schweizer}, {Sitarek}, {{\v S}nidari{\'c}}, {Sobczynska}, {Stamerra},
  {Strzys}, {Suri{\'c}}, {Tavecchio}, {Temnikov}, {Terzi{\'c}}, {Teshima},
  {Torres-Alb{\`a}}, {Tsujimoto}, {Vanzo}, {Vazquez Acosta}, {Vovk}, {Ward},
  {Will}, {Zari{\'c}}, {Ciprini}, {Fermi-LAT Collaboration}, {Desiante},
  {Barcewicz}, {Hovatta}, {Jormanainen}, {Takalo}, {Reinthal}, {Mankuzhiyil},
  {Wierda}, {L{\"a}hteenm{\"a}ki}, {Tammi}, {Tornikoski}, {Vera}, {Kiehlmann},
  {Max-Moerbeck}, \& {Readhead}}]{2018MNRAS.480..879M}
{MAGIC Collaboration}, {Ansoldi}, S., {Antonelli}, L.~A., {et~al.}
  2018{\natexlab{b}}, \mnras, 480, 879

\bibitem[{{Mannheim} \& {Biermann}(1989)}]{1989A&A...221..211M}
{Mannheim}, K. \& {Biermann}, P.~L. 1989, \aap, 221, 211

\bibitem[{{Maraschi} {et~al.}(1992){Maraschi}, {Ghisellini}, \&
  {Celotti}}]{1992ApJ...397L...5M}
{Maraschi}, L., {Ghisellini}, G., \& {Celotti}, A. 1992, \apjl, 397, L5

\bibitem[{{Maraschi} \& {Tavecchio}(2003)}]{2003ApJ...593..667M}
{Maraschi}, L. \& {Tavecchio}, F. 2003, \apj, 593, 667

\bibitem[{{Marscher} \& {Gear}(1985)}]{1985ApJ...298..114M}
{Marscher}, A.~P. \& {Gear}, W.~K. 1985, \apj, 298, 114

\bibitem[{{Max-Moerbeck} {et~al.}(2014{\natexlab{a}}){Max-Moerbeck}, {Hovatta},
  {Richards}, {King}, {Pearson}, {Readhead}, {Reeves}, {Shepherd}, {Stevenson},
  {Angelakis}, {Fuhrmann}, {Grainge}, {Pavlidou}, {Romani}, \&
  {Zensus}}]{2014MNRAS.445..428M}
{Max-Moerbeck}, W., {Hovatta}, T., {Richards}, J.~L., {et~al.}
  2014{\natexlab{a}}, \mnras, 445, 428

\bibitem[{{Max-Moerbeck} {et~al.}(2014{\natexlab{b}}){Max-Moerbeck},
  {Richards}, {Hovatta}, {Pavlidou}, {Pearson}, \&
  {Readhead}}]{2014MNRAS.445..437M}
{Max-Moerbeck}, W., {Richards}, J.~L., {Hovatta}, T., {et~al.}
  2014{\natexlab{b}}, \mnras, 445, 437

\bibitem[{{Moralejo} {et~al.}(2009){Moralejo}, {Gaug}, {Carmona}, {Colin},
  {Delgado}, {Lombardi}, {Mazin}, {Scalzotto}, {Sitarek}, {Tescaro}, \& {for
  the MAGIC collaboration}}]{2009arXiv0907.0943M}
{Moralejo}, A., {Gaug}, M., {Carmona}, E., {et~al.} 2009, ArXiv e-prints
  [\eprint[arXiv]{0907.0943}]

\bibitem[{{Nagai} {et~al.}(2014){Nagai}, {Haga}, {Giovannini}, {Doi},
  {Orienti}, {D'Ammando}, {Kino}, {Nakamura}, {Asada}, {Hada}, \&
  {Giroletti}}]{nagai14}
{Nagai}, H., {Haga}, T., {Giovannini}, G., {et~al.} 2014, \apj, 785, 53

\bibitem[{{Nalewajko} \& {Gupta}(2017)}]{2017A&A...606A..44N}
{Nalewajko}, K. \& {Gupta}, M. 2017, \aap, 606, A44

\bibitem[{{Nalewajko} {et~al.}(2014){Nalewajko}, {Sikora}, \&
  {Begelman}}]{2014ApJ...796L...5N}
{Nalewajko}, K., {Sikora}, M., \& {Begelman}, M.~C. 2014, \apjl, 796, L5

\bibitem[{{Nilsson} {et~al.}(2018){Nilsson}, {Lindfors}, {Takalo}, {Reinthal},
  {Berdyugin}, {Sillanp{\"a}{\"a}}, {Ciprini}, {Halkola}, {Hein{\"a}m{\"a}ki},
  {Hovatta}, {Kadenius}, {Nurmi}, {Ostorero}, {Pasanen}, {Rekola}, {Saarinen},
  {Sainio}, {Tuominen}, {Villforth}, {Vornanen}, \&
  {Zaprudin}}]{2018A&A...620A.185N}
{Nilsson}, K., {Lindfors}, E., {Takalo}, L.~O., {et~al.} 2018, \aap, 620, A185

\bibitem[{{Nilsson} {et~al.}(2007){Nilsson}, {Pasanen}, {Takalo}, {Lindfors},
  {Berdyugin}, {Ciprini}, \& {Pforr}}]{2007A&A...475..199N}
{Nilsson}, K., {Pasanen}, M., {Takalo}, L.~O., {et~al.} 2007, \aap, 475, 199

\bibitem[{{Nishiyama}(1999)}]{1999ICRC....3..370N}
{Nishiyama}, T. 1999, in International Cosmic Ray Conference, Vol.~3, 26th
  International Cosmic Ray Conference (ICRC26), Volume 3, 370

\bibitem[{{Ong}(2009)}]{2009ATel.2084....1O}
{Ong}, R.~A. 2009, The Astronomer's Telegram, 2084

\bibitem[{{Paiano} {et~al.}(2017){Paiano}, {Landoni}, {Falomo}, {Treves},
  {Scarpa}, \& {Righi}}]{2017ApJ...837..144P}
{Paiano}, S., {Landoni}, M., {Falomo}, R., {et~al.} 2017, \apj, 837, 144

\bibitem[{{Patel} {et~al.}(2018){Patel}, {Shukla}, {Chitnis}, {Dorner},
  {Mannheim}, {Acharya}, \& {Nagare}}]{2018A&A...611A..44P}
{Patel}, S.~R., {Shukla}, A., {Chitnis}, V.~R., {et~al.} 2018, \aap, 611, A44

\bibitem[{{Petry} {et~al.}(2000){Petry}, {B{\"o}ttcher}, {Connaughton},
  {Lahteenmaki}, {Pursimo}, {Raiteri}, {Schr{\"o}der}, {Sillanp{\"a}{\"a}},
  {Sobrito}, {Takalo}, {Ter{\"a}sranta}, {Tosti}, \&
  {Villata}}]{2000ApJ...536..742P}
{Petry}, D., {B{\"o}ttcher}, M., {Connaughton}, V., {et~al.} 2000, \apj, 536,
  742

\bibitem[{{Piner} \& {Edwards}(2004)}]{2004ApJ...600..115P}
{Piner}, B.~G. \& {Edwards}, P.~G. 2004, \apj, 600, 115

\bibitem[{{Piner} \& {Edwards}(2018)}]{2018ApJ...853...68P}
{Piner}, B.~G. \& {Edwards}, P.~G. 2018, \apj, 853, 68

\bibitem[{{Piner} {et~al.}(2008){Piner}, {Pant}, \&
  {Edwards}}]{2008ApJ...678...64P}
{Piner}, B.~G., {Pant}, N., \& {Edwards}, P.~G. 2008, \apj, 678, 64

\bibitem[{{Piner} {et~al.}(2010){Piner}, {Pant}, \&
  {Edwards}}]{2010ApJ...723.1150P}
{Piner}, B.~G., {Pant}, N., \& {Edwards}, P.~G. 2010, \apj, 723, 1150

\bibitem[{{Piner} {et~al.}(2009){Piner}, {Pant}, {Edwards}, \&
  {Wiik}}]{piner09}
{Piner}, B.~G., {Pant}, N., {Edwards}, P.~G., \& {Wiik}, K. 2009, \apjl, 690,
  L31

\bibitem[{{Poole} {et~al.}(2008){Poole}, {Breeveld}, {Page}, {Landsman},
  {Holland}, {Roming}, {Kuin}, {Brown}, {Gronwall}, {Hunsberger}, {Koch},
  {Mason}, {Schady}, {vanden Berk}, {Blustin}, {Boyd}, {Broos}, {Carter},
  {Chester}, {Cucchiara}, {Hancock}, {Huckle}, {Immler}, {Ivanushkina},
  {Kennedy}, {Marshall}, {Morgan}, {Pandey}, {de Pasquale}, {Smith}, \&
  {Still}}]{2008MNRAS.383..627P}
{Poole}, T.~S., {Breeveld}, A.~A., {Page}, M.~J., {et~al.} 2008, \mnras, 383,
  627

\bibitem[{{Prokoph} {et~al.}(2015){Prokoph}, {Schultz}, \& {Da
  Vela}}]{2015ICRC...34..864P}
{Prokoph}, H., {Schultz}, C., \& {Da Vela}, P. 2015, in International Cosmic
  Ray Conference, Vol.~34, 34th International Cosmic Ray Conference (ICRC2015),
  864

\bibitem[{{Pushkarev} {et~al.}(2005){Pushkarev}, {Gabuzda}, {Vetukhnovskaya},
  \& {Yakimov}}]{pushkarev05}
{Pushkarev}, A.~B., {Gabuzda}, D.~C., {Vetukhnovskaya}, Y.~N., \& {Yakimov},
  V.~E. 2005, \mnras, 356, 859

\bibitem[{{Pushkarev} {et~al.}(2012){Pushkarev}, {Hovatta}, {Kovalev},
  {Lister}, {Lobanov}, {Savolainen}, \& {Zensus}}]{2012A&A...545A.113P}
{Pushkarev}, A.~B., {Hovatta}, T., {Kovalev}, Y.~Y., {et~al.} 2012, \aap, 545,
  A113

\bibitem[{{Rees}(1967)}]{1967MNRAS.135..345R}
{Rees}, M.~J. 1967, \mnras, 135, 345

\bibitem[{{Reimer} {et~al.}(2005){Reimer}, {B{\"o}ttcher}, \&
  {Postnikov}}]{2005ApJ...630..186R}
{Reimer}, A., {B{\"o}ttcher}, M., \& {Postnikov}, S. 2005, \apj, 630, 186

\bibitem[{{Richards} {et~al.}(2011){Richards}, {Max-Moerbeck}, {Pavlidou},
  {King}, {Pearson}, {Readhead}, {Reeves}, {Shepherd}, {Stevenson},
  {Weintraub}, {Fuhrmann}, {Angelakis}, {Zensus}, {Healey}, {Romani}, {Shaw},
  {Grainge}, {Birkinshaw}, {Lancaster}, {Worrall}, {Taylor}, {Cotter}, \&
  {Bustos}}]{2011ApJS..194...29R}
{Richards}, J.~L., {Max-Moerbeck}, W., {Pavlidou}, V., {et~al.} 2011, \apjs,
  194, 29

\bibitem[{{Rovero} {et~al.}(2016){Rovero}, {Muriel}, {Donzelli}, \&
  {Pichel}}]{2016A&A...589A..92R}
{Rovero}, A.~C., {Muriel}, H., {Donzelli}, C., \& {Pichel}, A. 2016, \aap, 589,
  A92

\bibitem[{{Sahu} {et~al.}(2013){Sahu}, {Oliveros}, \&
  {Sanabria}}]{2013PhRvD..87j3015S}
{Sahu}, S., {Oliveros}, A.~F.~O., \& {Sanabria}, J.~C. 2013, \prd, 87, 103015

\bibitem[{{Savolainen} {et~al.}(2002){Savolainen}, {Wiik}, {Valtaoja},
  {Jorstad}, \& {Marscher}}]{2002A&A...394..851S}
{Savolainen}, T., {Wiik}, K., {Valtaoja}, E., {Jorstad}, S.~G., \& {Marscher},
  A.~P. 2002, \aap, 394, 851

\bibitem[{{Scarpa} {et~al.}(2000){Scarpa}, {Urry}, {Falomo}, {Pesce}, \&
  {Treves}}]{2000ApJ...532..740S}
{Scarpa}, R., {Urry}, C.~M., {Falomo}, R., {Pesce}, J.~E., \& {Treves}, A.
  2000, \apj, 532, 740

\bibitem[{{Schlafly} \& {Finkbeiner}(2011)}]{2011ApJ...737..103S}
{Schlafly}, E.~F. \& {Finkbeiner}, D.~P. 2011, \apj, 737, 103

\bibitem[{{Stern}(2003)}]{2003MNRAS.345..590S}
{Stern}, B.~E. 2003, \mnras, 345, 590

\bibitem[{{Stickel} {et~al.}(1991){Stickel}, {Padovani}, {Urry}, {Fried}, \&
  {Kuehr}}]{1991ApJ...374..431S}
{Stickel}, M., {Padovani}, P., {Urry}, C.~M., {Fried}, J.~W., \& {Kuehr}, H.
  1991, \apj, 374, 431

\bibitem[{{Stocke} {et~al.}(1991){Stocke}, {Morris}, {Gioia}, {Maccacaro},
  {Schild}, {Wolter}, {Fleming}, \& {Henry}}]{1991ApJS...76..813S}
{Stocke}, J.~T., {Morris}, S.~L., {Gioia}, I.~M., {et~al.} 1991, \apjs, 76, 813

\bibitem[{{Tagliaferri} {et~al.}(2008){Tagliaferri}, {Foschini}, {Ghisellini},
  {Maraschi}, {Albert}, {Aliu}, {Anderhub}, {Antoranz}, {Baixeras}, {Barrio},
  {Bartko}, {Bastieri}, {Becker}, {Bednarek}, {Bedyugin}, {Berger},
  {Bigongiari}, {Biland}, {Bock}, {Bordas}, {Bosch-Ramon}, {Bretz},
  {Britvitch}, {Camara}, {Carmona}, {Chilingarian}, {Coarasa}, {Commichau},
  {Contreras}, {Cortina}, {Costado}, {Curtef}, {Danielyan}, {Dazzi}, {De
  Angelis}, {Delgado}, {de los Reyes}, {De Lotto}, {Dorner}, {Doro}, {Errando},
  {Fagiolini}, {Ferenc}, {Fern{\'a}ndez}, {Firpo}, {Fonseca}, {Font}, {Fuchs},
  {Galante}, {Garc{\'{\i}}a-L{\'o}pez}, {Garczarczyk}, {Gaug}, {Giller},
  {Goebel}, {Hakobyan}, {Hayashida}, {Hengstebeck}, {Herrero}, {H{\"o}hne},
  {Hose}, {Huber}, {Hsu}, {Jacon}, {Jogler}, {Kosyra}, {Kranich}, {Kritzer},
  {Laille}, {Lindfors}, {Lombardi}, {Longo}, {L{\'o}pez}, {Lorenz}, {Majumdar},
  {Maneva}, {Mannheim}, {Mariotti}, {Mart{\'{\i}}nez}, {Mazin}, {Merck},
  {Meucci}, {Meyer}, {Miranda}, {Mirzoyan}, {Mizobuchi}, {Moralejo}, {Nieto},
  {Nilsson}, {Ninkovic}, {O{\~n}a-Wilhelmi}, {Otte}, {Oya}, {Panniello},
  {Paoletti}, {Paredes}, {Pasanen}, {Pascoli}, {Pauss}, {Pegna}, {Persic},
  {Peruzzo}, {Piccioli}, {Prandini}, {Puchades}, {Raymers}, {Rhode},
  {Rib{\'o}}, {Rico}, {Rissi}, {Robert}, {R{\"u}gamer}, {Saggion}, {Saito},
  {S{\'a}nchez}, {Sartori}, {Scalzotto}, {Scapin}, {Schmitt}, {Schweizer},
  {Shayduk}, {Shinozaki}, {Shore}, {Sidro}, {Sillanp{\"a}{\"a}}, {Sobczynska},
  {Spanier}, {Stamerra}, {Stark}, {Takalo}, {Tavecchio}, {Temnikov}, {Tescaro},
  {Teshima}, {Torres}, {Turini}, {Vankov}, {Venturini}, {Vitale}, {Wagner},
  {Wibig}, {Wittek}, {Zandanel}, {Zanin}, {Zapatero}, \& {MAGIC
  Collaboration``}}]{2008ApJ...679.1029T}
{Tagliaferri}, G., {Foschini}, L., {Ghisellini}, G., {et~al.} 2008, \apj, 679,
  1029

\bibitem[{{Tagliaferri} {et~al.}(2003){Tagliaferri}, {Ravasio}, {Ghisellini},
  {Tavecchio}, {Giommi}, {Massaro}, {Nesci}, \& {Tosti}}]{2003A&A...412..711T}
{Tagliaferri}, G., {Ravasio}, M., {Ghisellini}, G., {et~al.} 2003, \aap, 412,
  711

\bibitem[{{Tavecchio} {et~al.}(2011){Tavecchio}, {Becerra-Gonzalez},
  {Ghisellini}, {Stamerra}, {Bonnoli}, {Foschini}, \&
  {Maraschi}}]{2011A&A...534A..86T}
{Tavecchio}, F., {Becerra-Gonzalez}, J., {Ghisellini}, G., {et~al.} 2011, \aap,
  534, A86

\bibitem[{{Tavecchio} \& {Ghisellini}(2016)}]{2016MNRAS.456.2374T}
{Tavecchio}, F. \& {Ghisellini}, G. 2016, \mnras, 456, 2374

\bibitem[{{Tavecchio} {et~al.}(1998){Tavecchio}, {Maraschi}, \&
  {Ghisellini}}]{1998ApJ...509..608T}
{Tavecchio}, F., {Maraschi}, L., \& {Ghisellini}, G. 1998, \apj, 509, 608

\bibitem[{{Teshima}(2009)}]{2009ATel.2098....1T}
{Teshima}, M. 2009, The Astronomer's Telegram, 2098

\bibitem[{{The Fermi-LAT collaboration}(2019)}]{2019arXiv190210045T}
{The Fermi-LAT collaboration}. 2019, arXiv e-prints, arXiv:1902.10045

\bibitem[{{Tiet} {et~al.}(2012){Tiet}, {Piner}, \&
  {Edwards}}]{2012arXiv1205.2399T}
{Tiet}, V.~C., {Piner}, B.~G., \& {Edwards}, P.~G. 2012, arXiv e-prints
  [\eprint[arXiv]{1205.2399}]

\bibitem[{{Valtaoja} {et~al.}(1988){Valtaoja}, {Haarala}, {Lehto}, {Valtaoja},
  {Valtonen}, {Moiseev}, {Nesterov}, {Salonen}, {Terasranta}, {Urpo}, \&
  {Tiuri}}]{1988A&A...203....1V}
{Valtaoja}, E., {Haarala}, S., {Lehto}, H., {et~al.} 1988, \aap, 203, 1

\bibitem[{{Valtaoja} {et~al.}(1991){Valtaoja}, {Sillanpaa}, {Valtaoja},
  {Shakhovskoi}, \& {Efimov}}]{valtaoja91}
{Valtaoja}, L., {Sillanpaa}, A., {Valtaoja}, E., {Shakhovskoi}, N.~M., \&
  {Efimov}, I.~S. 1991, \aj, 101, 78

\bibitem[{{van den Berg} {et~al.}(2019){van den Berg}, {B{\"o}ttcher},
  {Dom{\'\i}nguez}, \& {L{\'o}pez-Moya}}]{2019ApJ...874...47V}
{van den Berg}, J.~P., {B{\"o}ttcher}, M., {Dom{\'\i}nguez}, A., \&
  {L{\'o}pez-Moya}, M. 2019, \apj, 874, 47

\bibitem[{{Villforth} {et~al.}(2010){Villforth}, {Nilsson}, {Heidt}, {Takalo},
  {Pursimo}, {Berdyugin}, {Lindfors}, {Pasanen}, {Winiarski}, {Drozdz},
  {Ogloza}, {Kurpinska-Winiarska}, {Siwak}, {Koziel-Wierzbowska}, {Porowski},
  {Kuzmicz}, {Krzesinski}, {Kundera}, {Wu}, {Zhou}, {Efimov}, {Sadakane},
  {Kamada}, {Ohlert}, {Hentunen}, {Nissinen}, {Dietrich}, {Assef}, {Atlee},
  {Bird}, {Depoy}, {Eastman}, {Peeples}, {Prieto}, {Watson}, {Yee}, {Liakos},
  {Niarchos}, {Gazeas}, {Dogru}, {Donmez}, {Marchev}, {Coggins-Hill},
  {Mattingly}, {Keel}, {Haque}, {Aungwerojwit}, \&
  {Bergvall}}]{2010MNRAS.402.2087V}
{Villforth}, C., {Nilsson}, K., {Heidt}, J., {et~al.} 2010, \mnras, 402, 2087

\bibitem[{{Virtanen} \& {Vainio}(2005)}]{2005ApJ...621..313V}
{Virtanen}, J. J.~P. \& {Vainio}, R. 2005, \apj, 621, 313

\bibitem[{{Welsh}(1999)}]{1999PASP..111.1347W}
{Welsh}, W.~F. 1999, \pasp, 111, 1347

\bibitem[{{Wood} {et~al.}(2017){Wood}, {Caputo}, {Charles}, {Di Mauro},
  {Magill}, {Perkins}, \& {Fermi-LAT Collaboration}}]{2017ICRC...35..824W}
{Wood}, M., {Caputo}, R., {Charles}, E., {et~al.} 2017, in International Cosmic
  Ray Conference, Vol. 301, 35th International Cosmic Ray Conference
  (ICRC2017), 824

\bibitem[{Zanin {et~al.}(2013)Zanin, Carmona, Sitarek, Colin, Frantzen, Gaug,
  Lombardi, Lopez, Moralejo, Satalecka, Scapin, \& Stamatescu}]{zanin13}
Zanin, R., Carmona, E., Sitarek, J., {et~al.} 2013, in {Proceedings of the 33rd
  International Cosmic Ray Conference (ICRC2013): Rio de Janeiro, Brazil, July
  2-9, 2013}, 0773

\end{thebibliography}
